\documentclass{egpubl_arxiv}
\usepackage{eg2022}
\usepackage{xcolor}
\usepackage[T1]{fontenc}
\usepackage{dfadobe}  

\usepackage{cite}  
\BibtexOrBiblatex
\electronicVersion
\PrintedOrElectronic
\ifpdf \usepackage[pdftex]{graphicx} \pdfcompresslevel=9
\else \usepackage[dvips]{graphicx} \fi

\usepackage{egweblnk} 
\usepackage{amsfonts}
\usepackage{csquotes}
\usepackage{pifont} %
\usepackage{booktabs}
\usepackage{multirow}
\usepackage{adjustbox}
\usepackage{fontawesome} 
\usepackage{amsmath} 
\usepackage{caption}
\usepackage{subcaption}

\usepackage{cleveref}
\usepackage{comment}
\usepackage{adjustbox}

\definecolor{cgray}{rgb}{0.5, 0.5, 0.5}
\newcommand{\darkgrayed}[1]{\textcolor{cgray}{#1}}
\makeatletter
\newcommand*\titleheader[1]{\gdef\@titleheader{#1}}
\AtBeginDocument{
  \let\st@red@title\@title
  \def\@title{
    \vskip-1.3em
    \bgroup\normalfont\small\centering\@titleheader\par\egroup
    \vskip1.5em\st@red@title}
}
\makeatother

\titleheader{\darkgrayed{This state of the art report is accepted at EUROGRAPHICS 2022.}}

\title[Advances in Neural Rendering]{Advances in Neural Rendering}

\begin{document}

\author[A. Tewari \& J. Thies \& B. Mildenhall \& P. Srinivasan et al.]
{
\parbox{\textwidth}
{\centering \vspace{-0.5cm}
A. Tewari$^{1,6\star}$~~J. Thies$^{2\star}$~~B. Mildenhall$^{3\star}$~~P. Srinivasan$^{3\star}$~~E. Tretschk$^{1}$~~W. Yifan$^{4,8}$~~C. Lassner$^{5}$~~V. Sitzmann$^{6}$~~R. Martin-Brualla$^{3}$ \\
S. Lombardi$^{5}$~~T. Simon$^{5}$~~C. Theobalt$^{1}$~~M. Nie{\ss}ner$^{7}$~~J. T. Barron$^{3}$~~G. Wetzstein$^{8}$~~M. Zollh{\"o}fer$^{5}$~~V. Golyanik$^{1}$
}
\\
\parbox{\textwidth}
{\centering
$^{1}$MPI for Informatics~~$^{2}$MPI for Intelligent Systems~~$^{3}$Google Research~~$^{4}$ETH Z\"urich~~$^{5}$Reality Labs Research 
\\$^{6}$MIT~~$^{7}$Technical University of Munich~~$^{8}$Stanford University~~$^{\star}$Equal contribution.
}
}

    \newcommand{\etal       }     {{et~al.}}
\newcommand{\apriori    }     {\textit{a~priori}}
\newcommand{\aposteriori}     {\textit{a~posteriori}}
\newcommand{\perse      }     {\textit{per~se}}
\newcommand{\eg         }     {{e.g.~}}
\newcommand{\Eg         }     {{E.g.~}}
\newcommand{\ie         }     {{i.e.~}}
\newcommand{\naive      }     {{na\"{\i}ve}}

\definecolor{darkred}{rgb}{0.7,0.1,0.1}
\definecolor{darkgreen}{rgb}{0.1,0.5,0.1}
\definecolor{cyan}{rgb}{0.7,0.0,0.7}
\definecolor{otherblue}{rgb}{0.1,0.4,0.8}
\definecolor{maroon}{rgb}{0.76,.13,.28}
\definecolor{burntorange}{rgb}{0.81,.33,0}
\definecolor{othergreen}{rgb}{0.29,0.49,0.07}
\definecolor{orange}{rgb}{1.0,0.65,0.0}

\newcommand*\rot{\rotatebox{90}}

\definecolor{cmarkcolor}{rgb}{0.49,0.74,0.49}
\definecolor{xmarkcolor}{rgb}{0.86,0.34,0.34}
\newcommand{\cmark}{\textcolor{cmarkcolor}{\ding{51}}}
\newcommand{\xmark}{\textcolor{xmarkcolor}{\ding{55}}}

\definecolor{opA}{rgb}{0.9,0.6,0.0}
\definecolor{opB}{rgb}{0.35,0.70,0.90}
\definecolor{opC}{rgb}{0.8,0.40,0.0}
\definecolor{opD}{rgb}{0.0,0.60,0.50} %
\definecolor{opE}{rgb}{0.8,0.6,0.7}
\definecolor{opF}{rgb}{0.,0.45,0.70} 

\newcommand{\tableopt}[2]{{\color{#1}\textbf{#2}}}

\newcommand {\opA}[1]{\hyperlink{in_out_options}{\tableopt{opA}{#1}}}
\newcommand {\opB}[1]{\hyperlink{contents_options}{\tableopt{opB}{#1}}}
\newcommand {\opC}[1]{\hyperlink{control_options}{\tableopt{opC}{#1}}}
\newcommand {\opD}[1]{\hyperlink{cg_options}{\tableopt{opD}{#1}}}

\newcommand {\opE}[1]{\hyperlink{control_options}{\tableopt{opE}{#1}}}
\newcommand {\opF}[1]{\hyperlink{cg_options}{\tableopt{opF}{#1}}}

\newcommand {\opG}[1]{\tableopt{cyan}{#1}}
\newcommand {\opH}[1]{\tableopt{orange}{#1}}
\newcommand {\opI}[1]{\tableopt{magenta}{#1}}
\newcommand {\opJ}[1]{\tableopt{brown}{#1}}
\newcommand {\opK}[1]{\tableopt{gray}{#1}}
\newcommand {\opL}[1]{\tableopt{lightgray}{#1}}
\newcommand {\opM}[1]{\tableopt{lime}{#1}}
\newcommand {\opN}[1]{\tableopt{darkgray}{#1}}

\newcommand {\opO}[1]{\tableopt{olive}{#1}}
\newcommand {\opP}[1]{\tableopt{pink}{#1}}
\newcommand {\opQ}[1]{\tableopt{purple}{#1}}
\newcommand {\opR}[1]{\tableopt{teal}{#1}}
\newcommand {\opS}[1]{\tableopt{violet}{#1}}
\newcommand {\opT}[1]{\tableopt{violet}{#1}}

\def\code#1{
    \ifx&#1&
        \xmark{}
    \else
        {\href{#1}{\faExternalLink}}
    \fi
}

    \teaser{
  \vspace{-1.2cm}
  \includegraphics[width=\linewidth]{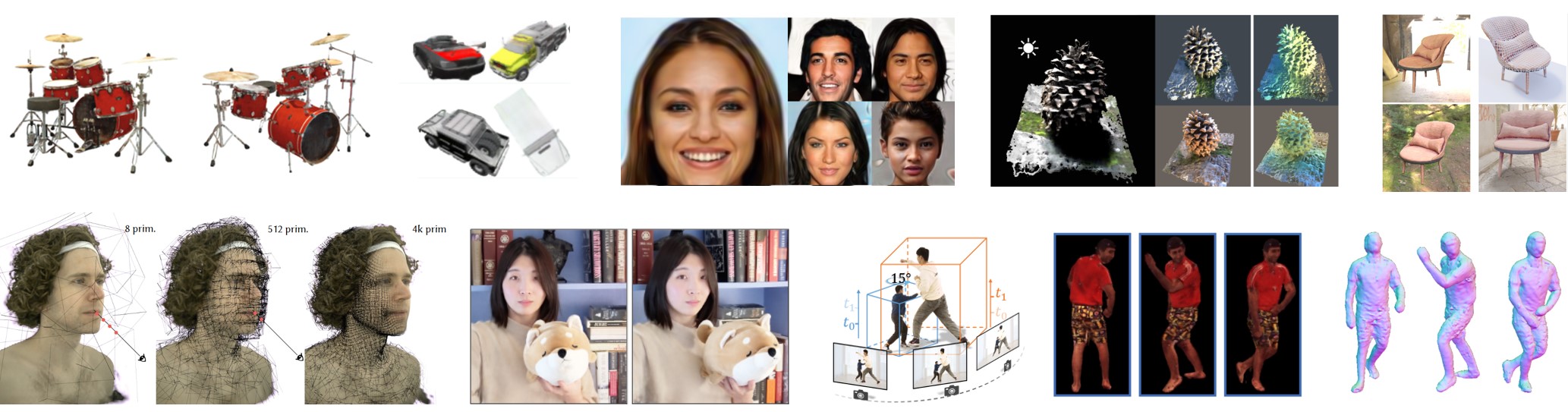}
  \centering
  \vspace{-0.8cm}
  \caption{
    This state-of-the-art report discusses a large variety of neural rendering methods which enable applications such as novel-view synthesis of static and dynamic scenes, generative modeling of objects, and scene relighting. 
    See \Cref{sec:applications} for more details on the various methods. Images adapted from \cite{Mildenhall_2020_NeRF,Trevithick2020,chanmonteiro2020piGAN,zhang2021nerfactor,boss2020nerd,Lombardi_2021_MVP,park2021nerfies,zhang2021stnerf,peng2021animatable} \textcopyright 2021 IEEE. 
  }
  \vspace{0.3cm}
  \label{fig:teaser}
}

    \maketitle
    \begin{abstract}
Synthesizing photo-realistic images and videos is at the heart of computer graphics and has been the focus of decades of research.
Traditionally, synthetic images of a scene are generated using rendering algorithms such as rasterization or ray tracing, which take specifically defined representations of geometry and material properties as input.
Collectively, these inputs define the actual scene and what is rendered, and are referred to as the \emph{scene representation} (where a scene consists of one or more objects). 
Example scene representations are triangle meshes with accompanied textures (e.g., created by an artist), point clouds (e.g., from a depth sensor), volumetric grids (e.g., from a CT scan), or implicit surface functions (e.g., truncated signed distance fields).
The reconstruction of such a scene representation from observations using  differentiable rendering losses is known as \emph{inverse graphics} or \emph{inverse rendering}.
Neural rendering is closely related, and combines ideas from  classical computer graphics and machine learning to create algorithms for synthesizing images from real-world observations. 
Neural rendering is a leap forward towards the goal of synthesizing photo-realistic image and video content.
In recent years, we have seen immense progress in this field through hundreds of publications that show different ways to inject learnable components into the rendering pipeline.
This state-of-the-art report on advances in neural rendering  focuses on methods that combine classical rendering principles with learned 3D scene representations, often now referred to as neural scene representations. 
A key advantage of these methods is that they are 3D-consistent by design, enabling applications such as novel viewpoint synthesis of a captured scene.
In addition to methods that handle static scenes, we cover neural scene representations for modeling non-rigidly deforming objects and scene editing and composition. 
While most of these approaches are scene-specific, we also discuss techniques that generalize across object classes and can be used for generative tasks.
In addition to reviewing these state-of-the-art methods, we provide an overview of fundamental concepts and definitions used in the current literature.
We conclude with a discussion on open challenges and social implications.
\end{abstract}

    \section{Introduction}
\label{sec:intro}
Synthesis of controllable and photo-realistic images and videos is one of the fundamental goals of computer graphics.
During the last decades, methods and representations have been developed to mimic the image formation model of real cameras, including the handling of complex materials and global illumination.
These methods are based on the laws of physics and simulate the light transport from light sources to the virtual camera for synthesis.
To this end, all physical parameters of the scene have to be known for the rendering process.
These parameters, for example, contain information about the scene geometry and material properties such as reflectivity or opacity.
Given this information, modern ray tracing techniques can generate photo-real imagery.
Besides the physics-based rendering methods, there is a variety of techniques that approximate the real-world image formation model.
These methods are based on mathematical approximations (e.g., a piece-wise linear approximation of the surface; i.e., triangular meshes) and heuristics (e.g., Phong shading) to improve the applicability (e.g., for real-time applications).
While these methods require fewer parameters to represent a scene, the achieved realism is also reduced.

While traditional computer graphics allows us to generate high-quality controllable imagery of a scene, all physical parameters of the scene, for example, camera parameters, illumination and  materials of the objects need to be provided as inputs. 
If we want to generate controllable imagery of a real-world scene, we would need to estimate these physical properties from existing observations such as images and videos. 
This estimation task is referred to as inverse rendering and is extremely challenging, especially when the goal is photo-realistic synthesis. 
In contrast, neural rendering is a rapidly emerging field which allows the compact representation of scenes, and rendering can be learned from existing observations by utilizing neural networks (see \Cref{fig:teaser}).
The main idea of neural rendering is to combine insights from classical (physics-based) computer graphics and recent advances in deep learning.
Similar to classical computer graphics, the goal of neural rendering is to generate photo-realistic imagery in a controllable way (c.f. definition of neural rendering in \cite{tewari2020neuralrendering}).
This, for example, includes novel viewpoint synthesis, relighting, deformation of the scene, and compositing.

\begin{figure}[t!]
     \centering
     \begin{subfigure}[b]{\linewidth}
         \centering
         \includegraphics[width=\linewidth]{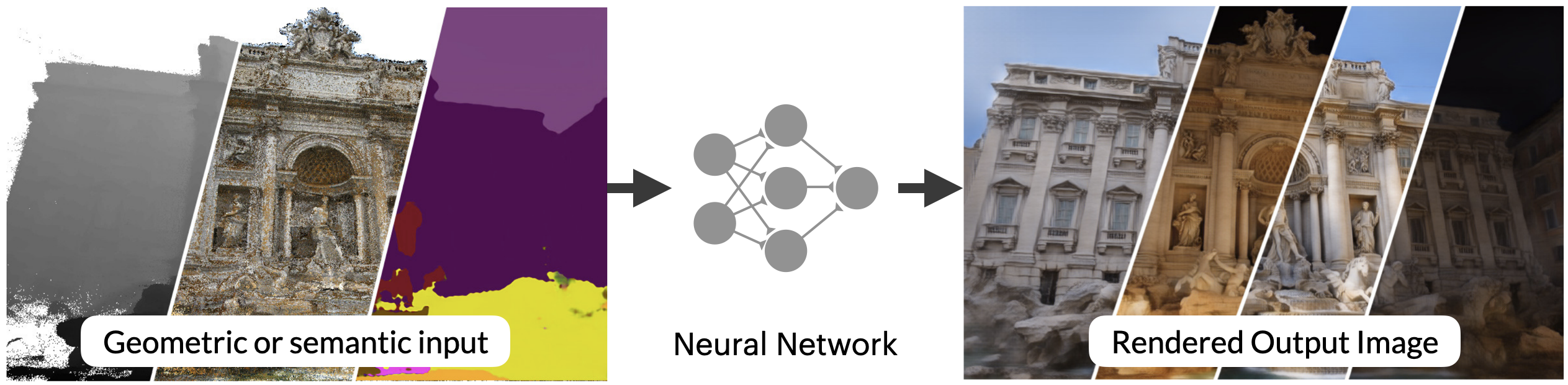}
         \caption{2D Neural Rendering, also known as neural refinement, neural re-rendering, or deferred neural rendering is based on 2D inputs that are generated for example using a classical renderer and \textit{learns to render a scene in 2D}. }
         \label{subfig:paradigm1}
     \end{subfigure}
     \begin{subfigure}[b]{\linewidth}
         \centering
         \includegraphics[width=\linewidth]{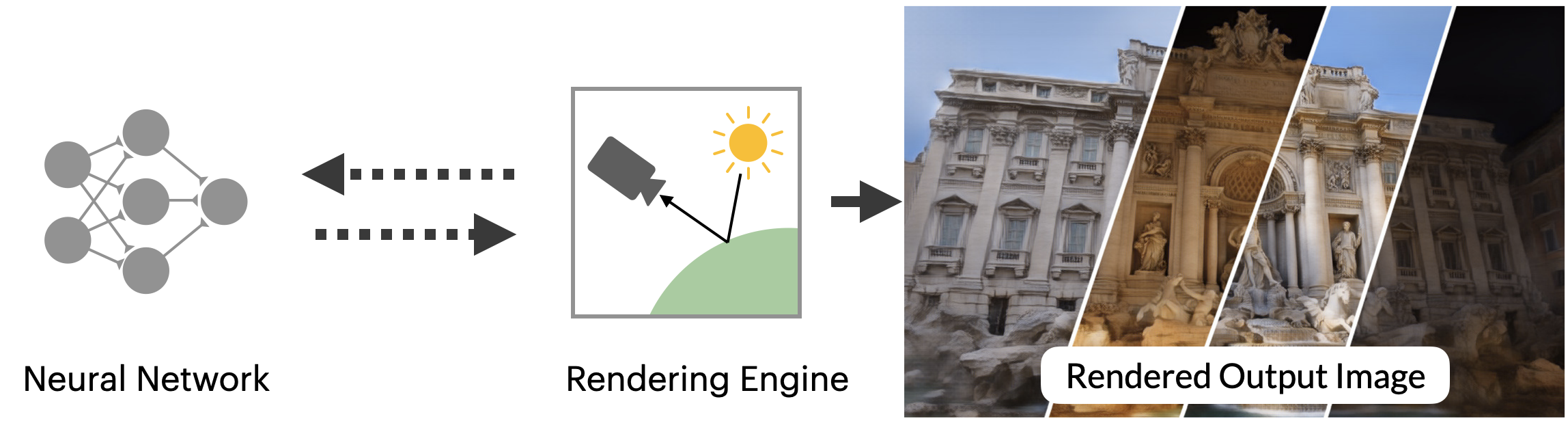}
         \caption{3D Neural Rendering \textit{learns to represent a scene in 3D} and uses fixed differentiable rendering schemes from computer graphics which are motivated by physics.
         }
         \label{subfig:paradigm2}
     \end{subfigure}
    \caption{
        The term ``Neural Rendering'' is often applied to what are two distinct concepts.
        The previous STAR report on neural rendering \cite{tewari2020neuralrendering} primarily focused on the paradigm shown in (\subref{subfig:paradigm1}), in which a neural network is trained to map from some 2D input signal (such as a semantic label or a rasterized proxy geometry) directly to the output image --- the neural network is trained to render.
        This report focuses on a newer emerging paradigm for neural rendering, shown in (\subref{subfig:paradigm2}) and well-exemplified by NeRF~\cite{Mildenhall_2020_NeRF}. Here, a neural network is supervised so as to represent the shape or appearance of a particular scene, and that the neural representation is rendered using a somewhat conventional graphics ``engine'' that is defined analytically, instead of being learned.
        Unlike the previous paradigm, here the neural network does not learn how to render --- it instead learns to represent a scene \emph{in 3D}, and that scene is then rendered according to the physics of image formation.
        Image adapted from ~\cite{meshry2019neural}.
    }
    \label{fig:paradigm}
\end{figure}

Early neural rendering approaches (covered in \cite{tewari2020neuralrendering}) used neural networks to convert scene parameters into the output images. 
The scene parameters are either directly given as one-dimensional inputs, or a classical computer graphics pipeline is used to generate two-dimensional inputs. 
The deep neural networks are trained on observations of real-world scenes and learn to model as well as render these scenes. 
A deep neural network can be seen as a universal function approximator.
Specifically, a network defines a family of functions based on its input arguments, model architecture, and trainable parameters.
Stochastic gradient descent is employed to find the function from this space that best explains the training set as measured by the training loss.
From this viewpoint, neural rendering aims to find the mapping $\mathbf{I} = \mathcal{M}(\mathbf{c})$ between control parameters $\mathbf{c} \in \mathbb{R}^{d_{in}}$ and the corresponding output image $\mathbf{I} \in \mathbb{R}^{H \times W \times 3}$, with $H$ and $W$ being image height and width.
This can be interpreted as a complex and challenging sparse data interpolation problem.
Thus, neural rendering, similar to classical function fitting, has to navigate the trade-off between under- and over-fitting, i.e., representing the training set well vs.~generalization to unobserved inputs.
If the representational power of the network is insufficient, the quality of the resulting images will be low, e.g., results are often blurry.
On the other hand, if the representational power is too large, the network overfits to the training set and does not generalize to unseen inputs at test time.
Finding the right network architecture is an art in itself. 
In the context of neural rendering, designing the right physically motivated inductive biases often requires a strong graphics background.
These physically motivated inductive biases act as regularizers and ensure that the found function is close to how 3D space and/or image formation works in our real world, thus leading to better generalization at test time.
Inductive biases can be added to the network in different ways.
For example, in terms of the employed layers, at what point in the network and in which form inputs are provided, or even via the integration of non-trainable (but differentiable) components from classical computer graphics.
One great example for this are recent neural rendering techniques that try to disentangle the modeling and rendering processes by only learning the 3D scene representation and relying on a rendering function from computer graphics for supervision.
For example, Neural Radiance Fields (NeRF)~\cite{Mildenhall_2020_NeRF} uses a multi-layer perceptron (MLP) to approximate the radiance and density field of a 3D scene. 
This learned volumetric representation can be rendered from any virtual camera using analytic differentiable rendering (i.e., volumetric integration). 
For training, observations of the scene from several camera viewpoints are assumed.
The network is trained on these observations by rendering the estimated 3D scene from these training viewpoints, and minimizing the difference between the rendered and observed images. 
Once trained, the 3D scene approximated by the neural network can be rendered from a novel viewpoint, enabling controllable synthesis. 
In contrast to approaches that use the neural network to learn the rendering function as well \cite{tewari2020neuralrendering}, NeRF uses knowledge from computer graphics more explicitly in the method, enabling better generalization to novel views due to the (physical) inductive bias: an intermediate 3D-structured representation of the density and radiance of the scene. 
As a result, NeRF learns physically meaningful color and density values in 3D space, which physics-inspired ray casting and volume integration can then render consistently into novel views. 
The achieved quality, as well as the simplicity of the method, led to an `explosion' of developments in the field.
Several advances have been made which improve the applicability, enable controllability, the capture of dynamically changing scenes as well as the training and inference times.
Within this report, we cover these recent advances in the field.
To foster a deep understanding of these methods, we discuss the fundamentals of neural rendering by describing the different components and design choices in detail in \Cref{sec:fundamentals}. 
Specifically, we clarify the definition of the different scene representations used in the current literature (surfaces and volumetric approaches), and describe ways to approximate them using deep neural networks. 
We also present the fundamental rendering functions from computer graphics that are used to train these representations. 
Since neural rendering is a very fast evolving field, with significant progress along many different dimensions, we develop a taxonomy of the recent approaches w.r.t. their application field to provide a concise overview of the developments.
Based on this taxonomy and the different application areas, we present the state-of-the-art methods in \Cref{sec:applications}.
The report is concluded with \Cref{sec:open_challenges} discussing the open challenges and \Cref{sec:social} discussing social implications of photo-realistic synthetic media.

 \vspace{1cm}
    \begin{figure*}[t!]
    \centering
    \includegraphics[width=\linewidth]{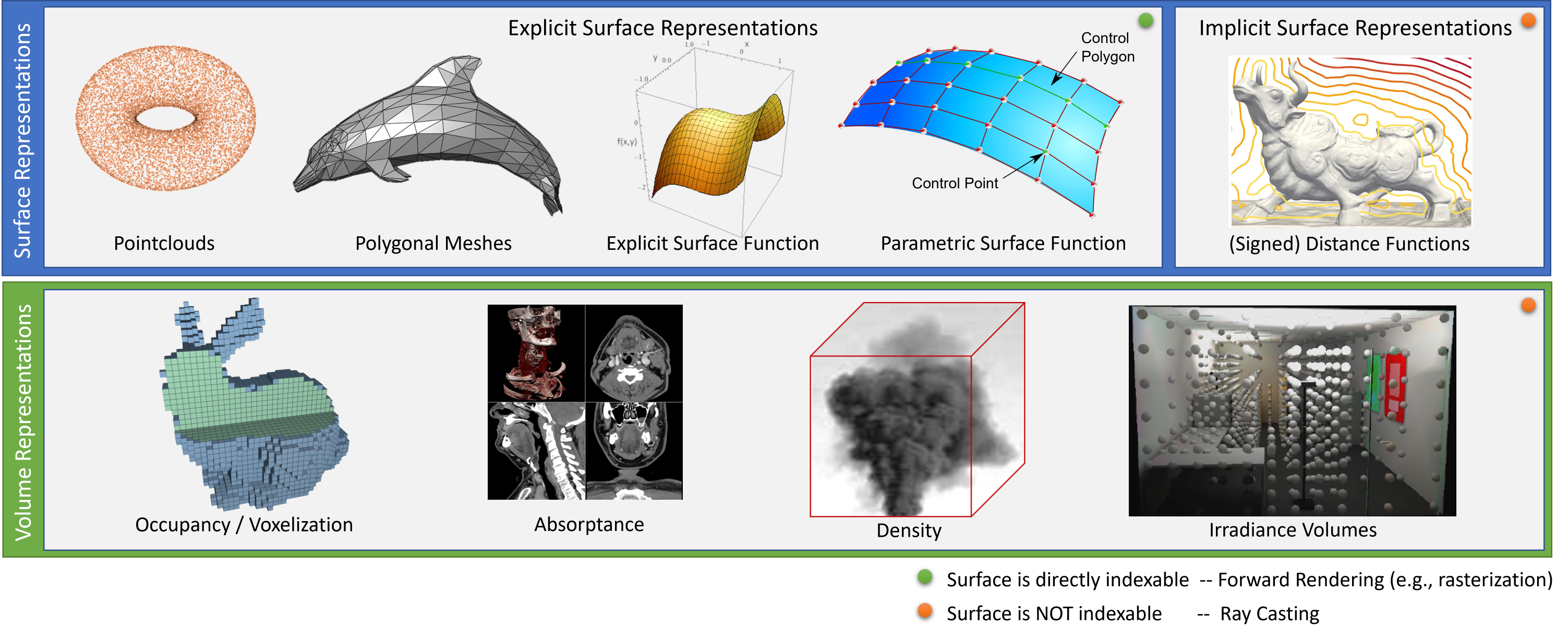}
    \caption{
    An overview of classical surface and volume representations. Images adapted from \cite{greger1998irradiance,voxelization,yariv2021volume_sdf_figure,Vladsinger2009,ChumpusRex2006}.
    }
    \label{fig:reps}
\end{figure*}

\vspace{-0.5cm}
\section{Scope of This STAR}
In this state-of-the-art report, we focus on advanced neural rendering approaches that combine classical rendering with learnable 3D representations (see \Cref{fig:paradigm}).
The underlying neural 3D representations are 3D-consistent by design and enable control over different scene parameters.
Within this report, we give a comprehensive overview of the different scene representations and detail the fundamentals of the components that are lent from classical rendering pipelines as well as machine learning.
We further focus on approaches that use Neural Radiance Fields~\cite{Mildenhall_2020_NeRF} and volumetric rendering. 
However, we do not focus on neural rendering methods that reason mostly in 2D screen space; we refer to~\cite{tewari2020neuralrendering} for a discussion on such approaches. 
We also do not cover neural super-sampling and denoising methods for ray-traced imagery~\cite{Chaitanya:2017,LBF}.

\textit{Selection scheme:}
As written above, the report concentrates mainly on the use of neural radiance fields and its derivatives.
It covers papers published in the proceedings of the major computer vision and computer graphics conferences (2020, 2021), as well as preprints published via arXiv.
The papers are selected by the authors of this report to fit the context of the survey and to give the reader an overview over a broad spectrum of different techniques.
Note that this report is intended to give a panoramic view of the field, to list literature to get up to speed with a specific research domain, and to standardize the notations and definitions.
This report covers a wide range of current literature. The authors do not claim completeness of the report and highly recommend the reader to study the cited works for in-depth details.

 \vspace{1cm}
    \vspace{-0.5cm}
\section{Fundamentals of Neural Rendering}
\label{sec:fundamentals}
Neural rendering, and especially 3D neural rendering is based on classical concepts of computer graphics (see \Cref{fig:paradigm}).
A neural rendering pipeline learns to render and/or represent a scene from real-world imagery, which can be an unordered set of images, or structured, multi-view images or videos.
It does so by mimicking the physical process of a camera that captures a scene.
A key property of 3D neural rendering is the disentanglement of the camera capturing process (i.e., the projection and image formation) and the 3D scene representation during this training.
This disentanglement has several advantages and leads especially to a high level of 3D consistency during the synthesis of images (e.g., for novel viewpoint synthesis).
To disentangle the projection and other physical processes from the 3D scene representation, 3D neural rendering methods rely on known image formation models from computer graphics (e.g., rasterization, point splatting, or volumetric integration).
These models are motivated by physics, especially the interaction of the light of an emitter with the scene as well as the camera itself.
This light transport is formulated using the rendering equation~\cite{kajiya1986rendering}.

The computer graphics field offers a variety of approximations to this rendering equation.
These approximations are dependent on the used scene representation and range from classical rasterization to path tracing and volumetric integration.
3D neural rendering exploits these rendering methods.
In the following, we will detail the scene representations (\Cref{sec:scene_rep}) as well as the rendering methods (\Cref{sec:diff_rendering}) used in common neural rendering methods.
Note that both the scene representation as well as the rendering method itself have to be differentiable in order to learn from real images (\Cref{sec:opt}).

\vspace{-0.5cm}
\subsection{Scene Representations}
\label{sec:scene_rep}

For decades, the computer graphics community has explored various primitives, including point clouds, implicit and parametric surfaces, meshes, and volumes (see \Cref{fig:reps}).
While these representations have clear definitions in the computer graphics field, there is often a confusion in the current literature of neural rendering, especially when it is about implicit and explicit surface representations and volumetric representations.
In general, volumetric representations can represent surfaces, but not vice versa.
Volumetric representations store volumetric properties such as densities, opacities or occupancies, but they can also store multidimensional features such as colors or radiance.
In contrast to volumetric representations, surface representations store properties w.r.t. the surface of an object.
They cannot be used to model volumetric matter, such as smoke (unless it is a coarse approximation).
For both surface and volumetric representations, there are continuous and discretized counterparts (see \Cref{fig:reps}).
The continuous representations are particularly interesting for neural rendering approaches since they can provide analytic gradients.
For surface representations, there are two different ways to represent the surface -- explicitly or implicitly.
The surface using an explicit surface function $f_\mathit{explicit}(.) \in \mathbb{R}$ in the Euclidean space is defined as:
\begin{equation}
    S_\mathit{explicit} = 
                    \left\{
                    \left(
                        \begin{array}{c}
                        x\\
                        y\\
                        f_\mathit{explicit}(x,y)\\
                        \end{array}
                    \right) ~\middle|~
                    \left(
                        \begin{array}{c}
                        x\\
                        y\\
                        \end{array}
                    \right) \in \mathbb{R}^2 
                    \right\}.
\end{equation}
Note that an explicit surface can also be represented as a parametric function $f_\mathit{parametric}(.) \in \mathbb{R}^3$, which generalizes $S_\mathit{explicit}$:
\begin{equation}
    S_\mathit{explicit}^{*} = 
                    \left\{
                        f_\mathit{parametric}(u,v)\\
                    ~\middle|~
                    \left(
                        \begin{array}{c}
                        u\\
                        v\\
                        \end{array}
                    \right) \in \mathbb{R}^2 
                    \right\}.
\end{equation}
The surface using an implicit surface function $f_\mathit{implicit}(\cdot) \in \mathbb{R}$ is defined as the zero-level set:
\begin{equation}
\label{eq:implicit}
    S_\mathit{implicit} = 
                    \left\{
                    \left(
                        \begin{array}{c}
                        x\\
                        y\\
                        z\\
                        \end{array}
                    \right) \in \mathbb{R}^3  ~\middle|~
                    f_\mathit{implicit}(x,y,z) = 0
                    \right\}.
\end{equation}
Whereas a volume representation defines properties in the entire space:
\begin{equation}
    V = 
                    \left\{
                    f_\mathit{vol}(x,y,z) ~\middle|~
                    \left(
                        \begin{array}{c}
                        x\\
                        y\\
                        z\\
                        \end{array}
                    \right) \in \mathbb{R}^3
                    \right\}.
\end{equation}
Note that the respective function domain can be restricted for all these representations.

In general, for all three scene representations, the underlying function can be any function that is capable to approximate the respective content.
For simple surfaces like a plane, the functions $f_{implicit}$, $f_{explicit}$ can be linear functions.
To handle more complex surfaces or volumes, polynomials (for example from a Taylor series) or multivariate Gaussians can be used. 
To increase the expressiveness further, these functions can be spatially localized and then combined into a mixture, for example multiple Gaussians can form a Gaussian mixture.
Radial basis function networks are such mixture models and can be used as an approximator for both, implicit surface and volume functions~\cite{RBF_reco}.
Note that these radial basis function networks can be interpreted as a single layer of a neural network.

Since neural networks and, especially, multi-layer perceptrons (MLPs) are universal function approximators, they can be used to `learn' the underlying functions ($f_\mathit{implicit}$, $f_\mathit{explicit}$, $f_\mathit{parametric}$,  or $f_\mathit{vol}$). 
(Similar to a Gaussian mixture, multiple localized, weaker MLPs can be combined into a mixture as well, e.g., \cite{Reiser2021}.)
In the context of neural rendering, a scene representation that is using a neural network to approximate the surface or volumetric representation function is called \textit{neural scene representation}.
Note that both surface and volumetric representations can be extended to store additional information, like color or view-dependent radiance.

In the following, we will discuss the different MLP-based function approximators that build the foundation of the recent neural surface and volumetric representations.

\subsubsection{MLPs as Universal Function Approximators}
\label{sec:coordinate-based}
Multi-Layer Perceptrons (MLPs) are known to act as Universal Function Approximators~\cite{HORNIK1989359}. %
Specifically, we use MLPs to represent surface or volumetric properties.
A multi-layer perceptron is a conventional fully-connected neural network.
In the context of scene repesentations, the MLP takes as input a \emph{coordinate} in space, and produces as output some value corresponding to that coordinate.
This type of network is also known as \emph{coordinate-based neural network} (and the resulting representation is called coordinate-based scene representation).
Note that the input coordinate space can be aligned with the Euclidean space, but it can also be embedded for example in the uv-space of a mesh (resulting in a neural parametric surface). %

A key finding to use ReLU-based MLPs for neural representation and rendering tasks is the usage of positional encoding.
Inspired by the positional encoding used in natural language processing (e.g., in Transformers~\cite{transformers}), the input coordinates are positionally encoded using a set of basis functions.
These basis functions can be fixed~\cite{Mildenhall_2020_NeRF} or they can be learned~\cite{tancik2020fourfeat}.
These spatial embeddings simplify the task of the MLP to learn the mapping from a location to a specific value, since through the spatial embedding, the input space is partitioned. 
As an example, the positional encoding used in NeRF~\cite{Mildenhall_2020_NeRF} is defined as:
\begin{align}
    \mathbf x &\mapsto [\cos(\mathbf M \mathbf x), \sin(\mathbf M\mathbf x)] \\
    \textrm{where } \mathbf M &= \begin{bmatrix} \mathbf I &  2 \mathbf I & 2^2 \mathbf I & \ldots & 2^{p-1} \mathbf I \end{bmatrix}^\top .
\end{align}
Here, $\mathbf x$ is the input coordinate and $p$ is a hyperparameter controlling the frequencies used (dependent on the target signal resolution). 
This ``soft'' binary encoding of the input coordinates makes it easier for the network to access higher frequencies of the input. 

As mentioned above, MLP-based function approximators can be used to represent a surface or volume (i.e., $f_\mathit{implicit}$, $f_\mathit{explicit}$, $f_\mathit{parametric}$,  or $f_\mathit{vol}$), but they can also be used to store other attributes like color.
For instance, there are hybrid representations composed of classical surface representations like point clouds or meshes with an MLP to store the surface appearance (e.g., texture field~\cite{Oechsle2019ICCV}).

MLP-based function approximators can employ different activation functions such as ReLU~\cite{Mildenhall_2020_NeRF}, sine activations~\cite{sitzmann2020siren} or Gaussians~\cite{ramasinghe2021unify}.

\subsubsection{Representing Surfaces}

\paragraph*{Point Clouds.}

A point cloud is a set of elements of the Euclidean space.
A continuous surface can be discretized by a point cloud - each element of the point cloud represents a sample point $(x,y,z)$ on the surface.
For each point, additional attributes can be stored such as normals or colors.
A point cloud that features normals is also referred to as oriented point cloud.
Besides simple points that can be seen as infinitesimally small surface patches, oriented point clouds with a radius can be used (representing a 2D disk that lies on the tangent plane of the underlying surface).
This representation is called surface elements, alias surfels~\cite{pfister2000surfelssurface}.
They are often used in computer graphics to render point clouds or particles from simulations.
The rendering of such surfels is called splatting, and recent work shows that it is differentiable~\cite{Yifan2019DSS}.
Using such a differentiable rendering pipeline, it is possible to directly back-propagate to the point cloud locations as well as the accompanied features (e.g., radius or color).
In Neural Point-based Graphics~\cite{aliev2020ngp} and SynSin~\cite{Wiles_2020_CVPR}, learnable features are attached to the points that can store rich information about the appearance and shape of the actual surface.
In ADOP~\cite{rueckert2021adop} these learnable features are interpreted by an MLP which can account for view-dependent effects.
Note that instead of storing explicitly features for specific points, one can also use an MLP to predict the features for the discrete positions.

As mentioned above, a point cloud is a set of elements of the Euclidean space, thus, besides surfaces, they can also represent volumes (e.g., storing additional opacity or density values). Using a radius for each point naturally leads to a full sphere-based formulation~\cite{Lassner_pulsar}.

\paragraph*{Meshes.}

Polygonal meshes represent a piece-wise linear approximation of a surface.
Especially, triangle and quad meshes are used in computer graphics as de facto standard representation for surfaces.
The graphics pipeline and graphic accelerators (GPUs) are optimized to process and rasterize billions of triangles per second.
The majority of graphics editing tools work with triangle meshes which makes this representation important for any content creation pipeline.
To be directly compatible with these pipelines, many 'classical' inverse graphics and neural rendering methods use this basic surface representation.
Using a differentiable renderer, the vertex positions as well as the vertex attributes (e.g., colors) can be optimized for to reproduce an image.
Neural networks can be trained to predict the vertex locations, e.g., to predict dynamically changing surfaces~\cite{burov2021dsfn}.
Instead of using vertex attributes, a common strategy to store surface attributes within the triangles are texture maps.
2D texture coordinates are attached to the vertices of the mesh which reference a location in the texture image.
Using barycentric interpolation, texture coordinates can be computed for any point in a triangle and the attribute can be retrieved from the texture using bilinear interpolation.
The concept of textures is also integrated into the standard graphics pipeline, with additional features such as mip-mapping which is needed to properly handle the sampling of the texture (c.f., sampling theorem).
Deferred Neural Rendering~\cite{thies2019deferred}, uses textures that contain learnable view-dependent features, so-called neural textures.
Specifically, a coarse mesh is used as underlying 3D representation, to rasterize these neural textures. A neural network interprets these rasterized features in image space.
Note that the network can for example be a pixel-wise MLP, then the neural texture represents the surface radiance.

In contrast to using discrete textures, continuous textures can be used.
The authors of texture fields~\cite{Oechsle2019ICCV} propose the usage of an MLP that predicts color values for each surface point.
In neural reflectance field textures (NeRF-Tex)~\cite{baatz2021nerftex} the idea of NeRF~\cite{Mildenhall_2020_NeRF} is combined with the idea of using a 2D neural texture and an underlying 3D mesh.
NeRF-Tex is conditioned on user-defined parameters that control the appearance, thus, being editable by artists.

\paragraph*{Implicit Surfaces.}
\label{sec:implicit_surfaces}

Implicit surfaces define the surface as the zero level-set of a function, see Eq.~\ref{eq:implicit}.
The most commonly used implicit surface representation is a signed distance function (SDF).
These SDF representations are used in numerous 3D scanning techniques that use volumetric fusion~\cite{curless1996volumetric} to incrementally reconstruct the surface of a static~\cite{izadi2011kinectfusion, niessner2013hashing} or dynamic object~\cite{newcombe2015dynamicfusion}.
Implicit surface representations offer many advantages as they avoid the requirement of defining a mesh template, thus, being able to represent objects with unknown topology or changing topology in a dynamic scenario.
The volumetric fusion approaches mentioned above use a discretized (truncated) signed distance function, i.e., using a 3D grid containing signed distance values.
Hoppe et al.~\cite{hoppe1992surface} propose piece-wise linear functions to model the signed distance function w.r.t. input surface point samples.
The seminal work of Carr et al.~\cite{carr2001reconstruction} uses a radial basis function network instead.
This radial basis function network represent a continuous implicit surface function and can be seen as the first 'neural' implicit surface representation.
Recent neural implicit surfaces representations are based on coordinate-based multi-layer perceptrons (MLPs), covered in \Cref{sec:coordinate-based}. 
Such representations have been gaining widespread popularity in neural scene representation and rendering.
They were proposed concurrently in~\cite{park2019deepsdf, chen2019learning} for shape modeling, where MLP architectures were used to  map continuous coordinates to signed distance values. %
The fidelity of signals represented by such coordinate networks, or neural implicit representation, is primarily limited by the capacity of the network.
Thus, compared to other aforementioned representations, implicit surfaces offer potential advantages in memory efficiency and, as a continuous representation, they can theoretically represent geometries at infinite resolution.
The initial proposals was ensued, with broad enthusiasm, by a variety of improvements of different focuses, including improving the training schemes~\cite{xu2020ladybird, duan2020curriculum, yifan2020iso}, leveraging global-local context~\cite{neurips2019_39059724, erler2020points2surf}, adopting specific parameterizations~\cite{genova2019learning,deng2020cvxnet,chen2020bsp,kellnhofer2021neural,yifan2021geometryconsistent} or spatial partitions~\cite{genova2020local, tretschk2020patchnets, chabra2020deep, takikawa2021nglod, martel2021acorn}.
As there is no requirement of pre-defining the mesh template or the object topology, neural implicit surfaces are well suited for modeling objects of varying topologies~\cite{park2019deepsdf,chen2019learning}. 
Analytic gradients of the output with respect to the input coordinates can be computed using backpropagation. 
This makes it possible to implement regularization terms on the gradients~\cite{gropp2020implicit}, in addition to  other geometrically motivated regularizers~\cite{gropp2020implicit,poursaeed2020coupling,yifan2020iso}.
These respresentations can be extended to also encode the radiance of the scene~\cite{kellnhofer2021neural,yenamandra2021i3dmm,saito2019pifu}. 
This is useful for neural rendering, where we want the scene representation to encode both the geometry and appearance of the scenes.

\subsubsection{Representing Volumes}

\paragraph*{Voxel Grids.}

As the pixel-equivalent in $\mathbb{R}^3$, voxels are commonly used to represent volumes.
They can store the geometry occupancy, or store the density values for a scene with volumetric effects such as transparency. %
In addition, the appearance of the scene can be stored~\cite{greger1998irradiance}.
Using trilinear interpolation these volume attributes can be accessed at any point within the voxel grid.
T his interpolation is especially used for sample-based rendering methods like ray casting.
While the stored attributes can have a specific semantic meaning (e.g., occupancy), the attributes can also be learned.
Sitzmann et al. propose the use of DeepVoxels~\cite{sitzmann2019deepvoxels}, where features are stored in a voxel grid.
The accumulation and interpretation of the features after the ray-casting rendering procedure is done using a deep neural network.
These DeepVoxels can be seen as volumetric neural textures, which can be directly optimized using backpropagation.
While dense voxel-based representations are fast to query, they are memory inefficient and 3D CNNs, potentially operating on these volumes, are computationally heavy. 
Octree data structures~\cite{laine2010efficient} can be used to represent the volume in a sparse manner. 
Sparse 3D convolution on octrees~\cite{wang2017cnn,riegler2017octnet} can help mitigate some problems, but these compact data structures cannot be easily updated on the fly.
Thus, they are difficult to integrate into learning frameworks.
Other approaches to mitigating the memory challenges of dense voxel grids include using object-specific shape templates~\cite{kanazawa2018learning}, multi-plane~\cite{zhou2018,Mildenhall:2019,flynn2019deepview,tucker2020single,Wizadwongsa2021NeX} or multi-sphere~\cite{Broxton:2020,Attal:2020:ECCV} images, which all aim at representing the voxel grid using a sparse approximation.

\paragraph*{Neural Volumetric Representations.}
Instead of storing features or other quantities of interest using a voxel grid, these quantities can also be defined using a neural network, similar to neural implicit surfaces (see \Cref{sec:implicit_surfaces}). 
MLP network architectures can be used to parameterize volumes, potentially in a more memory efficient manner than explicit voxel grids.
Still, these representations can be expensive to sample depending on the underlying network size because for each sample, an entire feedforward pass through the network has to be computed.
Most methods can be roughly classified as using global or local networks~\cite{genova2019learning,genova2020local,chen2019learning,michalkiewicz2019implicit,atzmon2020sal,saito2019pifu,sitzmann2019srns,Oechsle2019ICCV,gropp2020implicit,yariv2020multiview,davies2020overfit,sitzmann2020siren,Niemeyer2020CVPR,liu2020neural,jiang2020sdfdiff,liu2020dist,kohli2020inferring}.
Hybrid representations that use both grids and neural networks make a trade-off between computational and memory efficiency~\cite{peng2020convolutional,jiang2020local,chabra2020deep,martel2021acorn}.
Similar to neural implicit surfaces, neural volumetric representations allow for the computation of analytic gradients, which has been used to define regularization terms in ~\cite{sitzmann2020siren,tretschk2021nonrigid,park2021nerfies}.
Band-limited coordinate-based networks have been introduced in BACON~\cite{lindell2021bacon} which learn a smooth multi-scale decomposition of the surface.

\textit{General remark:}
The use of coordinate-based neural networks to model scenes volumetrically (as in NeRF) superficially resembles the use of coordinate networks to model surfaces implicitly (as in neural implicit surfaces). However, NeRF-like volumetric representations are not necessarily implicit --- because the output of the network is density and color, the geometry of the scene is parameterized by the network \emph{explicitly}, not implicitly. Despite this, it is common in the literature for these models to still be called ``implicit'', perhaps in reference to the fact that the geometry of the scene is defined ``implicitly'' by the weights of a neural network (a different definition of ``implicit'' than is used by the SDF literature). Also note that this is a distinct definition of ``implicit'' than what is commonly used by the deep learning and statistic communities, where ``implicit'' usually refers to models whose outputs are implicitly defined as fixed points of dynamic systems, and whose gradients are computed using the implicit function theorem~\cite{bai2019deep}.

\subsection{Differentiable Image Formation}
\label{sec:diff_rendering}

\begin{figure}[t!]
    \centering
     \begin{subfigure}[b]{\linewidth}
         \centering
         \includegraphics[width=\linewidth]{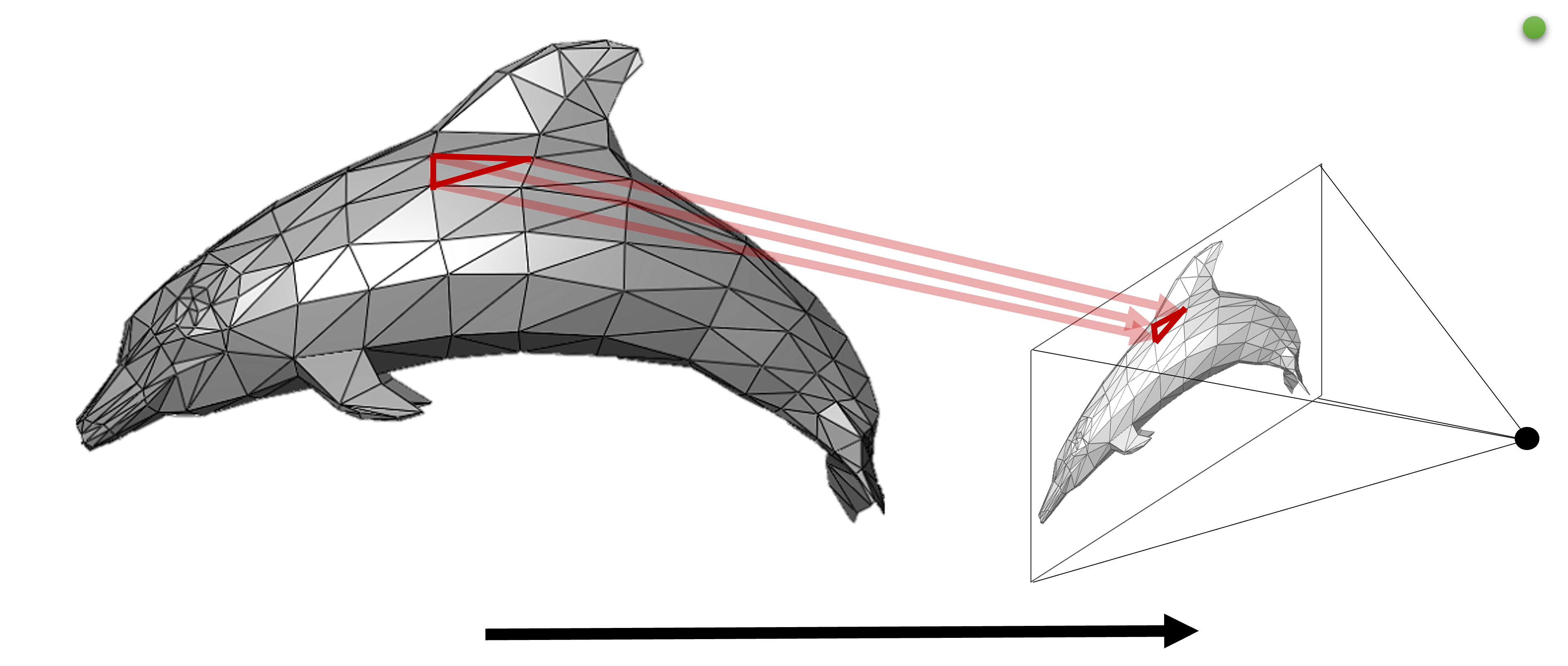}
         \caption{Forward Rendering (e.g., rasterization) -- the image is generated by projecting the 3D representation to the image plane.} %
     \end{subfigure}
     \begin{subfigure}[b]{\linewidth}
         \centering
         \includegraphics[width=\linewidth]{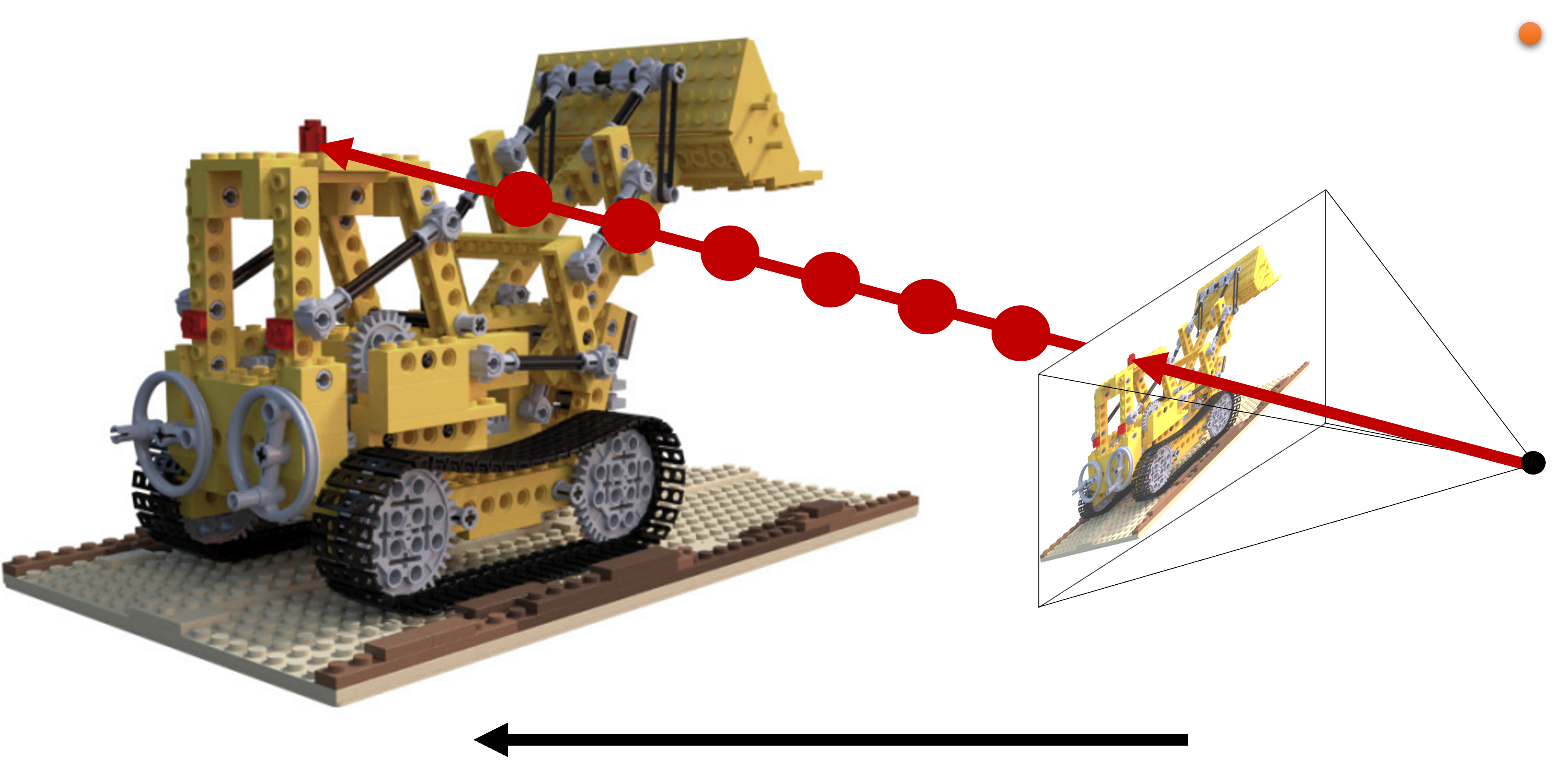}
         \caption{Ray Casting -- the image is generated by casting viewing rays, sampling the 3D representation and accumulating them. Image adapted from~\cite{Mildenhall_2020_NeRF}.}
     \end{subfigure}
     
    \caption{
    For explicit surfaces representations, the surface is directly indexable. This allows us to use forward rendering methods that project the surface to the image plane and to set a pixel accordingly (e.g., using rasterization or point splatting).
    Implicit surface representations and volumetric representations, do not provide direct information of the surface that would allow for forward rendering, instead, the 3D space seen from the virtual camera has to be sampled to generate an image (e.g., using ray marching).
    }
    \label{fig:rendering}
\end{figure}

The scene representations in the previous sections allow us to represent the 3D geometry and appearance of the scene. 
As a next step, we describe how images can be generated from such scene representations through rendering. 
There are two general approaches to rendering a 3D scene into a 2D image plane: ray casting and rasterization, see also \Cref{fig:rendering}. 
A rendered image of the scene can be computed by also defining the camera in the scene. 
Most methods use a pinhole camera, where all camera rays pass through a single point in space (focal point). 
With a given camera, rays from the camera origin can be cast towards the scene in order to calculate the rendered image.

\paragraph*{Ray Casting.} In the pinhole model, the basic intercept theorem can be used to describe how a point $\mathbf{p}\in\mathbb{R}^3$ in 3D is projected to the correct position $\mathbf{q}\in\mathbb{R}^2$ in the image plane. It is by definition a non-injective function and hard to invert---this puts it at the heart of the 3D reconstruction problem.

The Pinhole model has a single parameter matrix for this projection: the intrinsic matrix $\mathbf{K}$ contains the focal lengths normalized by pixel size $\mathbf{f}=[\alpha_x, \alpha_y]$, axis skew $\gamma$ and center point $\mathbf{c}=[c_x, c_y]$. Using the intercept theorem and assuming homogeneous coordinates $\mathbf{p}'=[x, y, z, 1]$, we find that the projected coordinates are $\mathbf{q}'=\mathbf{K}\cdot\mathbf{p}'$, with 
$$
\mathbf{K} =  
\begin{bmatrix}
\alpha_x & \gamma & c_x & 0\\
0 & \alpha_y & c_y & 0\\
0 & 0 & 1 & 0\\
\end{bmatrix}. 
$$
This assumes that the center of the projection is at the coordinate origin and that the camera is axis-aligned. To generalize this for arbitrary camera positions, an extrinsic matrix $\mathbf{R}$ can be used. This homogeneous $4\times 4$ matrix $\mathbf{E}$ is composed of
\vspace{10pt}
$$
\mathbf{E}=\begin{bmatrix}
\mathbf{R}_{3\times 3} & \mathbf{t}_{3\times 1} \\
\mathbf{0}_{1\times 3} & 1 \\
\end{bmatrix},
\vspace{10pt}
$$
where $\mathbf{R}$ is a rotation matrix and $\mathbf{t}$ is a translation vector, such that $\mathbf{R}\cdot \mathbf{p_w}+\mathbf{t}=\mathbf{p_c}$, where we use $\mathbf{p_w}$ to denote a point in world coordinates and $\mathbf{p_c}$ to denote it in camera coordinates. This definition of $\mathbf{R}$ and $\mathbf{t}$ is common in Computer Vision (for example, used by OpenCV) and referred to as `world-to-cam' mapping, whereas in Computer Graphics (for example, in OpenGL) a similar inverse `cam-to-world' mapping is more prevalent. Assuming the `world-to-cam' convention and using homogeneous coordinates, we can write the full projection of $\mathbf{p_w}$ to $\mathbf{q_p}$ as:

$$
\mathbf{q_p}'=\mathbf{K}\cdot\begin{bmatrix}
\mathbf{R} & \mathbf{t} \\
\mathbf{0}_{1\times 3} & 1 \\
\end{bmatrix}\cdot\mathbf{p_w}'.
$$

If the `cam-to-world' convention is used, the ray casting is similarly convenient. Whereas these equations are non-injective due to the depth ambiguity, they lend themselves very well for automatic differentiation and can be optimized end-to-end in image formation models.

To model current cameras correctly, there is one more component that has to be taken into account: the lens. Leaving aside effects such as depth-of-field or motion blur, which must be modeled in the image formation process, they add distortion effects to the projection function. Unfortunately, there is no single, simple model to capture all different lens effects. Calibration packages, such as the one in OpenCV, usually implement models with up to 12 distortion parameters. They are modeled through polynomials up to degree five, hence are not trivially invertible (which is required for raycasting as opposed to point projection). More modern approaches to camera calibration use many more parameters and achieve a higher accuracy~\cite{Schops_2020_CVPR} and could be made invertible and differentiable.

\paragraph*{Rasterization.} An alternative to ray casting is rasterization of geometric primitives. This technique does not try to emulate the image formation process of the real world, but instead exploits the geometric properties of objects to quickly create an image. It is mostly used with meshes, which are described by a set of vertices $\mathbf{v}$ and faces $\mathbf{f}$, connecting triples or quadruplets of vertices to define surfaces. One fundamental insight is that the geometric operations in 3D can solely work with the vertices: for example, we can use the same extrinsic matrix $\mathbf{E}$ to transform each point from the world to the camera coordinate system. After this transformation, the points outside of the view frustum or points with wrong normal orientation can be culled to reduce the amount of points and faces to be processed in the next steps. The location of the remaining points projected to image coordinates can again trivially be found by using the intrinsic matrix $\mathbf{K}$ as outlined above. The face information can be used to interpolate the depth on face primitives, and the top-most faces can be stored in a z-buffer.

This way of implementing the projection is often faster than ray casting: it mainly scales with the number of visible vertices in a scene, whereas ray-casting scales with the number of pixels and the numbers of primitives to intersect with. However, it is harder to capture certain effects using it (e.g., lighting effects, shadows, reflections). It can be made differentiable through `soft' rasterization. This has been implemented, for example, in \cite{liu2019soft,pytorch3d}.

\subsubsection{Surface Rendering}

\paragraph*{Point Cloud Rendering.}

In the computer graphics literature, point cloud rendering techniques are extensively used~\cite{KOBBELT2004801,SAINZ2004869}.
As mentioned before, point clouds are discrete samples of continuous surfaces or volumes.
Point cloud rendering corresponds to reconstructing the continuous signal, e.g., the appearance of a continuous surface, from irregularly distributed discrete samples then resampling the reconstructed signal in the image space at each pixel location.

This process can be done in two different ways.
The first approach is based on the theory of classic signal processing and can be seen as a `soft' point splatting (similar to the soft rasterizer in the mesh rendering section below).
It first constructs the continuous signal using continuous local reconstruction kernels \(r\left( \cdot \right)\), i.e., 
\(\mathbf{f} = \sum \mathbf{f}_{i} r\left( \mathbf{p}_{i} \right)\).
Essentially, this approach amounts to blending the discrete samples with some local deterministic blurring kernels~\cite{lin2018learning,insafutdinov2018unsupervised,roveri2018network}, such as EWA splatting~\shortcite{zwicker2001surface,zwicker2002ewa}, which is a spatially-variant reconstruction kernel that is designed to minimize aliasing.
In neural rendering, the discrete samples can store some learnable features.
Correspondingly, this aforementioned step effectively projects and blends the individual features into a 2D feature map.
If the features have a predefined semantic meaning (e.g., colors, normals), a fixed shading function or BRDF can be used to generate the final image.
If the features are learned neural descriptors, a 2D neural network is deployed to transform the 2D feature map to an RGB image.
Recent neural point rendering methods that adopt this approach include SinSyn and Pulsar~\cite{Wiles_2020_CVPR,Lassner_pulsar}.
They use spatially-invariant and isotropic kernels in the blending step for performance reasons.
While these simplified kernels can result in rendering artifacts such as holes, blurred edges and aliasing, these artifacts can be compensated in the neural shading step, and, in case of Pulsar, through optimization of the radii. \cite{KPLD21}~additionally uses strategies for camera selection and probabilistic depth testing and is able to tackle IBR, stylization, and harmonization in this framework. 

Alternative to the soft point splatting approach, one can use a conventional point renderer from OpenGL or DirectX.
Here, each point is projected to a single pixel (or a small area of pixels) resulting in a sparse feature map.
One can use a deep neural networks to reconstruct the signal directly in the image space~\cite{aliev2020neural}.
Note that this naive rendering approach does not provide gradients with respect to the point positions \(\mathbf{p}\), and only allows to differentiate the rendering function w.r.t. the (neural) features.
In contrast, the soft point splatting approaches provide point position gradients via the reconstruction kernel \(r\left( \mathbf{p} \right)\).

However, even in this case, the gradient is confined spatially within the support of the local reconstruction.
\shortcite{Yifan:DSS:2019} addressed this issue by approximating the gradient using finite difference, and successfully applied the renderer to surface denoising, stylization, and multiview shape reconstruction.
This idea was adopted in~\cite{ruckert2021adop} to optimize the geometry and camera poses jointly for novel view synthesis.

\paragraph*{Mesh Rendering.} 
There are a number of general-purpose renderers that allow meshes to be rasterized or otherwise rendered in a differentiable manner.
Among differentiable mesh rasterizers, Loper and Black~\shortcite{loper2014opendr} developed a differentiable rendering framework called OpenDR that approximates a primary renderer and computes the gradients via automatic differentiation. Neural mesh renderer (NMR)~\cite{kato2018neural} approximates the backward gradient for the rasterization operation using a handcrafted function for visibility changes. \cite{liu2018paparazzi} proposed Paparazzi, an analytic differentiable renderer for mesh geometry processing using image filters. Petersen et al.~\cite{petersen2019pix2vex} presented \emph{Pix2Vex}, a $C^{\infty}$  differentiable renderer via soft blending schemes of nearby triangles, and \cite{liu2019soft} introduced \emph{Soft Rasterizer}, which renders and aggregates the probabilistic maps of mesh triangles, allowing gradient flow from the rendered pixels to the occluded and far-range vertices.
While most rasterizers only support rendering based on direct illumination, \cite{lyu2021efficient} also supports differentiable rendering of soft shadows. 
In the domain of physics-based rendering, \cite{li2018differentiable} and \cite{azinovic2019inverse} introduced a differentiable ray tracer to implement the differentiability of physics-based rendering effects, handling camera position, lighting and texture.
In addition, Mitsuba 2~\cite{mitsuba2} and Taichi~\cite{hu2019taichi, difftaichi} are general-purpose physically based renderers that support differentiable mesh rendering via automatic differentiation, among many other graphics techniques.

\paragraph*{Neural Implicit Surface Rendering.}

When the input observations are in the form of 2D images, the network which implements the implicit surface is extended to not only produce geometry-related quantities, i.e., signed distance values, but also appearance-related quantities.
An implicit differentiable renderer~\cite{sitzmann2019srns,Niemeyer2020CVPR,liu2020dist,liu2019learning,yariv2020multiview,kellnhofer2021neural,bergman2021metanlr,takikawa2021nglod} can be implemented by first finding the intersection between a viewing ray and the surface using the geometric branch of the neural implicit function, and then obtaining the RGB value of this point from the appearance branch.
The search of surface intersection is typically based on some variant of the sphere tracing algorithm~\cite{hart1996sphere}. 
Sphere tracing iteratively samples the 3D space from the camera center in the direction of the view ray until the surface is reached.
Sphere tracing is an optimized ray marching approach that adjusts the step size by the SDF value sampled at the previous location, but still this iterative strategy can be computationally expensive.
Takikawa et al.~\shortcite{takikawa2021nglod} improved the rendering performance by adapting the ray-tracing algorithm to the sparse octree data structure.
A common problem for implicit surface rendering for joint geometry and appearance estimation from 2D supervision is the ambiguity of geometry and appearance.
In~\cite{Niemeyer2020CVPR,yariv2020multiview,kellnhofer2021neural,bergman2021metanlr}, foreground masks were extracted from the 2D images to provide additional supervision signals for the geometry branch.
Recently, \cite{oechsle2021unisurf} and \cite{yariv2021volume} addressed this issue by formulating the surface function into the volumetric rendering formulation (introduced below); on the other hand \cite{zhang2021learning} use off-the-shelf depth estimation methods to generate pseudo ground truth signed distance values to assist the training of the geometry branch.

\subsubsection{Volumetric Rendering}
\label{sec:volume_rendering}

Volumetric rendering is based on ray casting and has proven to be effective in neural rendering and, especially, in learning a scene representation from multi-view input data.
Specifically, the scene is represented as a continuous field of volume density or occupancy rather than a collection of hard surfaces.

This means that rays have some probability of interacting with the scene content at each point in space, rather than a binary intersection event. This continuous model works well as a differentiable rendering framework for machine learning pipelines that rely heavily on the existence of well-behaved gradients for optimization. 

Though fully general volumetric rendering does account for ``scattering'' events where rays can be reflected off of a volumetric particles~\cite{jarosz08thesis}, we will limit this summary to the basic model commonly used by neural volumetric rendering methods for view synthesis~\cite{levoy_lightfield_rendering, max_direct_vol_rendering}, which only accounts for ``emission'' and ``absorption'' events, where light is emitted or blocked by a volumetric particle.

Given a set of pixel coordinates, we can use the camera model previously described to calculate the corresponding ray through 3D space with origin $\mathbf p$ and direction $\omega_{\text{o}}$. The incoming light along this ray can be defined using a simple emission/absorption model as 
\begin{equation}
    L(\mathbf p, \omega_{\text{o}}) = \int_{t_0}^{t_1} T(\mathbf p, \omega_{\text{o}}, t_0,  t) \sigma(\mathbf p + t\omega_{\text{o}})  L_{\text{e}}(\mathbf p + t\omega_{\text{o}}, -\omega_{\text{o}}) \, dt \, ,
\end{equation}
where $\sigma$ is volume density at a point, $L_{\text{e}}$ is emitted light at a point and direction, and transmittance $T$ is a nested integral expression
\begin{equation}
    T(\mathbf p,\, \omega_{\text{o}},\, t_0, \, t) = \exp\left(- \int_{t_0}^t \sigma(\mathbf p + s\omega_{\text{o}})  \, ds \right) 
    \, .
\end{equation}
Density denotes the differential probability that a ray interacts with the volumetric ``medium'' of the scene at a particular point, whereas transmittance describes how much light will be attenuated as it travels back toward the camera from point $p + t\omega_{\text{o}}$.

These expression can only be evaluated analytically for simple density and color fields. In practice, we typically use quadrature to approximate the integrals, where $\sigma$ and $L_{\text{e}}$ are assumed to be piecewise-constant within a set of $N$ intervals $\{[t_{i-1}, t_i)\}_{i=1}^N$ that partition the length of the ray:
\begin{align}
    L(\mathbf p,\, \omega_{\text{o}}) \,&\approx \, 
    \sum_{i=1}^N T_i \alpha_i \, L_{\text{e}}^{(i)} \, , \\
    T_i &= \exp\left(- \sum_{j=1}^{i-1} \Delta_j \sigma_j \right)  \, , \\
    \alpha_i &= 1 - \exp(-\Delta_i \sigma_i) \, , \\
    \Delta_i &= t_i - t_{i-1} \, .
\end{align}
For a full derivation of this approximation, we refer the reader to Max and Chen~\cite{Max2010LocalAG}. Note that when written in this form, the expression for approximating $L$ exactly corresponds to alpha compositing the colors $L_{\text{e}}^{(i)}$ from back to front~\cite{porterduff}.

NeRF~\cite{Mildenhall_2020_NeRF} and related methods (e.g.,~\cite{martinbrualla2020nerfw,niemeyer2020giraffe,pumarola2020d,srinivasan2021nerv,zhang2020nerf,Neff2021}) use differentiable volume rendering to project the scene representations into 2D images. This allows these methods to be used in an ``inverse rendering'' framework, where a three- or higher-dimensional scene representation is estimated from 2D images. Volume rendering requires many samples to be processed along a ray, each requiring a full forward pass through the network. Recent work has proposed enhanced data structures~\cite{yu2021plenoctrees,hedman2021snerg,garbin2021fastnerf}, pruning~\cite{liu2020neural}, importance sampling~\cite{Neff2021}, fast integration~\cite{lindell2020autoint}, and other strategies to accelerate the rendering speed, although training times of these methods are still slow. Adaptive coordinate networks accelerate training using a multi-resolution network architecture that is optimized during the training phase by allocating available network capacity in an optimal and efficient manner~\cite{martel2021acorn}.

\subsection{Optimization}
\label{sec:opt}

At the heart of training neural networks lies a non-linear optimization which aims to apply the constraints of the training set in order to obtain a set of neural network weights.
As a result, the function which is approximated by the neural network is fit to the given training data.
Typically, optimization of the neural networks is gradient-based; more specifically SGD variants such as Momentum or Adam \cite{adam} are utilized, where the gradients are obtained by leveraging the backpropagation algorithm.
In the context of neural rendering, the neural network implements the 3D scene representation, and the training data consists of 2D observations of the scene. 
The renderings obtained using differentiable rendering of the neural scene representations is compared with the given observation using various loss functions. 
These reconstruction losses can be realized with per-pixel L1 or L2 terms, but also using perceptual~\cite{perceptual_loss} or even discriminator-based loss formulations \cite{Goodfellow_GAN}.
However, key is that the losses are directly coupled with the respective differentiable rendering formulation in order to update the scene representations, cf.~\Cref{sec:scene_rep}.
 \vspace{1cm}
    \vspace{-1.0cm}
\section{Applications} 
\label{sec:applications}
In this section, we discuss the specific applications of neural rendering and the underlying neural scene representations. 
We first discuss improvements to novel view synthesis of static content in \Cref{sec:static_nvp}. 
We then give an overview over methods that generalize across objects and scenes in \Cref{sec:generalization}. 
After that, \Cref{sec:dynamic_content} discusses non-static, dynamic scenes. 
We next turn to editing and composing scenes in \Cref{sec:compedit}. 
Then we provide an overview over relighting and material editing in \Cref{sec:relight}. 
Finally, we discuss several engineering frameworks in \Cref{sec:engineering}. 
We also develop a taxonomy of the different methods for each application. 
These are presented in \Cref{tbl:overview_novel_view}, \Cref{tbl:overview_generalization}, \Cref{tbl:overview_dynamic_content}, \Cref{tbl:overview_compedit}, and \Cref{tbl:overview_relight}, respectively.

\subsection{Novel View Synthesis of Static Content}
\label{sec:static_nvp}

\begin{table}
    \centering
    \begin{adjustbox}{max width=\linewidth}
    \begin{tabular}{lcccccc}
        \toprule
        Method &
        \rot{Required Data} &
        \rot{Requires Pre-trained NeRF} &
        \rot{3D Representation} & 
        \rot{Persistent 3D} & 
        \rot{Network Inputs} &          %
        \rot{Code} \\                   %
        
        \midrule
        Mildenhall~\etal~\cite{Mildenhall_2020_NeRF}
        & \opA{I}+\opB{P} & \xmark{} & \opB{V} & \opA{F} 
        & \opD{PE}(\opA{P})+\opD{PE}(\opF{V})  &  \code{https://github.com/bmild/nerf} 
        \\
        
        \cmidrule(lr){1-7}
        Sitzmann~\etal~\cite{sitzmann2019srns}
        & \opA{I}+\opB{P} & \xmark{} & \opC{S} & \opF{P}      
        & \opA{P} &  \code{https://github.com/vsitzmann/scene-representation-networks} 
        \\
        
        \cmidrule(lr){1-7}
        Niemeyer~\etal~\cite{Niemeyer2020CVPR}
        & \opA{I}+\opB{P}+\opD{M} & \xmark{} & \opD{O} & \opA{F}      
        & \opA{P} & \code{https://github.com/autonomousvision/differentiable_volumetric_rendering} 
        \\

        \cmidrule(lr){1-7}
        Chen~\etal~\cite{chen2019learning}
        & \opC{S} & \xmark{} & \opD{O} & \opA{F}      
        & \opA{P} &  \code{https://github.com/czq142857/implicit-decoder} 
        \\
        
        \cmidrule(lr){1-7} 
        Gu~\etal~\cite{liu2020neural} 
        & \opA{I}+\opB{P} & \xmark{} & \opA{G}+\opB{V} & \opA{F} 
        & \opD{PE}(\opA{P})+\opD{PE}(\opF{V}) &  \code{https://github.com/facebookresearch/NSVF} 
        \\
        
        \cmidrule(lr){1-7}
        Lindell~\etal~\cite{lindell2020autoint}
        & \opA{I}+\opB{P} & \xmark{} & \opB{V} & \opF{P}      
        & \opD{PE}(\opA{P})+\opD{PE}(\opF{V}) & \code{https://github.com/computational-imaging/automatic-integration}   
        \\
        
        \cmidrule(lr){1-7}
        Reiser~\etal~\cite{Reiser2021}
        & \opA{I}+\opB{P} & \cmark{} & \opA{G}+\opB{V} & \opA{F}     
        & \opD{PE}(\opA{P})+\opD{PE}(\opF{V}) & \code{https://github.com/creiser/kilonerf} 
        \\
        
        \cmidrule(lr){1-7}
        Garbin~\etal~\cite{garbin2021fastnerf}
        & \opA{I}+\opB{P} & \cmark{} & \opA{G} & \opA{F}      
        & \opA{P}+\opF{V} & \xmark{}       
        \\
        
        \cmidrule(lr){1-7}
        Hedman~\etal~\cite{hedman2021snerg}
        & \opA{I}+\opB{P} & \cmark{} & \opA{G} & \opA{F}      
        & \opA{P}+\opD{PE}(\opF{V}) & \code{https://github.com/google-research/google-research/tree/master/snerg}         
        \\
        
        \cmidrule(lr){1-7}
        Yu~\etal~\cite{yu2021plenoctrees}
        & \opA{I}+\opB{P} & \cmark{} & \opA{G} & \opA{F}      
        & \opA{P}+\opF{V} & \code{https://github.com/sxyu/plenoctree}
        \\
        
        \cmidrule(lr){1-7}
        Neff~\etal~\cite{Neff2021}
        & \opA{I}+\opB{P}+\opE{D} & \xmark{} & \opB{V} & \opA{F}      
        & \opD{PE}(\opA{P})+\opD{PE}(\opF{V}) & \code{https://github.com/facebookresearch/DONERF}     
        \\
        
        \cmidrule(lr){1-7}
        Sitzmann~\etal~\cite{sitzmann2021lfns}
        & \opA{I}+\opB{P} & \xmark{} & \xmark{} & \opC{N}      
        & \opC{L} & \code{https://github.com/vsitzmann/light-field-networks}         
        \\
            \end{tabular}
    \end{adjustbox}
    \caption{Selected methods for static scene view synthesis presented in \Cref{sec:static_nvp}. Although some of these representations are used for applications beyond static scene view synthesis, in this table we only classify such methods based on the use of their underlying 3D scene representation for static scene view synthesis. %
        \tableopt{opA}{I}:~Images,
        \tableopt{opB}{P}:~Camera poses (exact or approximate),
        \tableopt{opC}{S}:~3D shape,
        \tableopt{opD}{M}:~Object masks,
        \tableopt{opE}{D}:~Depth.
        \tableopt{opA}{G}:~Grid,
        \tableopt{opB}{V}:~Neural volumetric,
        \tableopt{opC}{S}:~Neural SDF,
        \tableopt{opD}{O}:~Neural occupancy.
        \tableopt{opA}{F}:~Fully,
        \tableopt{opF}{P}:~Partial,
        \tableopt{opC}{N}:~Not guaranteed.
        \tableopt{opA}{P}:~3D position,
        \tableopt{opF}{V}:~2D viewing direction,
        \tableopt{opC}{L}:~Light field ray coordinates.
        \tableopt{opD}{PE}():~Positinal encoding of argument.
    } 
    \label{tbl:overview_novel_view}
\end{table}

Novel view synthesis is the task of rendering a given scene from new camera positions, given a set of images and their camera poses as input. Most of the applications presented later in this section generalize the task of view synthesis in some way: in addition to being able to move the camera, they might allow moving or deforming objects within the scene, changing the lighting, and so on. 

View synthesis methods are evaluated on a few salient criteria. Clearly, output images should look as realistic as possible. However, this is not the whole story --- perhaps even more important is \emph{multiview 3D consistency}. Rendered video sequences must appear to portray consistent 3D content as the camera moves through the scene, without flickering or warping. As the field of neural rendering has matured, most methods have moved in the direction of producing a fixed 3D representation as output that can be used to render new 2D views, as explained in the scope. This approach automatically lends a degree of multiview consistency that has historically been hard to achieve when relying too heavily on black-box 2D convolutional networks as image generators or renderers. 

In \Cref{tbl:overview_novel_view}, we give an overview over the discussed methods.

\subsubsection{View Synthesis from a 3D Voxel Grid Representation}

We will briefly review the recent history of view synthesis using 3D voxel grids and a volumetric rendering model.

DeepStereo~\cite{flynn2016deepstereo} presented the first end-to-end deep learning pipeline for view synthesis. This work included many concepts that have now become commonplace. A convolutional neural network is presented with input images in the form of a plane sweep volume (PSV), where each nearby input is reprojected to a set of candidate depth planes, requiring the network to simply evaluate how well the reprojections match for each pixel at each candidate depth. The CNN's outputs are converted into a probability distribution over depths using a softmax, which is then used to combine a stack of proposed color images (one per depth plane). The final loss is only enforced on the pixel-wise difference between the rendered output and a heldout target image, with no intermediate heuristic losses required. 

A major drawback of DeepStereo is that it requires running a CNN to estimate depth probabilities and produce each output frame independently, resulting in slow runtime and a lack of multiview 3D consistency. Stereo Magnification~\cite{zhou2018} directly addresses this issue, using a CNN to process a plane sweep volume directly into an output persistent 3D voxel grid representation named a ``multiplane image,'' or MPI. Rendering new views simply requires using an alpha compositing to render the RGB-alpha grid from a new location. In order to achieve high image quality, Stereo Magnification heavily distorts the parameterization of its 3D grid to bias it to the frame of reference of one of the two input views. This significantly decreases storage requirements for the dense grid but means that new views can only be rendered in the direct neighborhood of the input stereo pair. This shortcoming was later addressed by improving the training procedure for a single MPI~\cite{srinivasan19}, providing many more than two input images to the network~\cite{flynn2019deepview}, or combining multiple MPIs together to represent a single scene~\cite{Mildenhall:2019}.

All methods mentioned above use a feed-forward neural network to map from a limited set of input images to an output image or 3D representation and must be trained on a large dataset of pairs of input/output views. In contrast, DeepVoxels~\cite{sitzmann2019deepvoxels} optimizes a 3D voxel grid of features jointly with a learned renderer using images of a \emph{single} scene, without requiring any external training data. Similarly, Neural Volumes~\cite{Lombardi_2019_NeuralVolumes} optimizes a 3D CNN to produce an output volumetric representation for a single scene of multiview video data This single-scene training paradigm has greatly increased in popularity recently, leveraging the unique ``self-supervised'' aspect of view synthesis: any input images can also be used as supervision via a rerendering loss. In comparison to MPI-based methods, DeepVoxels and Neural Volumes also use a 3D voxel grid parameterization that is not heavily skewed to one particular viewing direction, allowing novel views to be rendered observing the reconstructed scene from any direction. 

It is worth mentioning that a number of computer vision papers focused primarily on 3D shape reconstruction (rather than realistic image synthesis) adopted an alpha compositing volumetric rendering model in parallel with this view synthesis research~\cite{henzler18,kar17,tulsiani17}; however, these results were heavily constrained by the memory limitations of 3D CNNs and could not produce voxel grid outputs exceeding $128^3$ resolution.

\begin{figure}[t!]
    \centering
    \includegraphics[width=\linewidth]{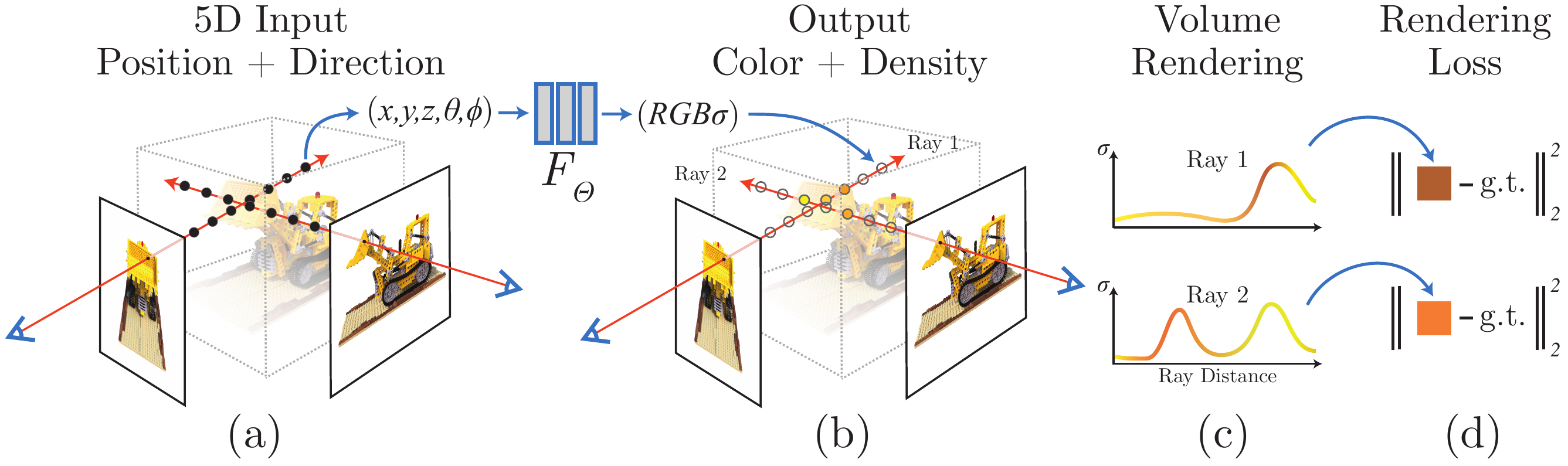}
    \caption{
    An overview of the neural radiance field (NeRF) scene representation and volume rendering procedure. NeRF synthesizes images by sampling 5D coordinates (location and viewing direction) along camera rays (a), feeding those locations into an MLP to produce color and volume density (b), and using volume rendering to composite these values into an image (c). Since this rendering function is differentiable, the NeRF scene representation MLP can be optimized by minimizing the residual between synthesized and ground truth observed images (d). 
    Digital zoom recommended. 
    Image adapted from~\cite{Mildenhall_2020_NeRF}. 
    }
    \label{fig:nerf_pipeline}
\end{figure}

\subsubsection{View Synthesis from a Neural Network Representation}

To address the resolution and memory limitations of voxel grids, Scene Representation Networks (SRNs)~\cite{sitzmann2019srns} combined a sphere-tracing based neural renderer with a multilayer perceptron (MLP) as a scene representation, focusing mainly on generalization across scenes to enable few-shot reconstruction. Differentiable Volumetric Rendering (DVR)~\cite{Niemeyer2020CVPR} similarly leveraged a surface rendering approach, but demonstrated that overfitting on single scenes enables reconstruction of more complex appearance and geometry.

Neural radiance fields (NeRF~\cite{Mildenhall_2020_NeRF} signified a breakthrough in the application of MLP-based scene representations to single-scene, photorealistic novel view synthesis, see \Cref{fig:nerf_pipeline}. Instead of a surface-based approach, NeRF directly applies the volume rendering model described in \Cref{sec:volume_rendering} to synthesize images from an MLP that maps from an input position and viewing direction to output volume density and color. A different set of MLP weights are optimized to represent each new input scene based on pixelwise rendering loss against the input images. 

This overall framework shares many similarities with the work described in the previous section. However, MLP-based scene representations can achieve higher resolution than discrete 3D volumes by virtue of effectively \emph{differentiably} compressing the scene during optimization. For example, a NeRF representation capable of rendering $800 \times 800$ resolution output images only required 5MB of network weights. In comparison, an $800^3$ RGBA voxel grid would consume close to 2GB of storage. 

This ability can be attributed to NeRF's use of a \emph{positional encoding} applied to the input spatial coordinates before passing through the MLP. In comparison to the previous work on using neural networks to represent implicit surfaces~\cite{park2019deepsdf,chen2019learning} or volumes~\cite{mescheder2019occupancy}, this allows NeRF's MLP to represent much higher frequency signals without increasing its capacity (in terms of number of network weights). 

The main drawback of switching from a discrete 3D grid to an MLP-based representation is rendering speed. Rather than directly querying a simple data structure, calculating the color and density value for a \emph{single point} in space now requires evaluating an entire neural network (hundreds of thousands of floating point operations). On a typical desktop GPU, an implementation of NeRF in a standard deep learning framework takes tens of seconds to render a single high resolution image.

\subsubsection{Improving Rendering Speed}

Several different methods have been proposed for speeding up volumetric rendering of MLP-based representations. Neural Sparse Voxel Fields~\cite{liu2020neural} builds and dynamically updates an octree structure while the MLP is optimized, allowing for aggressive empty space skipping and early ray termination (when the transmittance along the ray approaches zero). KiloNeRF~\cite{Reiser2021} combines empty space skipping and early termination with a dense 3D grid of MLPs, each with a much smaller number of weights than a standard NeRF network.

Three concurrent works recently proposed methods for caching the values various quantities learned by the NeRF MLP on a sparse 3D grid, allowing for realtime rendering once training is complete. Each method modifies the way in which view-dependent colors are predicted in order to facilitate faster rendering and smaller memory requirements for the cached representations. SNeRG~\cite{hedman2021snerg} stores volume density and a small spatially-varying feature vector in a sparse 3D texture atlas, using fast shader for compositing these values along a ray and running a tiny MLP decoder to produce view-dependent color for each ray. FastNeRF~\cite{garbin2021fastnerf} caches volume density along with weights for combining a set of learned spherical basis functions that produce view-varying colors at each point in 3D. PlenOctrees~\cite{yu2021plenoctrees} queries the MLP to produce a sparse voxel octree of volume density and spherical harmonic coefficients and further finetunes this octree representation using a rendering loss to improve its output image quality.

NeX-MPI~\cite{Wizadwongsa2021NeX} combines the multiplane image parameterization with an MLP scene representation, with view dependent effects parameterized as a linear combination of globally learned basis functions. Because the model is supervised directly on a 3D MPI grid of coordinates, this grid can be easily cached to render new views in real time once optimization is complete.

An alternative approach for accelerating rendering is to train the MLP representation itself to effectively precompute part or all of the volume integral along the ray. AutoInt~\cite{lindell2020autoint} trains a network to ``automatically integrate'' the output color value along ray segments by supervising the \emph{gradient} of the network to behave like a standard NeRF MLP. This allows the rendering step to break the integral along a ray into an order of magnitude fewer segments than the standard quadrature estimate (down to as few as 2 or 4 samples), trading off speed for a minor loss in quality. Light Field Networks~\cite{sitzmann2021lfns} takes this a step further, optimizing an MLP to directly encode the mapping from an input ray to an output color (the scene's light field). This enables rendering with only a \emph{single} evaluation of the MLP per ray, in contrast to hundreds of evaluations for volume- and surface-based renderers, and enables real-time novel view synthesis. These methods present a tradeoff between rendering speed and multiview consistency: reparameterizing the MLP representation as a function of rays rather than 3D points means that the scene is no longer guaranteed to appear consistent when viewed from different angles. In this case multiview consistency must be enforced through supervision, either by providing a very large number of input images or learning this property via generalization across a dataset of 3D scenes.

Recently, there has also been a flood of new approaches \cite{piala2021terminerf,fang2021neusample,yu2021plenoxels,wu2021diver,sun2021direct,kondo2021vaxnerf} that employ classical data structures, such as grids, sparse grids, trees, and hashes, for acceleration of rendering speed as well as faster training times.
Instant Neural Graphics Primitives~\cite{mueller2022ngp} enables the training of a NeRF in a few seconds exploiting a multi-resolution hash encoding instead of an explicit grid structure.

\subsubsection{Miscellaneous Improvements}

A variety of papers have augmented the rendering model, supervision data, or robustness of volumetric MLP scene representations.

\paragraph*{Depth Supervision.} DONeRF~\cite{Neff2021} trains an ``depth oracle'' network to predict sample locations along each ray, drastically reducing the number of samples sent through the NeRF MLP and allowing interactive rate rendering. However, this method is supervised with dense depths maps, which are challenging to obtain for real data. Depth-supervised NeRF~\cite{kangle2021dsnerf} directly supervises the output depths from NeRF (in the form of expected termination depth along each ray) using the sparse point cloud output which is a byproduct of estimating camera poses using structure-from-motion. 
NerfingMVS~\cite{wei2021nerfingmvs} uses a multistage pipeline for depth supervision, first finetuning a single-view depth estimation network on sparse multiview stereo depth estimates, then uses the resulting dense depth maps to guide NeRF sample placement.
Roessle et al.~\cite{roessle2021dense} directly applies a pretrained sparse-to-dense depth completion network to sparse structure-from-motion depth estimates, then uses depth (along with predicted uncertainty) to both guide sample placement and supervise the depth produced by NeRF.

\paragraph*{Optimizing Camera Poses.} NeRF-{}-~\cite{wang2021nerfmm} and Self-Calibrating Neural Radiance Fields  \cite{Jeong_2021_ICCV_Self_calibrating} jointly optimize the NeRF MLP and input camera poses, bypassing the need for structure-from-motion preprocessing for forward facing scenes. Bundle-Adjusting Neural Radiance Fields (BARF)~\cite{lin2021barf} extends this idea by applying a coarse-to-fine annealing schedule to each frequency component of the positional encoding function, providing a smoother optimization trajectory for joint reconstruction and camera registration. However, neither of these methods can optimize poses from scratch for wide-baseline 360 degree captures. GNeRF~\cite{Meng_2021_ICCV_GNeRF} achieves this by training a set of cycle consistent networks (a generative NeRF and a pose classifier) that map from pose to image patches and back to pose, optimizing until the classified pose of real patches matches that of sampled patches. They alternate this GAN training phase with a standard NeRF optimization phase until the result converges.

\paragraph*{Hybrid Surface/Volume Representations.} 
The Implicit Differentiable Renderer (IDR) from Yariv et al.~\cite{yariv2020multiview} combines a DVR-like implicit surface MLP with a NeRF-like view dependent branch which takes viewing direction, implicit surface normal, and the 3D surface point as inputs and predicts the view-varying output color. This work shows that including the normal vector as input to the color branch helps the representation disentangle geometry and appearance more effectively. It also demonstrates that camera pose can be jointly optimized along with the shape representation to recover from small miscalibration errors.

UNISURF~\cite{oechsle2021unisurf} proposes a hybrid MLP representation that unifies surface and volume rendering. To render a ray, UNISURF uses root finding to get a ``surface'' intersection point, treating the volume as an occupancy field, then distributes volume rendering samples only within an interval around that point. The width of this interval monotonically decreases over the course of optimization, allowing early iterations to supervise the whole training volume and later stages to more efficiently refine the surface with tightly spaced samples. Azinovic et al.~\cite{azinovic2021rgbdnerf} propose to use an SDF representation instead of volume densities to reconstruct scenes from RGB-D data. They convert the sdf values to densities that can be used in the NeRF formulation. NeuS~\cite{wang2021neus} ties the volume density to an signed distance field and reparameterizes the transmittance function such that it achieves its maximal slope precisely at the zero-crossing of this SDF, allowing an unbiased estimate of the corresponding surface. VolSDF~\cite{yariv2021volume} uses an alternate mapping from SDF to volume density, which allows them to devise a new resampling strategy to achieve provably bounded error on the approximated opacity in the volume rendering quadrature equation.
In \cite{Li_2021_ICCV}, the authors propose a method called MINE which is a hybrid between multi-plane images (MPI) and NeRF. They are able to reconstruct dense 3D reconstructions from single color images which they demonstrate on RealEstate10K, KITTI and Flowers Light Fields.

\paragraph*{Robustness and Quality.} NeRF++~\cite{zhang2020nerf} provides a ``inverted sphere'' parameterization of space that can allow NeRF to large-scale, unbounded 3D scenes. Points outside the unit sphere are inverted back into the unit sphere and passed through a separate MLP.

NeRF in the Wild~\cite{martinbrualla2020nerfw} adds additional modules to the MLP representation to account for inconsistent lighting and objects across different images. They apply their robust model to the PhotoTourism dataset~\cite{phototourism} (consisting of internet images of famous landmarks across the world) and are able to remove transient objects such as people and cars and capture time-varying appearance through use of a latent code embedding associated with each input image.
Ha-NeRF~\cite{chen2021hallucinated} extends the idea of NeRF in the Wild to hallucinate novel appearances.

\begin{figure}[t!] 
    \centering 
    \includegraphics[width=\linewidth]{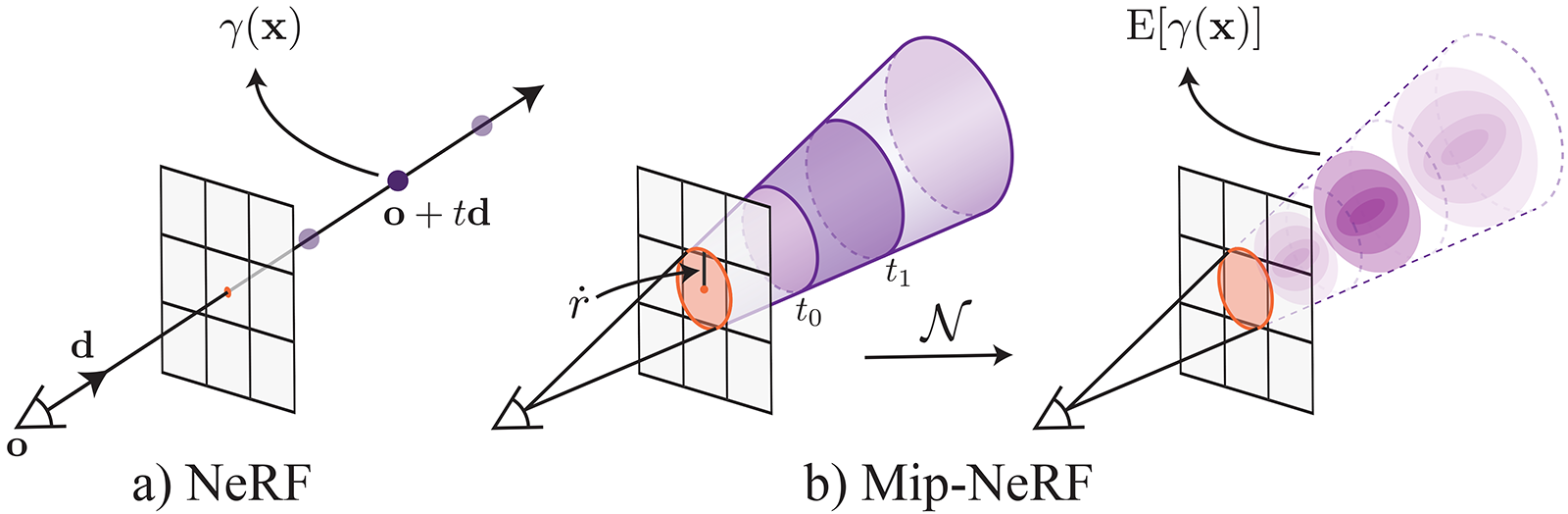} 
    \caption{Instead of sampling points $\mathbf{x}$ along the rays traced from the camera projection center (a), MipNeRF \cite{barron2021mipnerf} reasons about 3D canonical frustum per camera pixel (b). Image adapted from \cite{barron2021mipnerf}  \textcopyright 2021 IEEE.} 
    \label{fig:MipNeRF} 
\end{figure} 

MipNeRF~\cite{barron2021mipnerf} modifies the positional encoding applied to 3D points to incorporate the pixel footprint, see \Cref{fig:MipNeRF}. By pre-integrating the positional encoding over a conical frustum corresponding to each quadrature segment sampled along the ray, MipNeRF can be trained to encode a representation of the scene at multiple different scales (analogously to a mipmap of a 2D texture), preventing aliasing when rendering the scene from dramatically varying positions or resolutions. 
Mip-NeRF 360~\cite{barron2021mipnerf360} extends MipNeRF and addresses issues that arise when training on unbounded scenes (unbalanced detail of nearby and distant objects which leads to blurry, low-resolution renderings), where the camera rotates 360 degrees around a point. It leverages a non-linear scene parametrization, online distillation, and a novel distortion-based regularizer.

\vspace{-0.5cm}
\paragraph*{NeRF and Computational Imaging.} Several recent works combine NeRF with standard computational imaging tasks. Deblur-NeRF~\cite{ma2021deblurnerf} jointly optimizes a static NeRF representation along with per-ray offsets for every pixel in the training set that account for blur due to either camera motion or depth of field. Once optimization is complete, the NeRF can be rendered without applying the ray offsets to obtain sharp test views. NeRF in the Dark~\cite{mildenhall2021rawnerf} trains directly on raw linear camera data to achieve improved robustness to high levels of image noise, allowing reconstruction of dark nighttime scenes as well as recovery of full high dynamic range radiance values. HDR-NeRF~\cite{huang2021hdrnerf} similarly recovers a linear-valued HDR NeRF, but by using postprocessed variable-exposure images as input and solving for a nonlinear camera postprocessing curve that reproduces the inputs when applied to the optimized NeRF. NeRF-SR~\cite{wang2021nerfsr} averages multiple supersampled rays per pixel during training and also performs super-resolution on rendered image patches by merging them with similar patches from a high-resolution reference image of the scene using a CNN.

\paragraph*{Large-scale Scenes.}
A series of recent publications focus on large-scale neural radiance fields.
They enable the re-rendering of street-view data~\cite{rematas2021urban}, buildings, and entire cities~\cite{xiangli2021citynerf,turki2021meganerf}, or even earth-scale~\cite{xiangli2021citynerf}.
To handle such large scenes, the methods use localized NeRFs by decomposing the scene into spatial cells~\cite{turki2021meganerf} or on different scales~\cite{xiangli2021citynerf} (including a
progressive training scheme).
URF~\cite{rematas2021urban} exploits additional LIDAR data to supervise the depth prediction.
In these large-scale scenarios, special care must be taken to handle the sky and the highly varying exposure and illumination changes (cf.~NeRF-W~\cite{martinbrualla2020nerfw}). 
NeRF-W~\cite{martinbrualla2020nerfw} interpolates between learned appearances, but does not provide semantic control over it.
NeRF-OSR \cite{rudnev2021nerfosr} is the first method allowing joint editing of the camera viewpoints and the illumination for buildings and historical sites. 
For training, NeRF-OSR requires outdoor photo  collections shot in uncontrolled settings (see Section \ref{sec:relight} for further details).

\paragraph*{NeRF from Text.}
The NeRF formulation~\cite{Mildenhall_2020_NeRF} is an optimization-based framework, which also allows us to incorporate other energy terms during optimization. 
To manipulate or generate a NeRF by text inputs, one can employ a (pretrained) CLIP-based objective~\cite{radford2021learning}.
Dream Fields~\cite{jain2021dreamfields} combines NeRF with CLIP to generate diverse 3D objects solely from natural language descriptions, by optimizing the radiance field via multi-view constraints based on the CLIP scores on the image caption.
CLIP-NeRF~\cite{wang2021clipnerf} proposes a CLIP-based shape and appearance mapper to control a conditional NeRF.

\subsection{Generalization over Object and Scene Classes}
\label{sec:generalization}

\begin{table}
    \centering
    \begin{adjustbox}{max width=\linewidth}
    \begin{tabular}{lccccccc}
        \toprule
        Method &
        \rot{Conditioning} &           %
        \rot{Required Data} &       %
        \rot{3D Representation} &       %
        \rot{Class Specific Prior} &    %
        \rot{Generative Model} &        %
        \rot{Inference Type} &         %
        \rot{Code} \\                   %

        \cmidrule(lr){1-8}
        Yu~\etal~\cite{yu2020pixelnerf}
         & \tableopt{opA}{L}  & \tableopt{opA}{G} & \tableopt{opB}{V} & \xmark{} & \xmark{} & \tableopt{opA}{A} & \code{https://github.com/sxyu/pixel-nerf}         
         \\
         
        \cmidrule(lr){1-8}
        Raj~\etal~\cite{raj2020pva}
        & \tableopt{opA}{L}   & \tableopt{opC}{F} & \tableopt{opB}{V} & \xmark{} & \xmark{} & \tableopt{opA}{A} & \code{}         
         \\
        
        \cmidrule(lr){1-8}
        Trevithick~\etal~\cite{Trevithick2020}
         & \tableopt{opA}{L}  & \tableopt{opA}{G} & \tableopt{opB}{V} & \xmark{} & \xmark{} & \tableopt{opA}{A} & \code{https://github.com/alextrevithick/GRF}         
         \\
        
        \cmidrule(lr){1-8}
        Wang~\etal~\cite{Wang2021ibrnet}
         & \tableopt{opA}{L}  & \tableopt{opA}{G} & \tableopt{opB}{V} & \xmark{} & \xmark{} & \tableopt{opA}{A} & \code{}      
         \\
         
        \cmidrule(lr){1-8}
        Reizenstein~\etal~\cite{reizenstein21co3d}
        &  \tableopt{opA}{L}  & \tableopt{opA}{G} & \tableopt{opB}{V} & \xmark{} & \xmark{} & \tableopt{opA}{A} & \code{}         
         \\
        
        \cmidrule(lr){1-8}
        Sitzmann~\etal~\cite{sitzmann2019srns}
         & \tableopt{opB}{G}  & \tableopt{opA}{G} & \tableopt{opC}{S} & \cmark{} & \cmark{} & \tableopt{opB}{D} & \code{https://github.com/vsitzmann/scene-representation-networks}
        \\

        \cmidrule(lr){1-8}
        Kosiorek~\etal~\cite{Kosiorek2021}
         & \opB{G}  & \tableopt{opA}{G} & \tableopt{opB}{V} & \xmark{} & \cmark{} & \tableopt{opA}{A} & \code{}       
         \\
        
        \cmidrule(lr){1-8}
        Rematas~\etal~\cite{Rematas2021}
         & \opB{G}  & \tableopt{opA}{G} & \tableopt{opB}{V} & \cmark{} & \cmark{} & \tableopt{opB}{D} & \code{https://github.com/tensorflow/graphics/tree/master/tensorflow_graphics/projects/radiance_fields}        
         \\
        
        \cmidrule(lr){1-8}
        Xie~\etal~\cite{xie2021fignerf}
        & \opB{G}  & \tableopt{opA}{G} & \tableopt{opB}{V} & \xmark{} & \xmark{} & \tableopt{opA}{A} & \code{}         
         \\
        
        \cmidrule(lr){1-8}
        Tancik~\etal~\cite{tancik2020learned}
        & \opB{G}  & \tableopt{opA}{G} & \tableopt{opB}{V} & \xmark{} & \xmark{} & \tableopt{opC}{GB} & \code{https://github.com/tancik/learnit}        
         \\
        
        \cmidrule(lr){1-8}
        Gao~\etal~\cite{Gao-portraitnerf}
        & \opB{G}  & \tableopt{opC}{F} & \opB{V} & \xmark{} & \xmark{} & \tableopt{opC}{GB} & \code{}        
         \\
        
        \cmidrule(lr){1-8}
        Nguyen-Phuoc~\etal~\cite{nguyen2019hologan}
        & \opB{G}  & \tableopt{opA}{G} & \opB{V} & \cmark{} & \cmark{} & \xmark & \code{https://github.com/thunguyenphuoc/HoloGAN}          
         \\
        
        \cmidrule(lr){1-8}
        Schwarz~\etal~\cite{Schwarz2020GRAF}
        & \opB{G}  & \tableopt{opA}{G} & \opB{V} & \cmark{} & \cmark{} & \xmark & \code{}         
         \\
        
        \cmidrule(lr){1-8}
        Chan~\etal~\cite{chanmonteiro2020piGAN}
        & \opB{G}  & \tableopt{opA}{G} & \opB{V} & \cmark{} & \cmark{} & \xmark & \code{https://github.com/marcoamonteiro/pi-GAN}        
         \\
        
        \cmidrule(lr){1-8}
        Anonymous~\cite{anonymous2022stylenerf}
        & \tableopt{opB}{G}  & \tableopt{opA}{G} & \tableopt{opB}{V} & \cmark{} & \cmark{} & \xmark & \code{}          
         \\
        
        \cmidrule(lr){1-8}
        Niemeyer~\etal~\cite{Niemeyer2021campari}
        & \tableopt{opB}{G}  & \tableopt{opA}{G} & \opB{V} & \cmark{} & \cmark{} & \xmark & \code{}        
         \\

        \end{tabular}
    \end{adjustbox}
    \caption{Selected methods for generalization presented in \Cref{sec:generalization}. 
    \tableopt{opB}{G}:~Global,
    \tableopt{opA}{L}:~Local.
    \tableopt{opA}{I}:~Implicitly,
    \tableopt{opB}{E}:~Explicitly.
    \tableopt{opA}{G}:~General,
    \tableopt{opB}{B}:~Body,
    \tableopt{opC}{F}:~Face.
    \tableopt{opA}{A}:~Amortized/Encoder,
    \tableopt{opB}{D}:~Auto-Decoder,  
    \tableopt{opC}{GB}:~Gradient-based.
    \tableopt{opB}{V}:~Neural volumetric,
    \tableopt{opC}{S}:~Neural SDF.
    }
    \label{tbl:overview_generalization}
\end{table}

While a significant amount of prior work addresses generalization over multiple scenes and object categories for voxel-based, mesh-based, or non-3D structured neural scene representations, we focus this discussion on recent progress in generalization leveraging MLP-based scene representations. Where approaches that overfit a single MLP on a single scene~\cite{Mildenhall_2020_NeRF,yariv2020multiview} require a large number of image observations, the core objective of generalizing across scene representations is novel view synthesis given few or potentially only a single input view.
In \Cref{tbl:overview_generalization}, we give an overview over the discussed methods, classified by whether they leverage local or global conditioning, whether they can be used as unconditional generative models or not, what kind of 3D representation they leverage (volumetric, SDF, or occupancy), what kind of training data they require, and how inference is performed (amortized with an encoder, via the auto-decoder framework, or via gradient-based meta-learning).

\begin{figure}[t!] 
    \centering 
    \includegraphics[width=\linewidth]{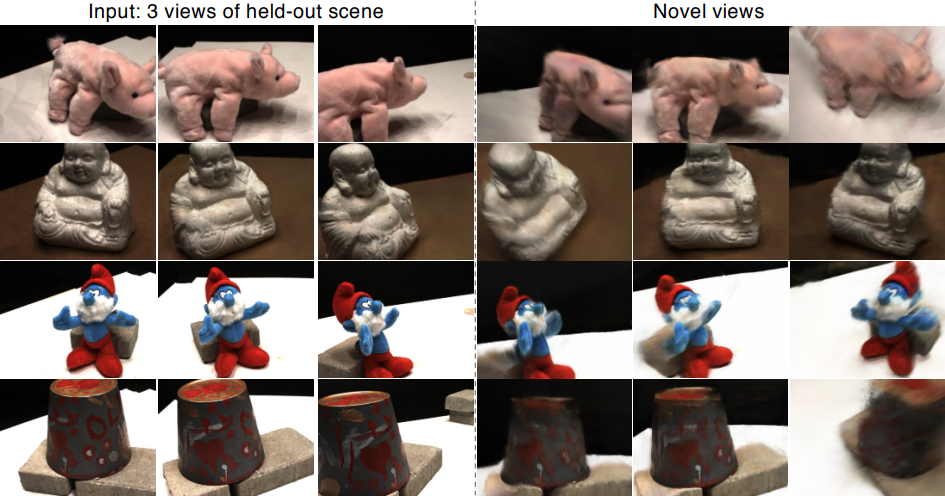} 
    \caption{Input images from DTU MVS dataset \cite{Jensen2014} and novel views obtained by PixelNeRF \cite{yu2020pixelnerf} with no test-time optimization. Furthermore, training and test sets do not share the same scenes. Image adapted from  \cite{yu2020pixelnerf}.} 
    \label{fig:PixelNeRF} 
\end{figure} 

We may differentiate two key approaches in generalizing across scenes.
One line of work follows an approach reminiscent of image-based rendering~\cite{chen1993view, shum2000review}, where multiple input views are warped and blended to synthesize a novel viewpoint. In the context of MLP-based scene representations, this is often implemented via \emph{local conditioning}, where the coordinate input to the scene representation MLP is concatenated with a locally varying feature vector, stored in a discrete scene representation, such as a voxel grid~\cite{peng2020convolutional}. PiFU~\cite{saito2019pifu} uses an image encoder to compute features on the input image and conditions a 3D MLP on these features via projecting 3D coordinates on the image plane - however, PiFU did not feature a differentiable renderer, and so required ground-truth 3D supervision. PixelNeRF~\cite{yu2020pixelnerf} (see \Cref{fig:PixelNeRF})  and Pixel-Aligned Avatars~\cite{raj2020pva} leverage this approach in a volume rendering framework where these features are aggregated over multiple views, and a MLP produces color and density fields that are rendered as in NeRF. When trained on multiple scenes, they learn scene priors for reconstruction, that enable high fidelity reconstruction of scenes from a few views. PixelNeRF can also be trained on specific object categories, enabling object instance 3D reconstruction from one or multiple posed images. GRF~\cite{Trevithick2020} uses a similar framework, with an additional attention module that reasons about the visibility of the 3D point in the different sampled input images. Stereo Radiance Fields~\cite{Chibane2021} similarly extracts features from several context views, but leverages learned correspondence matching between pairwise features across context images to aggregate features across context images instead of a simple mean aggregation. Finally, IBRNet~\cite{Wang2021ibrnet} and NeRFormer~\cite{reizenstein21co3d} introduce transformer networks across the ray samples that reason about visibility.
LOLNeRF~\cite{rebain2021lolnerf} learns a generalizable NeRF model on portrait images using only monocular supervision.
The generator network is conditioned on instance-specific latent vectors, which are jointly trained. 
Joint training on large datasets enable training without multi-view supervision.
GeoNeRF \cite{johari2021geonerf} constructs a set of cascaded cost volumes and employs transformers to infer geometry and appearance.

An alternative to such image-based approaches aims to learn a monolithic, global representation of a scene instead of relying on images or other discrete spatial data structures. This is accomplished by inferring a set of weights for the scene representation MLP that describes the whole scene, given a set of observations. One line of work accomplishes this by encoding a scene in a single, low-dimensional latent code that is then used to condition the scene representation MLP.  Scene Representation Networks (SRNs)~\cite{sitzmann2019srns} map low-dimensional latent codes to the parameters of a MLP scene representation via a hypernetwork, and subsequently render the resulting 3D MLP via ray-marching. To reconstruct an instance given a posed view, SRNs optimize the latent code so that its rendering matches the input view(s). Differentiable Volumetric Rendering~\cite{Niemeyer2020dvr} similarly uses surface rendering, but computes its gradients analytically and performs inference via a CNN encoder. Light Field Networks~\cite{sitzmann2021lfns} leverage low-dimensional latent codes to directly parameterize the 4D light field of the 3D scene, enabling single-evaluation rendering. NeRF-VAE embeds a NeRF in a variational auto-encoder, similarly representing the whole scene in a single latent code, but learning a generative model that enables sampling~\cite{Kosiorek2021}. Sharf~\cite{Rematas2021} uses a generative model of voxelized shapes of objects in a category, which in turn condition a higher resolution neural radiance field that is rendered using volume rendering for higher novel view synthesis fidelity. Fig-NeRF~\cite{xie2021fignerf} models an object category as a template shape conditioned on a latent code, that undergoes a deformation that is also conditioned on the same latent variable. This enables the network to explain certain shape variations as more intuitive deformations. Fig-NeRF focuses on retrieving an object category from real object scans, and also proposes using a learn background model to segment the object from its background. 
An alternative to representing the scene as a low-dimensional latent code is to quickly optimize the weights of an MLP scene representation in a few optimization steps via gradient-based meta-learning~\cite{sitzmann2019metasdf}. This can be used to enable fast reconstruction of neural radiance fields from few images~\cite{tancik2020learned}. The pre-trained models converge faster when trained on a novel scene, and require fewer views compared to standard neural radiance field training. PortraitNeRF~\cite{Gao-portraitnerf} proposes a meta-learning approach to recover a NeRF from a single frontal image of a person. To account for differences in pose between the subjects, it models the 3D portraits in a pose-agnostic canonical  reference frame, that is warped for each subject using 3D keypoints. Bergman et al.~\cite{bergman2021metanlr} leverage gradient-based meta-learning and local conditioning on image features to quickly recover a NeRF of a scene.

Instead of \emph{inferring} a low-dimensional latent code conditioned on a set of observations of the sought-after 3D scene, a similar approach can be leveraged to learn \emph{unconditional} generative models.
Here, a 3D scene representation equipped with a neural renderer is embedded in a generative adversarial network. Instead of inferring low-dimensional latent codes from a set of observations, we define a distribution over latent codes. In a forward pass, we sample a latent from that distribution, condition the MLP scene representation on that latent, and render an image via the neural renderer. This image can then be used in an adversarial loss. This enables learning of a \emph{3D} generative model of shape \& appearance of 3D scenes given only \emph{2D} images.
This approach was first proposed with 3D scene representations parameterized via voxelgrids~\cite{nguyen2019hologan}. GRAF~\cite{Schwarz2020GRAF} first leveraged a conditional NeRF in this framework and achieved significant improvements in photorealism. Pi-GAN~\cite{chanmonteiro2020piGAN} further improved on this architecture with a FiLM-based conditioning scheme~\cite{perez2018film} of a SIREN architecture~\cite{sitzmann2020siren}. 

Several recent approaches explore different directions for improving the quality and efficiency of these generative models. 
Computational cost and quality of geometric reconstructions can be improved by using a surface representation\cite{deng2021gram,orel2021styleSDF,xu2021generative}.
In addition to synthesizing multi-view images for the discriminator, ShadeGAN~\cite{pan2021shadegan} uses an explicit shading step to also generate the output image renderings under different illumination conditions for higher-quality geometry reconstructions. 
Many approaches have explored using a hybrid technique where an image-based CNN network is used to refine the output of the 3D generator~\cite{gu2021stylenerf,Chan2021,xu2021volumegan,zhou2021cips3d}. 
The image-space network enables training at higher resolutions with higher-fidelity outputs. 
Decomposing the generative model into  separate geometry and texture spaces has also been explored. 
Here, some approaches learn the texture in image space~\cite{sofgan,xu2021volumegan}, while others learn both geometry and texture in 3D~\cite{sun2021fenerf,Schwarz2020GRAF}.

While all these approaches do not require more than one observation per 3D scene and thus, also no ground-truth camera poses, they still require knowledge of the \emph{distribution} of camera poses (i.e., for portrait images, the distribution over camera poses must produce plausible portrait angles). CAMPARI~\cite{Niemeyer2021campari} addresses this constraint by jointly learning camera pose distribution and generative model.
GIRAFFE~\cite{niemeyer2020giraffe} proposes to learn a generative model of scenes composed of several objects by parameterizing a scene as a composition of several foreground (object) NeRFs and a single background NeRF. Latent codes are sampled for each NeRF separately, and a volume renderer composes them to a coherent 2D image.

\subsection{Learning to Represent and Render Non-static Content}
\label{sec:dynamic_content}

\begin{table}
    \centering
    \begin{adjustbox}{max width=\linewidth}
    \begin{tabular}{lccccc}
        \toprule
        Method &
        \rot{Data} &           %
        \rot{Deformation} &       %
        \rot{Class-Specific Prior} &          %
        \rot{Controllable Parameters} & %
        \rot{Code} \\                   %

        \cmidrule(lr){1-6} 
        Lombardi~\etal~\cite{Lombardi_2021_MVP} 
        & \opB{MV} & \opA{I}  & \opA{G} & \opA{V},\opB{R} &  \code{https://github.com/facebookresearch/neuralvolumes} 
        \\ 
        
        \midrule
        Li~\etal~\cite{li2021neural}
        & \opA{Mo} & \opA{I}+\opB{E}  & \opA{G}      
        & \opA{V},\opB{R} & \code{https://github.com/zl548/Neural-Scene-Flow-Fields}  
        \\
        
        \cmidrule(lr){1-6}
        Xian~\etal~\cite{xian2021space}
        & \opA{Mo} & \opA{I}  & \opA{G}      
        & \opA{V},\opB{R} & \code{}          
        \\
        
        \cmidrule(lr){1-6}
        Gao~\etal~\cite{Gao-freeviewvideo}
        & \opA{Mo} & \opA{I}  & \opA{G}      
        & \opA{V},\opB{R} & \code{}          
        \\
        
        \cmidrule(lr){1-6}
        Du~\etal~\cite{du2021nerflow}
        & \opA{Mo} & \opA{I}  & \opA{G}      
        & \opA{V},\opB{R} & \code{https://github.com/yilundu/nerflow}          
        \\
        
        \cmidrule(lr){1-6}
        Pumarola~\etal~\cite{pumarola2020d}
        & \opA{Mo} & \opB{E}  & \opA{G}      
        & \opA{V},\opB{R} & \code{https://github.com/albertpumarola/D-NeRF}          
        \\
        
        \cmidrule(lr){1-6}
        Park~\etal~\cite{park2021nerfies}
        & \opA{Mo} & \opB{E}  & \opA{G}      
        & \opA{V},\opB{R} & \code{https://github.com/google/nerfies}          
        \\
        
        \cmidrule(lr){1-6}
        Tretschk~\etal~\cite{tretschk2021nonrigid}
        & \opA{Mo} & \opB{E}  & \opA{G}      
        & \opA{V},\opB{R} & \code{https://github.com/facebookresearch/nonrigid_nerf}          
        \\
        
        \cmidrule(lr){1-6}
        Park~\etal~\cite{park2021hypernerf}
        & \opA{Mo} & \opA{I}+\opB{E}  & \opA{G}      
        & \opA{V},\opB{R} & \code{https://github.com/google/hypernerf}          
        \\
        
        \cmidrule(lr){1-6}
        Attal~\etal~\cite{attal2021torf}
        & \opA{Mo}+\opC{D} & \opA{I}  & \opA{G}      
        & \opA{V},\opB{R} & \code{https://imaging.cs.cmu.edu/torf/\#code}          
        \\
        
        \cmidrule(lr){1-6}
        Li~\etal~\cite{Li2021}
        & \opB{MV} & \opA{I}  & \opA{G} & \opA{V},\opB{R} & \code{}          
        \\
        
        \cmidrule(lr){1-6}
        Gafni~\etal~\cite{Gafni_2021_CVPR}
        & \opA{Mo} & \opB{E}  & \opC{F} &\opA{V},\opB{R},\opC{E} & \code{https://github.com/gafniguy/4D-Facial-Avatars}   
        \\
        
        \cmidrule(lr){1-6}
        Wang~\etal~\cite{wang2020learning}
        & \opB{MV} & \opA{I}  & \opC{F} & \opA{V},\opB{R} & \code{}             
        \\
        
        \cmidrule(lr){1-6}
        Guo~\etal~\cite{guo2021adnerf}
        & \opA{Mo} & \opA{I}  & \opC{F} & \opA{V},\opB{R},\opC{E} &  \code{https://github.com/YudongGuo/AD-NeRF} 
        \\
        
        \cmidrule(lr){1-6} 
        Noguchi~\etal~\cite{Noguchi2021} 
        & \opA{Mo}+\opD{3D} & \opB{E}  & \opB{B} & \opA{V},\opB{R},\opC{E} &  \code{https://github.com/nogu-atsu/NARF} 
        \\
        
        \cmidrule(lr){1-6}
        Su~\etal~\cite{su2021anerf}
        & \opA{Mo} & \opB{E}  & \opB{B} &\opA{V},\opB{R},\opC{E} & \code{}           
        \\ 
        
        \cmidrule(lr){1-6}
        Peng~\etal~\cite{peng2021animatable}
        & \opB{MV} & \opB{E}  & \opB{B} & \opA{V},\opB{R},\opC{E} &  \code{https://github.com/zju3dv/animatable_nerf}           
        \\ 
        
        \cmidrule(lr){1-6}
        Peng~\etal~\cite{Peng_2021_CVPR}
        & \opB{MV} &\opA{I}+ \opB{E} & \opB{B} & \opA{V},\opB{R} &  \code{https://github.com/zju3dv/neuralbody} 
        \\

        \cmidrule(lr){1-6}
        Liu~\etal~\cite{liu2021neural}
        & \opB{MV} & \opB{E}  & \opB{B} & \opA{V},\opB{R},\opC{E} & \code{} 
        \\ 
        
        \cmidrule(lr){1-6}
        Xu~\etal~\cite{Xu2021}
        & \opB{MV}+\opA{Mo}  & \opA{I}  & \opB{B} & \opA{V},\opB{R},\opC{E} & \code{} 
        \\ 
        
    \end{tabular}
    \end{adjustbox}
    \caption{Selected methods for non-static, dynamic scenes presented in \Cref{sec:dynamic_content}. 
            \tableopt{opB}{MV}:~Multi-View,
            \tableopt{opA}{Mo}:~Monocular,
            \tableopt{opC}{D}:~Depth,
            \tableopt{opD}{3D}:~3D pose.
            \tableopt{opA}{I}:~Implicitly,
            \tableopt{opB}{E}:~Explicitly.
            \tableopt{opA}{G}:~General,
            \tableopt{opB}{B}:~Body,
            \tableopt{opC}{F}:~Face/Head.
            \tableopt{opA}{V}:~Viewpoint,
            \tableopt{opB}{R}:~Replay,
            \tableopt{opC}{E}:~Editing.
    }
    \label{tbl:overview_dynamic_content}
\end{table}     

While the original neural radiance fields~\cite{Mildenhall_2020_NeRF} are used to represent static scenes and objects, there are approaches that can additionally handle dynamically changing content.
In \Cref{tbl:overview_dynamic_content}, we give an overview over the discussed methods.

\begin{figure}[t!]
    \centering
    \vspace{10pt} 
    \includegraphics[width=\linewidth]{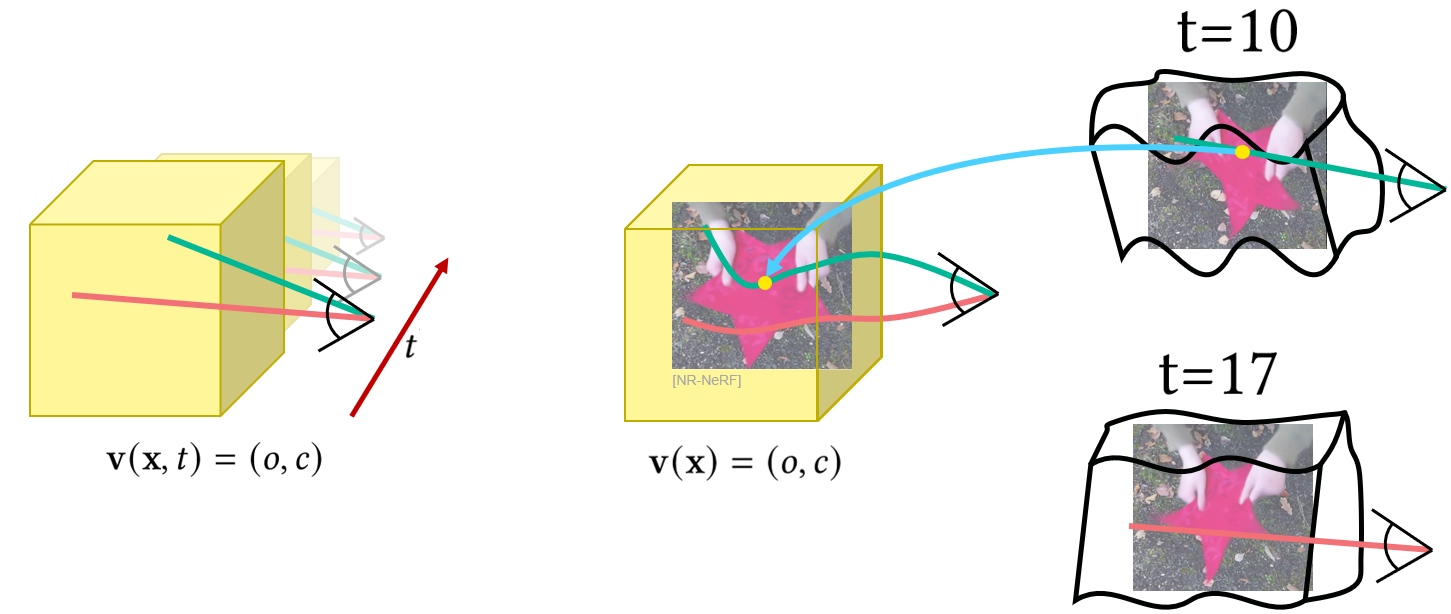}
    \caption{
    Current methods for modelling deformations with neural radiance fields fall into two categories. Left: Implicitly, by conditioning the radiance field, $\mathbf{v}$, on the deformation (here: time $t$). Right: Explicitly, by warping space with a separate deformation MLP that regresses offsets (blue arrow) from the deformed space (black) into static canonical space (yellow). This bends straight rays into the canonical radiance field.
    Image adapted from \cite{siggraph_course_21}.
    }
    \label{fig:dynamic_nerf_deformations}
\end{figure}

These approaches can be categorized in time-varying representations that allow to do novel viewpoint synthesis of a dynamically changing scene as an unmodified playback (e.g., to produce a bullet-time effect), or in techniques that also give control over the deformation state, thus, allowing for novel-view point synthesis and editing of the content.
The deforming neural radiance field can be achieved implicitly or explicitly, see \Cref{fig:dynamic_nerf_deformations}:
\begin{itemize}
    \item Implicitly, by conditioning the NeRF on a representation of the deformation state (e.g., a time input)
    \item Explicitly, by using a separate deformation field that can map from the deformed space to a canonical space where the NeRF is embedded.
\end{itemize}

\subsubsection{Time-varying Neural Radiance Fields}\label{ssec:time_varying_NeRF}

\begin{figure}[t!] 
    \centering 
    \includegraphics[width=\linewidth]{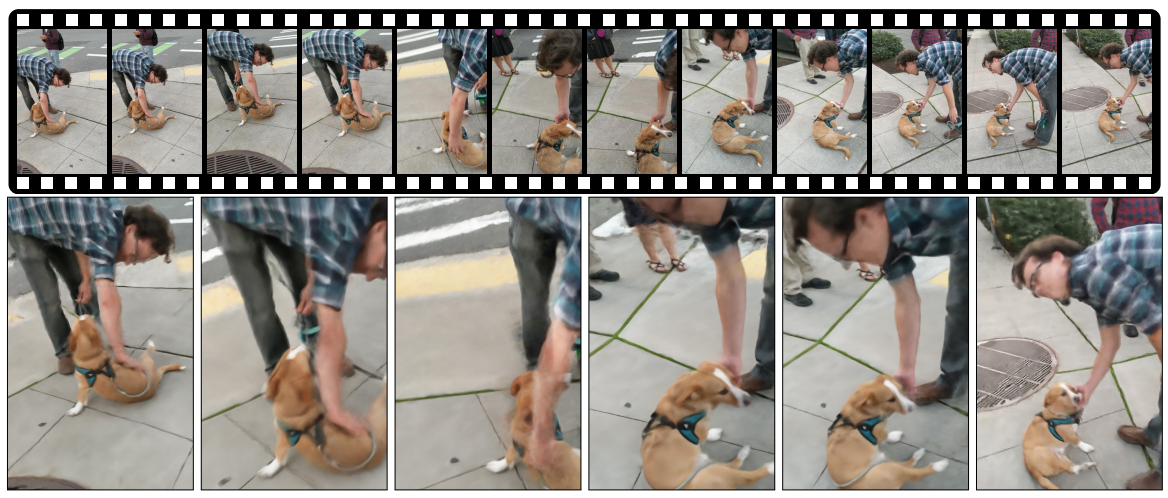} 
    \caption{An input monocular video of a general deformable scene and novel view rendering thereof by space-time neural irradiance fields of Xian et al.~\cite{xian2021space}. Image adapted from \cite{xian2021space}.} 
    \label{fig:STIF} 
\end{figure}

Time-varying neural radiance fields allow to playback a video  with novel view points, see \Cref{fig:STIF}. 
Since they forego control, these methods do not rely on a specific motion model and can thus handle general objects and scenes. 

Several extensions of NeRF for non-rigid scenes were proposed concurrently. We first discuss methods that model deformations implicitly \cite{li2021neural, xian2021space, Gao-freeviewvideo, du2021nerflow}. While the original NeRF is static and takes as input only a 3D spatial point, it can be extended in a straightforward manner to become time-varying: the volume can additionally be conditioned on a vector that represents the deformed state. In current methods, this conditioning takes the form of a time input (potentially positionally encoded) \cite{xian2021space, li2021neural, Gao-freeviewvideo, du2021nerflow, pumarola2020d} or an auto-decoded latent code per time step \cite{park2021nerfies, tretschk2021nonrigid, park2021hypernerf}. 

Since handling non-rigid scenes without prior knowledge of object type or 3D shape is an ill-posed problem, methods of this class adopt various geometric regularizers and condition learning on additional data modalities. To encourage consistency of reflectance and opacity across time, several methods learn scene-flow mappings between temporally neighboring time steps \cite{li2021neural, xian2021space, Gao-freeviewvideo, du2021nerflow}. Since this is restricted to small temporal neighborhoods, artifact-free novel-view synthesis is predominantly demonstrated on spatio-temporal camera trajectories that are close to the spatio-temporal input camera trajectories. The scene-flow mapping can be trained with reconstruction losses that warp the scene from other time steps into the current time step \cite{li2021neural, du2021nerflow}, by encouraging consistency between estimated optical flow and the 2D projection of the scene flow \cite{li2021neural, Gao-freeviewvideo}, or by tracking backprojected keypoints in 3D \cite{du2021nerflow}. The scene flow is often constrained with additional regularization losses \cite{li2021neural, xian2021space, Gao-freeviewvideo, du2021nerflow}, e.g., to encourage spatial or temporal smoothness or forward-backward cycle consistency. Unlike the other methods mentioned, Neural Radiance FLow (NeRFlow) of Du et al.~\cite{du2021nerflow} models deformations with infinitesimal displacements that need to be integrated with Neural ODE \cite{NeuralODE2018} to obtain offsets.

In addition, several methods use estimated depth maps to supervise the geometry estimation \cite{li2021neural, xian2021space, Gao-freeviewvideo, du2021nerflow}. One limitation of this regularization is that the accuracy of the reconstruction depends on the accuracy of monocular depth estimation methods. As a result, artefacts of monocular depth estimation methods are recognizable in the novel views \cite{xian2021space}.  

Finally, the static background is often handled separately, allowing it to exploit multi-view clues from monocular input recordings across time. To that end, some methods estimate a second static volume that is not conditioned on the deformation \cite{li2021neural, Gao-freeviewvideo} or introduce soft regularization losses to constrain static scene content \cite{xian2021space}. Gao \textit{et al.}~\cite{Gao-freeviewvideo}, a follow-up to Xian \textit{et al.}'s work ~\cite{xian2021space}, train the static NeRF on observations that do not contain moving and deforming parts with the help of a binary segmentation mask (one of the inputs to the model and user-provided).

One advantage of Guo et al.'s method is that it produces the most accurate quantitative and qualitative results on the challenging dataset of Yoon \textit{et al.}~\cite{yoon2020dynamic} (compared to Tretschk et al.~\cite{tretschk2021nonrigid} and Li et al.~\cite{li2021neural}). The latter dataset was initially introduced for novel-view synthesis from a comparably sparse set of input monocular views of dynamic scenes with moderate changes in the camera poses. Limitations of the method include strong reliance on optical flow and handling of arbitrary non-rigid deformations (in contrast to scenes composed of independent rigidly moving objects).

Finally, NeRFlow \cite{du2021nerflow} can be used to de-noise and super-resolve views of pre-trained scenes. Limitations of NeRFlow, which the authors mention, include difficulty in preserving static backgrounds, handling complex scenes (non-piecewise-rigid deformations and motions) and rendering novel views at substantially different camera trajectories compared to the input ones. 

The methods discussed so far model deformations implicitly by conditioning the scene representation on the deformation. This makes controllability of the deformation cumbersome and difficult. Other works instead disentangle the deformations from the geometry and appearance: they factor out the deformations into a separate function on top of a static canonical scene, a crucial step towards controllability. The deformations are accomplished by shooting straight rays into deformed space and then bending them into the canonical scene, usually by regressing per-point offsets for points on the straight ray using a coordinate-based MLP that is conditioned on the deformation. This can be thought of as space warping or scene flow. In contrast to implicit modelling, these methods share geometry and appearance information across time by construction via the static canonical scene, thereby providing hard correspondences, which do not drift. Due to that hard constraint, unlike implicit methods, current methods with explicit deformations cannot handle topological changes and only demonstrate results on scenes with significantly smaller motion than implicit methods. 

D-NeRF \cite{pumarola2020d} uses an unregularized ray-bending MLP to model deformations of a single or multiple synthetic objects segmented from the background and observed by virtual cameras. It assumes a pre-defined set of multi-view images given, though, at training time, only a single view chosen arbitrarily is used for supervision at any time. Thus, D-NeRF can be considered an intermediate step between techniques with multi-view supervision and truly monocular approaches.

Several works show results on real-world scenes observed by a moving monocular camera. The core application of Deformable NeRF of Park \textit{et al.}~\cite{park2021nerfies} %
is the creation of Nerfies, i.e., free-viewpoint selfies. Deformable NeRF conditions deformations and appearance with an auto-decoded latent code per input view. The bent rays are regularized using an as-rigid-as-possible term (also known as elastic energy term) that penalizes deviations from piece-wise rigid scene configurations. Thus, Deformable NeRF works well on articulated scenes (e.g., a hand holding a tennis racket) and scenes such as human heads (where the head is moving w.r.t. the torso). Still, small non-rigid deformations are handled well (such as smiling), as the regularizers are soft. Another important innovation of this work is using a coarse-to-fine scheme which allows learning low-frequency components first and avoiding local minima due to overfitting to high-frequency details. 

HyperNeRF \cite{park2021hypernerf} is an extension of Deformable NeRF  \cite{park2021nerfies} using a canonical hyperspace instead of a single canonical frame. This allows tackling scenes with topological changes such as opening and closing the mouth. In HyperNeRF, the bending network (MLP) of Deformable NeRF is augmented with an ambient slicing surface network (likewise an MLP) that selects a canonical subspace for every input RGB view by indirectly conditioning the canonical scene on the deformation. As such it is a hybrid that combines both explicit and implicit deformation modelling, which allows it to handle topological changes by sacrificing hard correspondences.

Non-rigid NeRF (NR-NeRF) \cite{tretschk2021nonrigid} %
models a time-varying scene appearance using a per-scene canonical volume, per-scene rigidity flag (an MLP) and per-frame ray bending operator (an MLP).
NR-NeRF shows that no additional supervisory cues such as depth maps or scene flows are required to handle scenes with small  non-rigid deformations and motions, in contrast to \cite{park2021nerfies, xian2021space, li2021neural}. %
Moreover, the observed deformations are regularized by a divergence operator, which imposes a volume-preserving constraint and stabilizes occluded areas with respect to supervising monocular input views. In this regard, it has similarities with the elastic regularizer of Nerfies penalizing deviations from piece-wise rigid deformations. This regularization makes it possible for the camera trajectory of novel views to differ significantly from the input camera trajectory.
While controllability is still severely limited, NR-NeRF demonstrates several simple edits of the learned deformation field, such as motion exaggeration or removal of dynamic scene content. 

Other works do not restrict themselves to the case of monocular RGB input video, but instead consider other inputs. 

Time-of-Flight Radiance Fields (T\"oRF) method  \cite{attal2021torf} replaces data-driven priors for reconstructing dynamic contents with depth maps from a depth sensor. In contrast to the vast majority of computer vision works, T\"oRF uses raw ToF sensor measurements (so-called phasors), which brings advantages when handling weakly-reflecting regions and other limitations of modern depth sensors (e.g., restricted working depth range). Integration of measured scene depths in the learning of NeRF reduces the requirement on the number of input views leading to sharp and detailed models. The depth cue also enables superior accuracy compared to NSFF \cite{li2021neural} and space-time neural irradiance fields \cite{xian2021space}. 

Neural 3D Video Synthesis \cite{Li2021} uses a multi-view RGB setup and models deformations implicitly. The method exploits temporal smoothness by first training on keyframes. It also exploits that the cameras remain static and that the scene content is predominantly static by sampling rays in a biased manner for training. The results are sharp even for dynamic content that is small.

\subsubsection{Controllable Dynamic Neural Radiance Fields}
To allow controllability of the deformation of the neural radiance field, method use class specific motion models as underlying representation of the deformation state (e.g., a morphable model for the human face or a skeletal deformation graph for the human body).

\begin{figure}[t!]
    \centering
    \includegraphics[width=\linewidth]{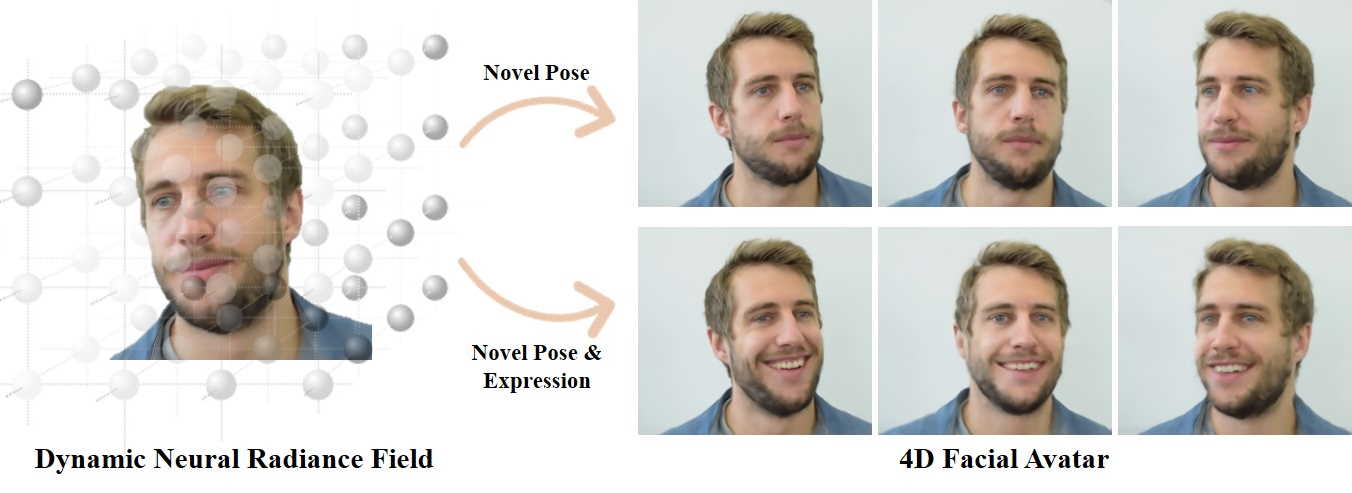}
    \caption{
    Dynamic Neural Radiance Field to synthesize novel views and expressions of humans.
    Image adapted from ~\cite{Gafni_2021_CVPR}.
    }
    \label{fig:nerface}
\end{figure}

NeRFace~\cite{Gafni_2021_CVPR} is the first approach that uses a morphable model to implicitly control a neural radiance field (see \Cref{fig:nerface}). They use a face tracker~\cite{thies2016face} to reconstruct the face blendshape parameters as well as the camera pose in the training views (monocular video). The MLP is trained on these views with the blendshape parameters and a learnable per-frame latent codes as conditioning. In addition, they assume a known static background such that the radiance field only stores the information about the face. The latent codes are used to compensate missing tracking information (i.e., the shoulders of a person) as well as errors in the tracking). Once trained the radiance field can be controlled via the blendshape parameters, thus, allowing reenactment and expression editing.
While NeRFace uses a global deformation code based on a morphable model, Wang et al.~\cite{wang2020learning} generate local animation codes. Specifically, they extract a global animation code from multi-view inputs which is mapped to local codes using 3D convolutional neural network. These are used to condition the fine-level radiance field which are represented as MLPs. In contrast to NeRFace, the method does not allow direct control over expressions of the face, but an encoder has to be trained that for example can generate the animation codes from facial keypoints.
Guo et al.~\cite{guo2021adnerf} propose an audio driven neural radiance field (AD-NeRF) which is inspired by NeRFace.
But instead of using expression coefficients, they map audio features extracted using DeepSpeech~\cite{deepspeech,thies2020nvp} to a feature which serves as a conditioning to the MLP that represents the radiance field. While the expression is controlled implicitly via an audio signal, they provide explicit control over the rigid pose of the head. To synthesize the portrait view of a person, they employ two separate radiance fields, one for the head and one for the torso.

,,I M Avatar''~\cite{zheng2021i} extends NeRFace based on skinning fields~\cite{chen2021snarf} which are used to deform the canonical NeRF volume given novel expression and pose parameters.
CoNeRF~\cite{kania2021conerf} presents a method to disentangle attribute/expression combinations leveraging sparse mask annotations in the training images.
They rely on a locality assumption, i.e., one attribute affects only a specific region. These localized attribute masks are treated as latent codes within their framework.

Besides these subject-specific training methods, HeadNerf~\cite{hong2021headnerf} and MoFaNeRF~\cite{zhuang2021mofanerf} propose a generalized model to represent faces under different views, expressions and illumination.
Similar to NeRFace, they condition the NeRF MLP on additional parameters that control the shape of the person, the expression, albedo, and illumination.
Both methods, require a refinement network (2D network) to improve the coarse results of the volumetric rendering based on this conditioned NeRF MLP.

\begin{figure}[t!] 
    \centering 
    \includegraphics[width=\linewidth]{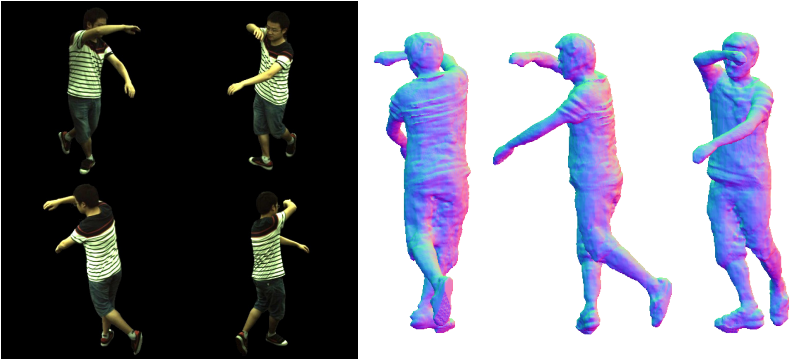} 
    \caption{Neural Body \cite{Peng_2021_CVPR} approach recovers 3D models of humans with fine appearance details from sparse multi-view video. 
    } 
    \label{fig:NeuralBody} 
\end{figure} 

While the afore mentioned approaches show promising results in a portrait scenario, they are not applicable to highly non-rigid deformations, especially, for articulated motion of a human body captured from a single view.
Therefore, methods leverage the human skeleton embedding explicitly.
Neural Articulated Radiance Field (NARF) \cite{Noguchi2021} is trained via pose-annotated images.
An articulated object is decomposed into several rigid object parts with their local coordinate systems and global shape variations on top.
The converged NARF can be used to render novel views by manipulating the poses, estimate depth maps and perform body parts segmentation. 
In contrast to NARF, A-NeRF \cite{su2021anerf} learns actor-specific volumetric neural body models from monocular footage in a self-supervised manner.
The method combines a dynamic NeRF volume with the explicit controllability of an articulated human skeleton embedding and reconstructs both the pose and radiance field in an analysis-by-synthesis way.
Once trained, the radiance field can be used for novel view point synthesis as well as motion retargeting.
They show the benefits of using a learned surface-free model which improves the accuracy of human pose estimation from monocular videos with the help of a photometric reconstruction loss.
While A-NeRF is trained on monocular videos, Animatable Neural Radiance Fields (ANRF) \cite{peng2021animatable} is a skeleton-driven approach for human model reconstruction from multi-view videos.
Its core component is a new motion representation, i.e., the neural blend weight field, that is combined with 3D human skeletons for deformation field generation.
Similarly to several general non-rigid NeRFs \cite{park2021nerfies, tretschk2021nonrigid}, ANRF maintains a canonical space and estimates two-way correspondences between the multi-view inputs and the canonical frame.
The reconstructed animatable human models can be used for free-viewpoint rendering and re-rendering under novel poses.
Human meshes can also be extracted from ANRF by running marching cubes on volume densities at the discretized canonical space points.
The method achieves high visual accuracy for the learned human models, and the authors suggest that handling complex non-rigid deformations on the observed surfaces (such as those due to loose clothes) can be improved in future work.

The Neural Body approach of Peng and colleagues \cite{Peng_2021_CVPR} enables novel view synthesis of human performances from sparse multi-view videos (e.g., only four synchronized views), see \Cref{fig:NeuralBody} for  exemplary inputs and the result. 
Their method uses conditioning by the parametric human shape model SMPL \cite{SMPL2015} as a shape proxy prior.
It assumes that the recovered neural representation at different frames has the same set of latent codes anchored to a deformable mesh. General-purpose baselines such as rigid NeRF \cite{Mildenhall_2020_NeRF} (applied per timestamp) or NeuralVolumes \cite{Lombardi_2019_NeuralVolumes} assume much denser input image sets and, consequently, cannot compete with Neural Body in its ability to render novel views of moving humans from a few synchronized input images.
The method also favourably compares to human mesh reconstruction techniques such as PIFuHD \cite{Saito2020}, which strongly depends on training 3D data when it comes to the 3D reconstruction of fine appearance details (e.g., rarely-worn or unique garments). 
Similar to the Neural Body approach, Neural Actor (NA)~\cite{liu2021neural} and HVTR~\cite{hu2021hvtr} use the SMPL model to represent the deformation states.
They leverage the proxy to explicitly unwarp the surrounding 3D space into a canonical pose, where the NeRF is embedded.
To improve the recovery of high fidelity details in geometry and appearance, they use additional 2D texture maps defined on the SMPL surface, which are used as an additional conditioning to the NeRF MLP.
H-NeRF~\cite{Xu2021} is another technique for temporal 3D reconstructions of humans with conditioning using an human body model. 
Similarly to Neural Body \cite{Peng_2021_CVPR}, they require a sparse set of  videos from synchronized and calibrated cameras. 
In contrast to it, H-NeRF uses a structured implicit body model with signed distance fields \cite{Alldieck_2021_ICCV}, which results in sharper renderings and more complete geometry for challenging subjects.
Similar to H-NeRF, DD-NeRF~\cite{yao2021ddnerf} builds on top of a signed distance field to render entire human bodies. Given multi-view input images and a reconstructed SMPL body, they regress SDF and radiance values which are accumulated using volumetric rendering. 
HumanNeRF~\cite{zhao2021humannerf} is also based on multi-view images as input, but learns a generalized neural radiance field for free view-point rendering which can be fine-tuned for a specific actor.
Another work called HumanNeRF~\cite{weng2022humannerf} shows how to train a neural radiance field for a specific actor based on monocular input data, using a skeleton-driven motion field which is refined by a general non-rigid motion field.

Mixture of Volumetric Primitives~\cite{Lombardi_2021_MVP} is a model for rendering dynamic, animatable virtual humans in real time. The main idea is to model a scene or object with a set of volumetric primitives that can dynamically change position and content. These primitives model components of the scene like a parts-based model. Each volumetric primitive is a voxel grid produced by a decoder network from a latent code. The code defines the configuration of the scene (e.g., a facial expression, in the case of human faces) which is used by the decoder network to produce primitive locations and voxel values (which contain RGB color and opacity). To render, a raymarching procedure is used to accumulate color and opacity values along the rays corresponding to each pixel. Similar to other dynamic NeRF methods, multi-view video is used as training data. The method is capable of creating extremely high-quality realtime renderings that look realistic even on challenging materials, like hair and clothing.
E-NeRF~\cite{lin2021efficient} demonstrates an efficient NeRF rendering scheme based on depth-guided sampling. They show realtime rendering on moving humans as well as on static objects using multi-view images as input.

\subsection{Compositionality and Editing} \label{sec:compedit} 
    
\begin{table}
    \centering
    \begin{adjustbox}{max width=\linewidth}
    
    \centering
    \begin{tabular}{lccccc}
        \toprule
        Method &
        \rot{Required Data} &           %
        \rot{3D Representation} &       %
        \rot{Controllable Parameters} & %
        \rot{Generative Model} &        %
        \rot{Code} \\                   %

        \midrule
        Nguyen-Phuoc \etal~\cite{nguyen2019hologan} 
        & \opC{MVI}+\opD{UIC} & \opA{V}  & \opA{P},\opB{S},\opC{T}  & \cmark{}  & \code{https://github.com/thunguyenphuoc/HoloGAN}  
        \\
        \cmidrule(lr){1-6}
        Liu~\etal~\cite{liu2021editing}
        & \opC{MVI} & \opA{V}  & \opB{S},\opE{C}   & \xmark{}  & \code{https://github.com/stevliu/editnerf}            
        \\
        
        \cmidrule(lr){1-6}
        Jang and Agapito~\cite{Jang_2021_ICCV}
        & \opD{UIC} & \opA{V}  & \opA{P},\opB{S},\opC{T}  & \xmark{}  & \code{https://github.com/wayne1123/code-nerf}        
        \\
        \midrule
        Ost~\etal~\cite{Ost2021}
        & \opA{VID} & \opB{V-O}  & \opA{P},\opB{S},\opC{T},\opD{O}  & \cmark{}  & \code{https://github.com/princeton-computational-imaging/neural-scene-graphs}  
        \\
        \cmidrule(lr){1-6}
        Zhang~\etal~\cite{zhang2021stnerf}
        & \opB{MV-VID} & \opB{V-O}  & \opB{S},\opE{C}  & \cmark{}  & \code{https://github.com/DarlingHang/ST-NeRF}          
        \\
        \cmidrule(lr){1-6}
        Niemeyer and Geiger~\cite{niemeyer2020giraffe}
        & \opD{UIC} & \opC{NFF}  & \opA{P},\opB{S},\opC{T} & \cmark{}  & \code{https://github.com/autonomousvision/giraffe} 
        \\
        
   \end{tabular}
   \end{adjustbox}
    
    \caption{Selected methods for editing and scene compositionality presented in \Cref{sec:compedit}. 
    \tableopt{opA}{VID}:~Monocular videos, 
    \tableopt{opB}{MV-VID}:~Synchronised multi-view videos, 
    \tableopt{opC}{MVI}:~Multi-view image collections, 
    \tableopt{opD}{UIC}:~Unstructured 2D image collections.
    \tableopt{opA}{V}:~Neural volumetric, 
    \tableopt{opB}{V-O}:~Per-object+background neural volumetric, 
    \tableopt{opC}{NFF}:~Neural feature fields.
    \tableopt{opA}{P}:~Camera pose,
    \tableopt{opB}{S}:~Shape, 
    \tableopt{opC}{T}:~Texture/appearance, 
    \tableopt{opD}{O}:~Opacity,
    \tableopt{opE}{C}:~Color.
    }
    \label{tbl:overview_compedit}
\end{table}

The methods discussed so far allow reconstructing volumetric representations of static or dynamic scenes and render novel views of them, perhaps from a few input images. They keep the observed scene unchanged, except for comparably straightforward modifications (e.g., foreground removal). Several recent methods also allow editing the reconstructed 3D scenes, i.e., rearranging and affine-transforming the objects and altering their structure and appearance. 
In \Cref{tbl:overview_compedit}, we give an overview of the discussed methods.

Conditional NeRF \cite{liu2021editing} can alter the color and shape of rigid objects observed in 2D images from manual user edits (e.g., it is possible to remove some object parts). This functionality is enabled by a single NeRF trained on multiple object instances of the same category. During editing, the network parameters are adjusted to match the shape and color of a newly observed instance. One of the contributions of this work is finding a subset of tunable parameters which can successfully propagate user edits for novel view generation. This avoids expensive modifications of the entire network. CodeNeRF \cite{Jang_2021_ICCV} represents shape and texture variations across an  object class. Similar to pixelNeRF, CodeNeRF can synthesize novel views of unseen  objects. It learns two different embeddings for the shape and texture. At test time, it estimates a camera pose, 3D shape and texture of the object from a single image, and both can be continuously modified by altering their latent codes. CodeNeRF achieves comparable performance to  previous methods for single-image 3D reconstruction, while not assuming known camera poses.

\begin{figure}[t!] 
    \centering 
    \includegraphics[width=\linewidth]{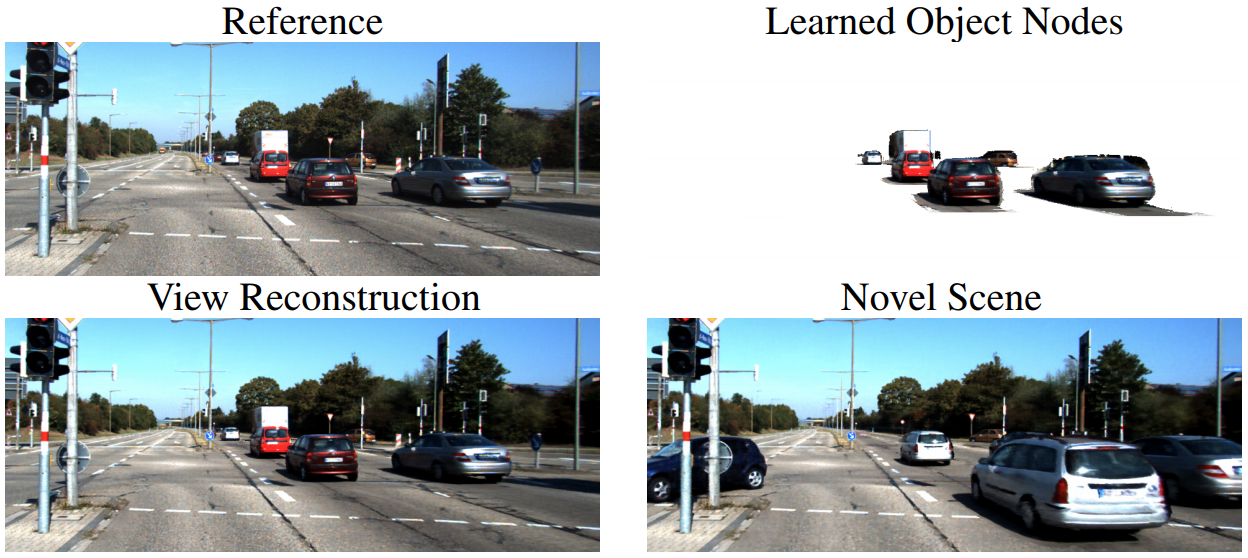} 
    \caption{A reconstructed view, learned object nodes and novel scene  renderings by the Neural Scene Graphs approach  \cite{Ost2021}.} 
    \label{fig:NSG} 
\end{figure}

Neural Scene Graphs (NSG) \cite{Ost2021} is a recent method for novel view synthesis from monocular videos recorded while driving (ego-vehicle views). This technique decomposes a dynamic scene with multiple independent rigidly moving objects into a learned scene graph that encodes individual object transformations and radiances. Thus, each object and the background are encoded by different neural networks. In addition, the sampling of the static node is restricted to layered planes (which are parallel to the image plane) for increased efficiency, i.e., a 2.5D  representation. NSG requires annotated tracking data for each rigidly moving object of interest over the set of input frames, and each object class (e.g., a car or bus) shares a single volumetric prior. The neural scene graph can then be used to render novel views of the same (i.e., observed) or edited (i.e., by rearranging the objects) scene. Applications of NSG include background-foreground decomposition, enriching training datasets for automotive perception, and improved object detection and scene understanding (see \Cref{fig:NSG}). 

Another layered representation for editable free-viewpoint videos is  introduced in Zhang \textit{et al.}~\cite{zhang2021stnerf}. Their spatially and temporally consistent NeRF (ST-NeRF) relies on bounding boxes for all  independently moving and articulated objects---resulting in multiple  layers---and disentangles their positions, deformations and appearance. The input to ST-NeRF is a set of 16 synchronized videos from the cameras placed at regular intervals in a half-circle, along with human-background segmentation masks. The method's name suggests that space-time coherence constraints are reflected in its architecture, i.e., as a space-time deformation module and a NeRF module of the canonical space. ST-NeRF also accepts timestamps to account for the appearance evolving in time. While rendering novel views, the sampling rays are cast through multiple scene layers, which results in accumulated densities and colors. ST-NeRF can be used for neural scene editing such as rescaling, shifting, duplication or removing of the performers, and temporal rearrangements. As promising directions for future work, the authors name reducing the number of input views and enabling non-rigid scene editing. 

Note that some of the methods~\cite{niemeyer2020giraffe, nguyen2019hologan} discussed in \Cref{sec:generalization} can be used for scene  editing as well.
E.g., GIRAFFE~\cite{niemeyer2020giraffe} can rotate an object of a known class observed in a single monocular image, change its appearance and translate it along the depth channel. 
See \Cref{tbl:overview_compedit} for a comparison of the methods discussed in this section.

\subsection{Relighting and Material Editing}
\label{sec:relight}

\begin{table}
    \centering
    \begin{adjustbox}{max width=\linewidth}
    \begin{tabular}{lcccccc}
        \toprule
        Method &
        \rot{Required Data} &           
        \rot{3D Representation} &       
        \rot{Controllable Parameters} & 
        \rot{Models Light Visibility} &  
        \rot{Models Indirect Illumination} &  
        \rot{Code} \\                   %
        
        \midrule
        Bi~\etal~\cite{bi2020nrf}
        & \opA{I}+\opB{L} & \opA{V} & \opA{L}+\opB{M} & \cmark{} & \xmark{} & \xmark{}
        \\
        \cmidrule(lr){1-7}
        Zhang~\etal~\cite{zhang2021physg}
        & \opA{I}+\opC{M} & \opB{S} & \opA{L}+\opB{M} & \xmark{} & \xmark{} & \code{https://github.com/Kai-46/PhySG}       
        \\
        
        \cmidrule(lr){1-7}
        Boss~\etal~\cite{boss2020nerd}
        & \opA{I}+\opC{M} & \opA{V} & \opA{L}+\opB{M} & \xmark{} & \xmark{} &    \code{https://github.com/cgtuebingen/NeRD-Neural-Reflectance-Decomposition}   
        \\
        
        \cmidrule(lr){1-7}
        Srinivasan~\etal~\cite{srinivasan2021nerv}
        & \opA{I}+\opB{L} & \opA{V} & \opA{L}+\opB{M} & \cmark{} & \cmark{} & \xmark{}           
        \\
        
        \cmidrule(lr){1-7}
        Zhang~\etal~\cite{zhang2021nerfactor}
        & \opA{I} & \opA{V} & \opA{L}+\opB{M} & \cmark{} & \xmark{} & \code{https://github.com/google/nerfactor} 
        \\
        
        \cmidrule(lr){1-7}
        Xiang~\etal~\cite{xiang2021neutex}
        & \opA{I}+\opC{M} & \opA{V} & \opC{T} & N/A & N/A & \xmark{}         
        \\

    \end{tabular}
    \end{adjustbox}
    \caption{Selected methods for relighting presented in \Cref{sec:relight}.
            \tableopt{opA}{I}:~Images,
            \tableopt{opB}{L}:~Lighting parameters for input images,
            \tableopt{opC}{M}:~Object masks.
            \tableopt{opA}{V}:~Neural volumetric,
            \tableopt{opB}{S}:~Neural SDF.
            \tableopt{opA}{L}:~Lighting,
            \tableopt{opB}{M}:~Materials,
            \tableopt{opC}{T}:~Texture map.
    }
    \label{tbl:overview_relight}
\end{table}

\begin{figure}[t!] 
    \centering 
    \includegraphics[width=\linewidth]{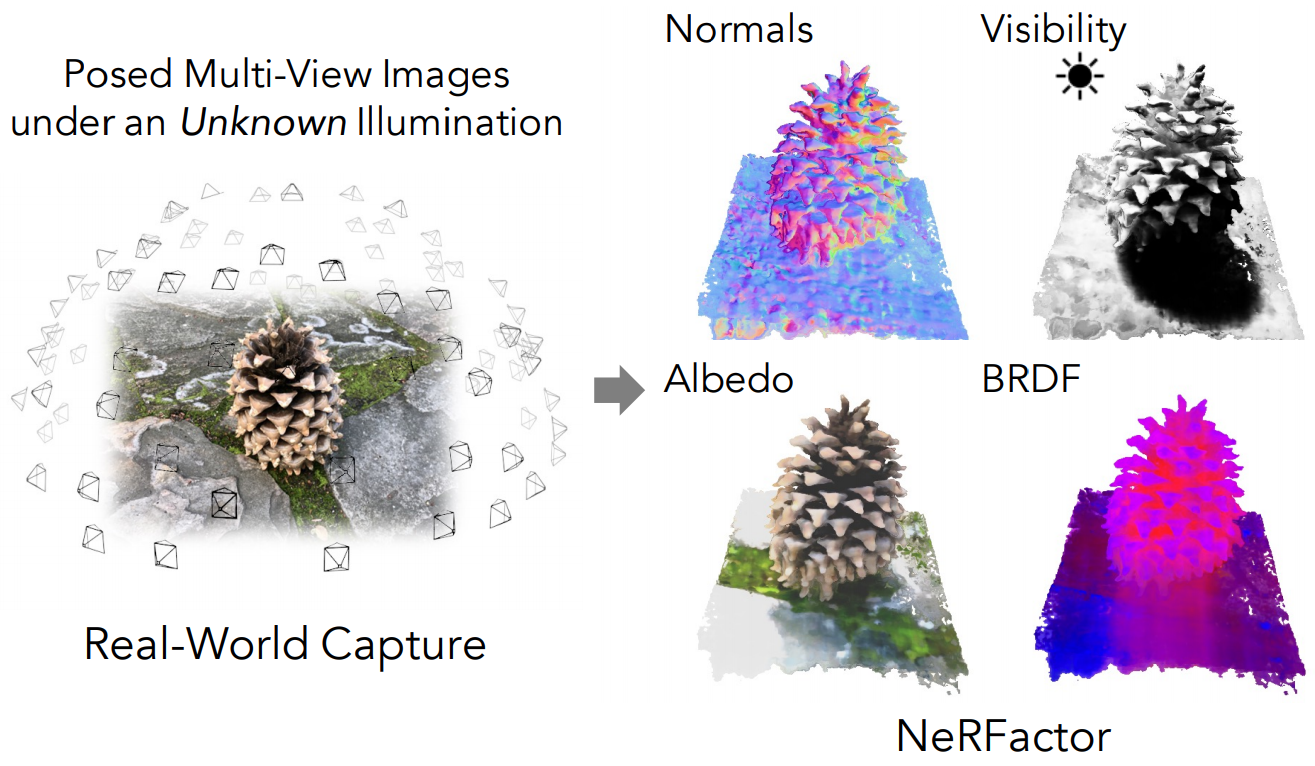} 
    \caption{NeRFactor \cite{zhang2021nerfactor} decomposes a scene captured under an unknown illumination into 3D neural fields of surface normals, albedo, BRDF and shading. This enables  free-viewpoint relighting and material editing. 
    Image adapted from \cite{zhang2021nerfactor}.} 
    \label{fig:NeRFactor1} 
\end{figure}

The applications we have presented so far are based on the simplified absorption-emission volumetric rendering model discussed in \Cref{sec:volume_rendering}, in which the scene is modeled as a volume of particles that block and emit light. While this model is sufficient for rendering images of the scene from novel viewpoints, it is unable to render images of the scene under different lighting conditions. Enabling relighting requires a scene representation that can simulate the transport of light through the volume, including the scattering of light by particles with various material properties. 
In \Cref{tbl:overview_relight}, we give an overview over the discussed methods.

Neural Reflectance Fields~\cite{bi2020nrf} proposed the first extension of NeRF to enable relighting. Instead of representing a scene as a field of volume density and view-dependent emitted radiance, as in NeRF, Neural Reflectance Fields represent a scene as a field of volume density, surface normals, and bi-directional reflectance distribution functions (BRDFs). This allows for rendering the scene under arbitrary lighting conditions by using the predicted surface normals and BRDFs at each 3D location to evaluate how much incoming light is reflected off particles at that location towards the camera. However, evaluating the visibility from each point along the camera ray to each light source is extremely computationally intensive for neural volumetric rendering models. Even when just considering direct lighting, the MLP must be evaluated at densely-sampled locations between each point along the camera ray and each light source in order to compute the incident lighting to render that ray. Neural Reflectance Fields sidesteps this issue by only training with images of objects illuminated by a single point light that is co-located with the camera, so the MLP only needs to be evaluated along the camera ray. 

Other recent works that recover relightable models have avoided the difficulty of computing light source visibility by simply ignoring self-occlusions and assuming that all light sources in the upper hemisphere above any surface are fully visible. Both PhySG~\cite{zhang2021physg} and NeRD~\cite{boss2020nerd} assume full light source visibility, and further accelerate rendering by representing the environment lighting and scene BRDFs as mixtures of spherical Gaussians, which enables the hemispherical integral of the incoming light multiplied by the BRDF to be computed in closed form. Assuming full light source visibility can work well for objects that are mostly convex, but this strategy is unable to simulate effects such as cast shadows that are due to the occlusion of light sources by scene geometry.

Neural Reflectance and Visibility Fields~\cite{srinivasan2021nerv} (NeRV) trains an MLP to approximate the light source visibility for any input 3D location and 2D incoming light direction. Instead of querying an MLP at densely-sampled points along each light ray, the visiblity MLP only needs to be queried a single time for each incoming light direction. This enables NeRV to recover relightable models of scenes from images with significant shadows and self-occlusion effects.

Instead of optimizing a relightable representation from scratch, as done in the previously discussed methods, NeRFactor~\cite{zhang2021nerfactor} starts with a pre-trained NeRF model. NeRFactor then recovers a relightable model by simplifying the pre-trained NeRF's volumetric geometry into a surface model, optimizing MLPs to represent the light source visibility and surface normals at any point on the surface, and finally optimizing a representation of the environment lighting and the BRDF at any surface point; see \Cref{fig:NeRFactor1} for an example decomposition. This results in a relightable model that is more efficient when rendering images, since the volumetric geometry has been simplified into a single surface and light-source visibility at any point can be computed by a single MLP query. 

The NeROIC technique~\cite{kuang2021neroic} also uses a multi-stage pipline to recover a relightable NeRF-like model from images of an object captured under multiple unconstrained lighting environments. The first stage recovers geometry while explaining appearance variations due to lighting with latent appearance embeddings, the second stage extracts normal vectors from this recovered geometry, and the third stage estimates BRDF properties and a spherical harmonic representation of lighting.

In contrast to the approaches described above, which focus on recovering relightable representations of objects, NeRF-OSR~\cite{rudnev2021nerfosr} recovers NeRF-like relightable models of large-scale buildings and historical sites. NeRF-OSR assumes a Lambertian model, and decomposes scenes into diffuse albedo, surface normals, a spherical harmonics representation of lighting, and shadows, which can be combined to relight the scene under novel environment illumination.

The relightable models described above represent scene materials as a continuous 3D field of BRDFs. This enables some basic amount of material editing since the recovered BRDFs can be changed before rendering. NeuTex~\cite{xiang2021neutex} enables more intuitive material editing by introducing a surface parameterization network that learns a mapping from 3D coordinates in the volume to 2D texture coordinates. After a NeuTex model of a scene is recovered, the 2D texture can easily be edited or replaced.

Ref-NeRF~\cite{verbin2021refnerf} focuses on improving NeRF's ability to represent and render specular surfaces. Although Ref-NeRF cannot be used for relighting as it does not disentangle incoming light from reflectance properties, it structures outgoing light into physically-meaningful components (diffuse and specular colors, normal vectors, and roughness) that enable intuitive material editing applications.

Guo et al.~\cite{guo2021nerfren} extends NeRF to handle reflections and propose to split a scene into transmitted and reflected components which are represented as separate neural radiance fields. While it does not allow scene editing, it is able to handle reflections from glas and mirrors.

\subsection{Light Fields}

Volume rendering, sphere-tracing, and other 3D rendering forward models can yield photo-realistic results. However, for a given ray, they all require the sampling of the underlying 3D scene representation at whatever 3D coordinate that ray first intersects the scene's geometry. As this intersection point is not known a-priori, ray-marching algorithms first have to discover that surface point. Ultimately, this yields a time and memory complexity that scales with the geometric complexity of the scene, where more and more points have to be sampled to render more and more complex scenes. In practice, these are hundreds or even thousands of points  per ray. Moreover, accurately rendering reflections and second-order lighting effects requires \emph{multi-bounce} ray-tracing, such that for every pixel, many rays have to be traced instead of only a single one. This yields a high computational burden. While in the regime of reconstructing a \emph{single} scene (overfitting), this may be circumvented by smart data structures, hashing, and expert low-level engineering, in the regime of reconstructing a 3D scene given just \emph{few observations or even just a single} image, such data structures hinder the application of learned reconstruction algorithms, such as inferring the parameters of the 3D scene from a single image using convolutional neural networks.

A pair of concurrent works~\cite{sitzmann2021lfns,liu2021nelf} thus introduced the idea of parametrizing light fields via coordinate-based networks. Specifically, Light Field Networks~\cite{sitzmann2021lfns} %
paramaterize a 3D scene not via its 3D radiance field, but instead via its 360-degree \emph{light field}, i.e., a function that maps \emph{every oriented ray} directly to the color observed by that ray. Concurrently, Liu et al.~\cite{liu2021nelf} proposed to parameterize a fronto-parallel light field for novel view synthesis of forward-facing scenes as a neural field. Representing a scene via its light field obviates the need for ray-marching, as to render a single pixel, the light field can be sampled by the corresponding camera ray and directly yields the pixel color. It further obviates the need for multi-bounce ray-tracing, as reflections are similarly absorbed by the light field. On the flip-side, this loses the guarantee of multi-view consistency: where a 3D renderer is guaranteed to map a single 3D coordinate to single value, a neural light field may map two rays that hit the same point in the scene to two different colors. This has to be addressed by additional means.

Sitzmann et al.~\cite{sitzmann2021lfns} propose to generalize across scenes by conditioning the neural light field on a latent code, thus learning a space of multi-view consistent light fields, however constrained to simple scenes due to their use of global conditioning. Sajjadi et al.~\cite{sajjadi2021scene} follow the prior-based inference approach and use a transformer to parameterize the 360-degree light fields of scenes, inferred from few image observations, achieving novel view synthesis for complex, real-world scenes. Attal et al.~\cite{attal2021learning}, Ost et al.~\cite{ost2021neural}, Liu et al.~\cite{liu2021nelf} and Suhail et al.~\cite{suhail2021light} instead investigate the same paradigms as NeRF, i.e., reconstruction and novel view synthesis of a single 3D scene. To ensure multi-view consistency, Suhail et al.~\cite{suhail2021light} leverage pixel-aligned CNN features as in PixelNeRF~\cite{yu2020pixelnerf} for 3D points along a ray, which are accumulated with a transformer. This yields significant improvements over NeRF, but still requires sampling in 3D and is several times slower. Ost et al.~\cite{ost2021neural} relies on a coarse 3D reconstruction of the 3D scene in form of a point cloud to parameterize point-wise light fields. Attal et al.~\cite{attal2021learning} rely on storing features in a voxel grid, where every feature parameterizes the local light field of rays intersecting that voxel, and render via volume rendering. Liu et al.~\cite{liu2021nelf} leverage regularization to ensure multi-view consistency.

\subsection{Engineering Frameworks}\label{sec:engineering}

Working with neural rendering models poses notable engineering challenges for practitioners: 
large amounts of image and video data must be processed in a highly non-sequential manner, and the models often require differentiation of large and complex computational graphs. Developing efficient operators often requires working with low-level languages which at the same time makes it harder to use automatic differentiation. In this section, we will discuss recent advances in tools that can help to overcome problems across the entire software stack relevant for neural rendering.

\subsubsection{Storage}

Saturating a GPU with data in particular for neural rendering is challenging: often, each pixel of images or videos is treated as a separate data point. Methods require random iteration over the entire pool of pixels in the dataset, in case of temporal reconstruction often across the entire sequence for a single batch. Flexible storage solutions should take this into account.

NVIDIA AIStore~\cite{Aizman2019} is a general purpose storage solution that allows to monitor throughput per drive and implements tiered architectures for loading and shuffling, while abstracting these layers away from the user. Independent of the storage backend, sharding has tremendous benefits through 1) allowing to shuffle data in memory while 2) using mostly sequential reads within the shards. Tensorflow~\cite{tensorflow} has built-in support sharded storage through the \texttt{tfrecord} file format, whereas \texttt{webdataset}~\code{https://github.com/webdataset/webdataset} offers similar convenient features for PyTorch~\cite{pytorch}.

\subsubsection{Hyperparameter Search and Experiments}

With long runtimes and complex configuration hierarchies, neural rendering experiments require good techniques for experiment management and hyperparameter search. Hydra~\cite{hydra} excels at configuring even the most complex experiments and offers integrated support for hyperparameter search, for example using the AX adaptive experimentation framework\code{https://ax.dev}.
However, running all experiments for a sweep until convergence, even for smartly picked parameters using Bayesian hyperparameter search, might be too time consuming. Ray tune~\cite{tune} has implementations of algorithms like ASHA~\cite{asha} and Hyperband~\cite{hyperband}, which can dynamically assign computational and time budgets to experiments for a faster hyperparameter search.

\subsubsection{Differentiable Rendering and Autodiff}

Neural Rendering has high demands towards differentiability: complex computational graphs need to be built and, depending on the application, be executed either on large inputs vectorized (macro AD---for brevity we refer to auto-differentiation as AD throughout this section) or on large amounts of small inputs (micro AD). Depending on the application, the AD package might have to be used low level (e.g., in CUDA), or high level (e.g., in Python). A powerful AD library for C++ is STAN~\cite{stan}. We refer to the accompanying paper for a comprehensive overview and evaluation of AD libraries until its publication in 2015, which is beyond the scope of this article. A noteworthy more recent AD package for C++17 is the \texttt{autodiff}~\code{https://autodiff.github.io}
package. Enzyme AD~\cite{enzyme,enzymegpu} is taking a particularly versatile approach for low-level AD: it leverages the LLVM ecosystem as a whole. This is particularly powerful, because of the concept of frontends, the LLVM IR and backends. In broad strokes, LLVM frontends translate a language, for example C++, to the LLVM \emph{intermediate representation} (IR). This representation is an abstract, language-agnostic representation of low-level commands, and it is the same for all frontends. This is where Enzyme comes in: it is an extension that can create derivatives of functions in this IR. That means that it works for all languages that LLVM supports. LLVM backends emit code from the IR: this could be for x86, ARM or GPU processors. This means, that Enzyme supports a variety of processors, including GPUs. Another C++ package specifically for processing images and graphics is Halide~\cite{halide}. Its standout feature is flexible scheduling for parallel processing of pixels.

Difftaichi~\cite{difftaichi} offers differentiable programming in Python for physical simulation with applications in rendering. Enoki~\cite{enoki} is a very versatile and high performance AD component for physically-based differentiable rendering and is the core component of the Mitsuba 2 renderer~\cite{mitsuba2}. Jax~\cite{jax} is a Python framework for differentiable and accelerated linear algebra with compilation options for GPUs and TPUs. JaxNeRF~\code{https://github.com/google-research/google-research/tree/master/jaxnerf}
is a reference implementation for NeRF using Jax. The Swift programming language provides AD as a first class use case~\code{https://github.com/apple/swift/blob/main/docs/DifferentiableProgramming.md}, and was heavily used for developing a Tensorflow integration~\code{https://github.com/tensorflow/swift}.

\subsubsection{Raycasting and Rendering}

Several packages exist for providing high-level rendering and aggregation primitives. NVIDIA OptiX~\code{https://developer.nvidia.com/optix}
is a high performance library for ray-casting and ray-intersection and provides to date the only possibility to use the hardware acceleration on NVIDIA RTX hardware for ray intersection. Teg~\cite{teg} is a differentiable programming language which provides primitives for optimizing integrals with discontinuous integrands, as frequently found in rendering. Redner~\cite{redner} is a framework for differentiable ray tracing; Mitsuba 2~\cite{mitsuba2} provides an even more general framework for physically based differentiable rendering and path tracing. psdr-cuda~\cite{psdrcuda} improves over Redner by using better gradient calculation techniques and sampling strategies. PyTorch3D~\cite{pytorch3d} offers a broad suite of tools around differentiable rendering and graphics, tightly integrated with PyTorch. Tensorflow Graphics~\cite{tensorflowgraphics} has a similar goal for Tensorflow.

 \vspace{1cm}
    \vspace{-1.0cm}
\section{Open Challenges} 
\label{sec:open_challenges}

After covering a wide variety of computer graphics and vision problems to which neural volumetric representations can be successfully applied, we now take a look at problems where only classical representations have been used. Thus, there are various avenues for future research. We further discuss multiple open challenges in the field. Many of the points discussed in the following are related to each other. 

\noindent\textbf{Seamless Integration and Usage.} Most computer graphics algorithms and techniques developed over more than half a century assume meshes or point clouds as 3D scene representations for rendering and editing. In contrast, neural rendering is such a young field that this notion was used for the first time just a few years ago in 2018 \cite{eslami2018neural}. Thus, inevitably, there is still a gap between the spectrum of available methods that can operate on classical 3D representations and those that are applicable to neural representations. Furthermore, many methods exist to edit classical representations, e.g., widely-used tools such as Blender \cite{Blender2018} and Maya \cite{Maya2021} support meshes and texture maps, whereas their counterparts for neural representations have to be developed from scratch. On the other hand, it is foreseeable that this gap will decrease with further improvement in the field and more and more widespread adoption and integration of neural representations. Moreover, modern hardware accelerators are designed for classical computer graphics and could in the future be similarly tailored to neural representations.

Another related challenge is interpretability of the learned representations, which concerns deep learning in general. Thus, learned neural network weights are notoriously hard to interpret in terms of the target quantities (e.g., point colors and opacities in the 3D space). At the same time, they aim to replace the graphics pipeline, which is well understood and relies on analytically derived steps. 
Ultimately, to improve controllability and enable seamless integration of learned volumetric models in computer graphics tools, we would like to be able to modify the scene parametrization to change the scene in a desired direction. While this is likely not tractable for arbitrary scenes parametrized by global MLPs, composing a full scene out of local neural representations might make it tractable by opening up the intriguing possibility of re-introducing aspects of classical graphics.

\begin{figure}[t!] 
    \centering 
    \includegraphics[width=\linewidth]{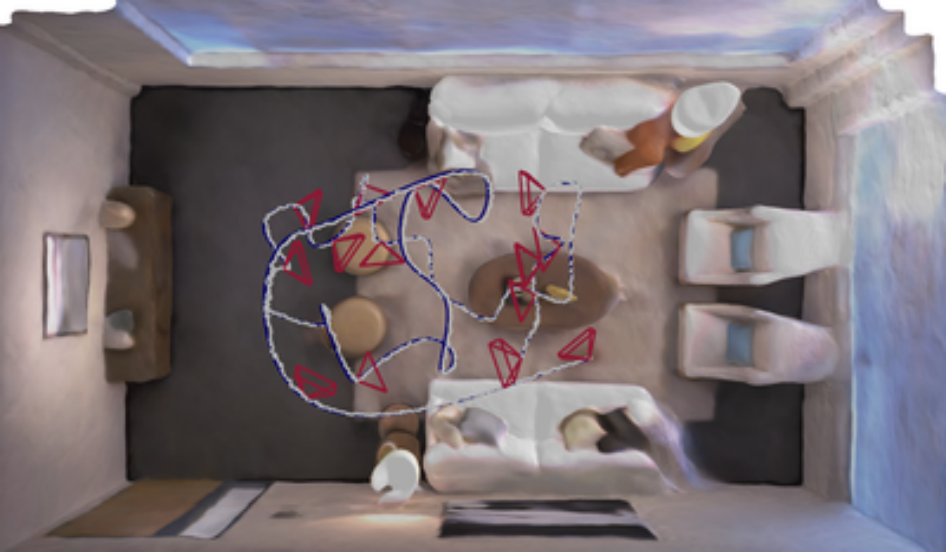} 
    \caption{3D reconstruction for a room with keyframes (shown in red) obtained by iMAP \cite{Sucar2021}. 
    iMAP is a real-time SLAM system for a single handheld RGB-D camera that can efficiently fill in occluded regions. Despite the first successful steps, learning neural representations for large-scale scenes has still many open challenges. Image adapted from \cite{Sucar2021} \textcopyright 2021 IEEE.} 
    \label{fig:iMAP} 
\end{figure}

\noindent\textbf{Scalability.} Most of the works on volumetric neural rendering focus on single objects and relatively simple composite scenes (e.g., a human and a background, several humans in the same environment, a street with moving cars) with or without background. Learning neural representations for large-scale scenes---which can only be partially observed in each input frame---is still challenging. Although the first successful and impressive steps in this direction have been made (we refer here to Nerf in the Wild \cite{martinbrualla2020nerfw}, NeRF-OSR  \cite{rudnev2021nerfosr} and the neural SLAM system iMAP  \cite{Sucar2021}, see \Cref{fig:iMAP}), many open challenges  remain. For instance, the approaches for scene editing, relighting, and compositionality developed for single objects cannot be straightforwardly extended to handle large-scale scenes.  Moreover, a global representation for large-scale environments becomes unfeasible starting from some scene size, even when applying space partitioning policies such as those used in PlenOctrees \cite{yu2021plenoctrees}. Thus, a new generation of storage and retrieval techniques need to be developed for efficient neural models for large-scale scenes, along the lines of VoxelHashing \cite{niessner2013hashing} for TSDFs. First, they should make the scene completion more efficient (i.e., without the need to constantly recompute the entire model from scratch) and, second, enable easy retrieval of partial contents. Both these points are related to the open challenge of interpretability discussed above. 

\noindent\textbf{Generalizability.} Only a few initial but promising methods exist for generalizable and instantiable volumetric neural representations. For example, StereoNeRF \cite{Chibane2021} uses only a dozen spread-out views to generate novel views of a rigid scene with the visual accuracy comparable to the original NeRF \cite{Mildenhall_2020_NeRF} after fine-tuning, while pixelNeRF \cite{yu2020pixelnerf} can infer volumetric models of rigid scenes unseen at training time just from a single image. This class of approaches is data-driven and requires large-scale multi-view datasets with a sufficiently wide baseline. Consequently, these methods can produce views at arbitrary novel viewpoints if the datasets provide sufficient viewpoint coverage. Reducing this strong dependency is an exciting direction for future work. Another open challenge is the generalizability of instantiable approaches to scenes with non-rigid deformations. The inputs can be sparse sets of spatiotemporal observations or even single images at the extreme (in this case, the task becomes scene animation from a single image). One straightforward direction towards such techniques would be relying on multi-view datasets of deformable scenes, which would likely increase required dataset sizes by a multitude. Another possible way would be to disentangle deformation modes and scene shapes and appearances at rest. 
Furthermore, while there exists some work on generating neural scene representations (e.g., using hypernets \cite{sitzmann2021lfns}), there is less progress on designing neural operators that take neural scene representations as input to work on them, for example to complete a partial scene or to regress semantic labels for an existing representation. No operator analogous to mesh convolutions for meshes or 3D convolutions for voxel grids exists. Such an operator would ideally be trained only once and then be generally applicable.

\noindent\textbf{Multi-Modal Learning.}
Multi-modal learning means going beyond visual signals and incorporating other data types such as semantics, textual descriptions and sound. 
For example, telepresence and augmented reality would highly benefit from a method that can not only render novel views of dynamically interacting and talking humans but also synthesize the corresponding novel sounds;  
existing work can, for instance, synthesize stereo audio from mono audio inputs  \cite{Richard2021binaural}. 
Synthesizing textual descriptions and semantics of the scene (e.g., semantic segmentation labels) can be very useful for downstream applications based on volumetric neural representations. While some prior work addresses this goal~\cite{kohli2020inferring,zhi2021place}, this remains an open challenge. 
More detailed and scenario-specific modeling could take into account such information as the  camera capture system (e.g., as already shown in TöRF \cite{attal2021torf} for depth cameras), whether the camera is using rolling or global shutter, or if there is motion blur in the input images. Other sensors like IMUs, Lidar, or event streams could all potentially be modeled in a continuous fashion. (Ultrasound and x-rays could be continuously modeled at arbitrary resolution for medical imaging.) It is also conceivable to optimize for certain capture properties that are not trivial to measure like color calibrations of a multi-view capture setup.
This extends to physical simulation in general, where neural scene representations offer an exciting venue to ``learn less and know more" by incorporating differentiable physics simulators; e.g., for physically motivated deformation models or  physically correct light transport. 

\noindent\textbf{Other Questions.} 
Can we increase quality? Reconstructing objects with many high-frequency details, shading, and view-dependent appearance remains a largely unsolved problem. 
Can we decrease training time? 
Although there has been progress on very fast inference for novel view synthesis at test time, improving the training time remains a big challenge. 
The work by Lange and Kutz \cite{LangeKutz2021} is one of  the first  encouraging steps towards this goal. It introduces fast computational neural network layers that rely on series expansions and fast summation algorithms. Thus, the proposed integral-implicit layer reduces the required computational performance to train a NeRF model by two orders of magnitude (${\sim}150$x reduction in FLOPs per epoch). 
Are fewer input images sufficient? Fewer input views might be sufficient to reach a similar visual fidelity as fully converged models requiring hundreds of views. Currently, partial observations (e.g., parts of the scene observed only in a subset of images) tend to be blurrier than the rest of the scene. 

Beyond the immediate use case of AR/VR, there is little research on using neural scene representations in other contexts like robotics, with the notable exception of the real-time SLAM system iMAP \cite{Sucar2021} supporting single-room environments (see \Cref{fig:iMAP}). 
How can we obtain, incorporate, and predict object affordances or other annotations like temperature? Are there advantages to using neural scene representations for motion prediction or planning?

The list of future directions discussed in this section does not aim for completeness. We expect to see many improvements on more aspects of coordinate-based neural volumetric representations already in the near future.

 \vspace{1cm}
    \vspace{-0.5cm}
\section{Social Implications} 
\label{sec:social}

Neural approaches discussed in this state-of-the-art report achieve a very high degree of realism for synthesized novel views. 
Rapid developments in the field already influence and will continue influencing society in many positive and potentially negative ways which we discuss in this section.

\noindent\textbf{Research and Industry.} The fields which are starkly impacted by the new volumetric neural representations are computer vision, computer graphics as well as augmented and virtual reality, which can benefit from increased photo-realism of rendered environments. The fact that the state-of-the-art volumetric models rely on well-understood and elegant principles lowers the barrier to entry for research on photogrammetry and 3D reconstruction. Moreover, this effect is magnified by the ease of use of the methods and publicly available
codebases and datasets. 
 
Since neural rendering is still not mature and well understood, end-user tools like Blender do not yet exist, putting these novel methods out of reach for both 3D hobbyists and industry as of now. However, more widespread understanding of the technology inevitably impacts developed products and applications. With that, we foresee decreased effort in content creation for games and special effects for movies. The possibility to render photo-realistic novel views of a scene from a few input images is a significant advantage compared to existing technology. This can potentially reshape the entire established pipeline for content design used in the visual effects (VFX) industry. %

\noindent\textbf{Trustworthiness.} However, at the same time, photo-realism creates the possibility to misuse the technology and create synthetic content that malicious actors may falsely claim to be real, in particular, when neural rendering approaches focus on human faces~\cite{thies2016face,thies2019deferred,Gafni_2021_CVPR}.
In response to these potential misuses, methods to automatically detect such fake content are being developed by the research community \cite{cozzolino2021id,roessler2019faceforensics++}
and security measures including encryption and block chain measures are being explored.
And there are a number of other mitigations that could be explored to minimize these risks.
For example, while there are cases where we expect users not to object to seeing synthetic photo-real content (e.g., when watching movies), synthetic content could be labelled or otherwise identified as such to inform users.
Further user studies could investigate people's judgement of the need to label synthetic content in different contexts.
On the collection side, people could provide explicit and informed consent that their identity can be used for creating synthetic content in a specified context.

\noindent\textbf{Environment.} Since current neural volumetric scene representations are deep-learning-based, the GPUs used for training them consume a sizable amount of energy. 
Since more and more laboratories are working on neural rendering, the use of high-end and multi-GPU systems increases accordingly.
If the resources for manufacturing and the electricity for operating the GPU clusters are not taken predominantly from renewable sources, training volumetric neural representations can negatively influence the environment and global climate in the long term. In an attempt to soften the need for computational resources and hence electricity and hardware, there are many architectures that require less compute power for training than NeRF-based methods.
Last but not least, high GPU demand potentially implies that not all groups can afford to contribute on equal footing as experimenting with volumetric representations is not the most lightweight task.  %

 \vspace{1cm}
    \vspace{-0.3cm}
\section{Conclusion}

In this state-of-the-art report, we have reviewed the recent trends on neural rendering techniques. 
The methods covered learn 3D neural scene representations based on 2D observations as inputs for training, and enable synthesis of photo-realistic imagery with control over different scene parameters. 
The field of neural rendering has seen rapid progress during the last few years and continues to grow fast. 
Its applications range from free-viewpoint videos of rigid and non-rigid scenes to shape and material editing, relighting, and human avatar generation, among many others. 
These applications have been discussed in detail in this report. 

At the same time, we believe that neural rendering is still an emerging field with many open challenges that can be addressed. 
To this end, we identify and discuss multiple directions for future research.
In addition, we discuss social implications, which arise from the democratization of neural rendering along with its capability to synthesize photo-realistic image content.
Overall, we conclude that neural rendering is an exciting field, which is inspiring thousands of researchers across many communities to tackle some of computer graphics' hardest problems, and we look forward to seeing further developments on the topic.

    \section{Acknowledgements} 
A.~Tewari, V.~Golyanik, and C.~Theobalt are supported in part by the ERC Consolidator Grant 4DReply (770784).
E.~Tretschk is supported by a Reality Labs Research grant.
M.~Nie{\ss}ner is supported by the ERC Starting Grant Scan2CAD (804724).

    \bibliographystyle{eg-alpha-doi} 
    \bibliography{main}       

\newcommand{\etalchar}[1]{$^{#1}$}
\begin{thebibliography}{\uppercase{PCPMMN21}}

\bibitem[AAB{\etalchar{*}}15]{tensorflow}
\textsc{Abadi M., Agarwal A., Barham P., Brevdo E., Chen Z., Citro C., Corrado
  G.~S., Davis A., Dean J., Devin M., Ghemawat S., Goodfellow I., Harp A.,
  Irving G., Isard M., Yangqing J., Jozefowicz R., Kaiser L., Kudlur M.,
  Levenberg J., Man{\'{e}} D., Monga R., Moore S., Murray D., Olah C., Schuster
  M., Shlens J., Steiner B., Sutskever I., Talwar K., Tucker P., Vanhoucke V.,
  Vasudevan V., Vi{\'{e}}gas F., Vinyals O., Warden P., Wattenberg M., Wicke
  M., Yu Y., Zheng X.}:
\newblock {{TensorFlow}: Large-Scale Machine Learning on Heterogeneous
  Systems}.
\newblock \url{http://tensorflow.org/}, 2015.

\bibitem[AHZ{\etalchar{*}}21]{attal2021learning}
\textsc{Attal B., Huang J.-B., Zollhoefer M., Kopf J., Kim C.}:
\newblock Learning neural light fields with ray-space embedding networks.
\newblock \emph{arXiv preprint arXiv:2112.01523} (2021).

\bibitem[AL20]{atzmon2020sal}
\textsc{Atzmon M., Lipman Y.}:
\newblock Sal: Sign agnostic learning of shapes from raw data.
\newblock In \emph{Proceedings of the IEEE/CVF Conference on Computer Vision
  and Pattern Recognition} (2020), pp.~2565--2574.

\bibitem[ALG{\etalchar{*}}20]{Attal:2020:ECCV}
\textsc{Attal B., Ling S., Gokaslan A., Richardt C., Tompkin J.}:
\newblock {MatryODShka}: Real-time {6DoF} video view synthesis using
  multi-sphere images.
\newblock In \emph{Proc. ECCV} (Aug. 2020).
\newblock URL: \url{https://visual.cs.brown.edu/matryodshka}.

\bibitem[ALG{\etalchar{*}}21]{attal2021torf}
\textsc{Attal B., Laidlaw E., Gokaslan A., Kim C., Richardt C., Tompkin J.,
  O'Toole M.}:
\newblock T\"orf: Time-of-flight radiance fields for dynamic scene view
  synthesis.
\newblock In \emph{Neural Information Processing Systems (NeurIPS)} (2021).

\bibitem[ALKN19]{azinovic2019inverse}
\textsc{Azinovic D., Li T.-M., Kaplanyan A., Nie{\ss}ner M.}:
\newblock Inverse path tracing for joint material and lighting estimation.
\newblock In \emph{Proceedings of the IEEE/CVF Conference on Computer Vision
  and Pattern Recognition} (2019), pp.~2447--2456.

\bibitem[AMB19]{Aizman2019}
\textsc{Aizman A., Maltby G., Breuel T.}:
\newblock {High Performance I/O For Large Scale Deep Learning}.
\newblock \emph{IEEE International Conference on Big Data} (2019), 5965--5967.

\bibitem[AMBG{\etalchar{*}}21]{azinovic2021rgbdnerf}
\textsc{Azinovic D., Martin-Brualla R., Goldman D.~B., Nie{\ss}ner M., Thies
  J.}:
\newblock Neural rgb-d surface reconstruction.

\bibitem[ASK{\etalchar{*}}20a]{aliev2020ngp}
\textsc{Aliev K.-A., Sevastopolsky A., Kolos M., Ulyanov D., Lempitsky V.}:
\newblock Neural point-based graphics.
\newblock \href {http://arxiv.org/abs/2110.06635} {\path{arXiv:2110.06635}}.

\bibitem[ASK{\etalchar{*}}20b]{aliev2020neural}
\textsc{Aliev K.-A., Sevastopolsky A., Kolos M., Ulyanov D., Lempitsky V.}:
\newblock Neural point-based graphics.
\newblock In \emph{Computer Vision--ECCV 2020: 16th European Conference,
  Glasgow, UK, August 23--28, 2020, Proceedings, Part XXII 16} (2020),
  Springer, pp.~696--712.

\bibitem[{Aut}]{Maya2021}
\textsc{{Autodesk, INC.}}:
\newblock Maya.
\newblock URL: \url{https://autodesk.com/maya}.

\bibitem[AXS21]{Alldieck_2021_ICCV}
\textsc{Alldieck T., Xu H., Sminchisescu C.}:
\newblock imghum: Implicit generative models of 3d human shape and articulated
  pose.
\newblock In \emph{International Conference on Computer Vision (ICCV)} (2021).

\bibitem[BBJ{\etalchar{*}}21]{boss2020nerd}
\textsc{Boss M., Braun R., Jampani V., Barron J.~T., Liu C., Lensch H. P.~A.}:
\newblock {NeRD}: Neural reflectance decomposition from image collections.
\newblock \emph{ICCV} (2021).

\bibitem[BFH{\etalchar{*}}18]{jax}
\textsc{Bradbury J., Frostig R., Hawkins P., Johnson M.~J., Leary C., Maclaurin
  D., Necula G., Paszke A., Vander{P}las J., Wanderman-{M}ilne S., Zhang Q.}:
\newblock {JAX}: composable transformations of {P}ython+{N}um{P}y programs,
  2018.
\newblock URL: \url{http://github.com/google/jax}.

\bibitem[BFO{\etalchar{*}}20]{Broxton:2020}
\textsc{Broxton M., Flynn J., Overbeck R., Erickson D., Hedman P., Duvall M.,
  Dourgarian J., Busch J., Whalen M., Debevec P.}:
\newblock Immersive light field video with a layered mesh representation.
\newblock \emph{ACM Trans. Graph. (SIGGRAPH) 39}, 4 (2020).

\bibitem[BGP{\etalchar{*}}21]{baatz2021nerftex}
\textsc{Baatz H., Granskog J., Papas M., Rousselle F., Nov\'{a}k J.}:
\newblock Nerf-tex: Neural reflectance field textures.
\newblock In \emph{Eurographics Symposium on Rendering} (June 2021), The
  Eurographics Association.

\bibitem[BKK19]{bai2019deep}
\textsc{Bai S., Kolter J.~Z., Koltun V.}:
\newblock Deep equilibrium models.
\newblock \emph{NeurIPS} (2019).

\bibitem[BKW21]{bergman2021metanlr}
\textsc{Bergman A.~W., Kellnhofer P., Wetzstein G.}:
\newblock Fast training of neural lumigraph representations using meta
  learning.
\newblock In \emph{Proceedings of the IEEE International Conference on Neural
  Information Processing Systems (NeurIPS)} (2021).

\bibitem[BMM{\etalchar{*}}21]{teg}
\textsc{Bangaru S., Michel J., Mu K., Bernstein G., Li T.-M., Ragan-Kelley J.}:
\newblock Systematically differentiating parametric discontinuities.
\newblock \emph{ACM Trans. Graph. 40}, 107 (2021), 107:1--107:17.

\bibitem[BMT{\etalchar{*}}21]{barron2021mipnerf}
\textsc{Barron J.~T., Mildenhall B., Tancik M., Hedman P., Martin-Brualla R.,
  Srinivasan P.~P.}:
\newblock Mip-nerf: A multiscale representation for anti-aliasing neural
  radiance fields.
\newblock \emph{ICCV} (2021).

\bibitem[BMV{\etalchar{*}}21]{barron2021mipnerf360}
\textsc{Barron J.~T., Mildenhall B., Verbin D., Srinivasan P.~P., Hedman P.}:
\newblock Mip-nerf 360: Unbounded anti-aliased neural radiance fields.
\newblock \emph{arXiv} (2021).

\bibitem[BNT21]{burov2021dsfn}
\textsc{Burov A., Nie{\ss}ner M., Thies J.}:
\newblock Dynamic surface function networks for clothed human bodies.

\bibitem[BXS{\etalchar{*}}20]{bi2020nrf}
\textsc{Bi S., Xu Z., Srinivasan P.~P., Mildenhall B., Sunkavalli K., Hašan
  M., Hold-Geoffroy Y., Kriegman D., Ramamoorthi R.}:
\newblock Neural reflectance fields for appearance acquisition.
\newblock \href {http://arxiv.org/abs/2008.03824} {\path{arXiv:2008.03824}}.

\bibitem[CBC{\etalchar{*}}01a]{RBF_reco}
\textsc{Carr J.~C., Beatson R.~K., Cherrie J.~B., Mitchell T.~J., Fright W.~R.,
  McCallum B.~C., Evans T.~R.}:
\newblock Reconstruction and representation of 3d objects with radial basis
  functions.
\newblock In \emph{Proceedings of the 28th Annual Conference on Computer
  Graphics and Interactive Techniques} (New York, NY, USA, 2001), SIGGRAPH '01,
  Association for Computing Machinery, p.~67–76.
\newblock URL: \url{https://doi.org/10.1145/383259.383266}, \href
  {https://doi.org/10.1145/383259.383266} {\path{doi:10.1145/383259.383266}}.

\bibitem[CBC{\etalchar{*}}01b]{carr2001reconstruction}
\textsc{Carr J.~C., Beatson R.~K., Cherrie J.~B., Mitchell T.~J., Fright W.~R.,
  McCallum B.~C., Evans T.~R.}:
\newblock Reconstruction and representation of 3d objects with radial basis
  functions.
\newblock In \emph{Proceedings of the 28th annual conference on Computer
  graphics and interactive techniques} (2001), pp.~67--76.

\bibitem[CBLPM21]{Chibane2021}
\textsc{Chibane J., Bansal A., Lazova V., Pons-Moll G.}:
\newblock Stereo radiance fields (srf): Learning view synthesis from sparse
  views of novel scenes.
\newblock In \emph{Computer Vision and Pattern Recognition (CVPR)} (2021).

\bibitem[CHB{\etalchar{*}}15]{stan}
\textsc{Carpenter B., Hoffman M.~D., Brubaker M., Lee D., Li P., Betancourt
  M.}:
\newblock {The Stan Math Library: Reverse-Mode Automatic Differentiation in
  C++}.
\newblock URL: \url{http://arxiv.org/abs/1509.07164}, \href
  {http://arxiv.org/abs/1509.07164} {\path{arXiv:1509.07164}}.

\bibitem[Chu06]{ChumpusRex2006}
\textsc{ChumpusRex}:
\newblock Craniale computertomographie, 2006.
\newblock URL:
  \url{https://de.wikipedia.org/wiki/Computertomographie##/media/Datei:Ct-workstation-neck.jpg}.

\bibitem[CKS{\etalchar{*}}17]{Chaitanya:2017}
\textsc{Chaitanya C. R.~A., Kaplanyan A.~S., Schied C., Salvi M., Lefohn A.,
  Nowrouzezahrai D., Aila T.}:
\newblock Interactive reconstruction of monte carlo image sequences using a
  recurrent denoising autoencoder.
\newblock \emph{ACM Trans. Graph. 36}, 4 (July 2017), 98:1--98:12.
\newblock URL: \url{http://doi.acm.org/10.1145/3072959.3073601}, \href
  {https://doi.org/10.1145/3072959.3073601}
  {\path{doi:10.1145/3072959.3073601}}.

\bibitem[CL96]{curless1996volumetric}
\textsc{Curless B., Levoy M.}:
\newblock A volumetric method for building complex models from range images.
\newblock In \emph{Proceedings of the 23rd annual conference on Computer
  graphics and interactive techniques} (1996), pp.~303--312.

\bibitem[CLC{\etalchar{*}}22]{Chan2021}
\textsc{Chan E.~R., Lin C.~Z., Chan M.~A., Nagano K., Pan B., Mello S.~D.,
  Gallo O., Guibas L., Tremblay J., Khamis S., Karras T., Wetzstein G.}:
\newblock Efficient geometry-aware {3D} generative adversarial networks.
\newblock In \emph{Proceedings of the IEEE/CVF Conference on Computer Vision
  and Pattern Recognition} (2022).

\bibitem[CLI{\etalchar{*}}20]{chabra2020deep}
\textsc{Chabra R., Lenssen J.~E., Ilg E., Schmidt T., Straub J., Lovegrove S.,
  Newcombe R.}:
\newblock Deep local shapes: Learning local sdf priors for detailed 3d
  reconstruction.
\newblock In \emph{European Conference on Computer Vision (Proceedings of the
  European Conference on Computer Vision)} (2020).

\bibitem[CLX{\etalchar{*}}21]{sofgan}
\textsc{Chen A., Liu R., Xie L., Chen Z., Su H., Jingyi Y.}:
\newblock Sofgan: A portrait image generator with dynamic styling.
\newblock \emph{ACM Trans. Graph. 41}, 1 (2021).
\newblock URL: \url{https://doi.org/10.1145/3470848}, \href
  {https://doi.org/10.1145/3470848} {\path{doi:10.1145/3470848}}.

\bibitem[CMK{\etalchar{*}}21]{chanmonteiro2020piGAN}
\textsc{Chan E., Monteiro M., Kellnhofer P., Wu J., Wetzstein G.}:
\newblock pi-gan: Periodic implicit generative adversarial networks for
  3d-aware image synthesis.
\newblock In \emph{CVPR} (2021).

\bibitem[Com18]{Blender2018}
\textsc{Community B.~O.}:
\newblock \emph{Blender - a 3D modelling and rendering package}.
\newblock Blender Foundation, Stichting Blender Foundation, Amsterdam, 2018.
\newblock URL: \url{http://www.blender.org}.

\bibitem[CRBD18]{NeuralODE2018}
\textsc{Chen R. T.~Q., Rubanova Y., Bettencourt J., Duvenaud D.~K.}:
\newblock Neural ordinary differential equations.
\newblock In \emph{Advances in Neural Information Processing Systems} (2018),
  vol.~31.

\bibitem[CRT{\etalchar{*}}21]{cozzolino2021id}
\textsc{Cozzolino D., Rossler A., Thies J., Nie{\ss}ner M., Verdoliva L.}:
\newblock Id-reveal: Identity-aware deepfake video detection.
\newblock In \emph{Proceedings of the IEEE/CVF International Conference on
  Computer Vision} (2021), pp.~15108--15117.

\bibitem[CTZ20]{chen2020bsp}
\textsc{Chen Z., Tagliasacchi A., Zhang H.}:
\newblock Bsp-net: Generating compact meshes via binary space partitioning.
\newblock In \emph{Proceedings of the IEEE/CVF Conference on Computer Vision
  and Pattern Recognition} (2020), pp.~45--54.

\bibitem[CW93]{chen1993view}
\textsc{Chen S.~E., Williams L.}:
\newblock View interpolation for image synthesis.
\newblock In \emph{SIGGRAPH} (1993), pp.~279--288.

\bibitem[CZ19]{chen2019learning}
\textsc{Chen Z., Zhang H.}:
\newblock Learning implicit fields for generative shape modeling.
\newblock In \emph{Proceedings of the IEEE/CVF Conference on Computer Vision
  and Pattern Recognition} (2019), pp.~5939--5948.

\bibitem[CZB{\etalchar{*}}21]{chen2021snarf}
\textsc{Chen X., Zheng Y., Black M.~J., Hilliges O., Geiger A.}:
\newblock Snarf: Differentiable forward skinning for animating non-rigid neural
  implicit shapes, 2021.
\newblock \href {http://arxiv.org/abs/2104.03953} {\path{arXiv:2104.03953}}.

\bibitem[CZL{\etalchar{*}}21]{chen2021hallucinated}
\textsc{Chen X., Zhang Q., Li X., Chen Y., Feng Y., Wang X., Wang J.}:
\newblock Hallucinated neural radiance fields in the wild, 2021.
\newblock \href {http://arxiv.org/abs/2111.15246} {\path{arXiv:2111.15246}}.

\bibitem[DGY{\etalchar{*}}20]{deng2020cvxnet}
\textsc{Deng B., Genova K., Yazdani S., Bouaziz S., Hinton G., Tagliasacchi
  A.}:
\newblock Cvxnet: Learnable convex decomposition.
\newblock In \emph{Proceedings of the IEEE/CVF Conference on Computer Vision
  and Pattern Recognition} (2020), pp.~31--44.

\bibitem[DLZR21]{kangle2021dsnerf}
\textsc{Deng K., Liu A., Zhu J.-Y., Ramanan D.}:
\newblock Depth-supervised nerf: Fewer views and faster training for free.
\newblock \emph{arXiv preprint arXiv:2107.02791} (2021).

\bibitem[DNJ20]{davies2020overfit}
\textsc{Davies T., Nowrouzezahrai D., Jacobson A.}:
\newblock Overfit neural networks as a compact shape representation, 2020.
\newblock \href {http://arxiv.org/abs/2009.09808} {\path{arXiv:2009.09808}}.

\bibitem[DYXT21]{deng2021gram}
\textsc{Deng Y., Yang J., Xiang J., Tong X.}:
\newblock Gram: Generative radiance manifolds for 3d-aware image generation.
\newblock In \emph{arXiv} (2021).

\bibitem[DZW{\etalchar{*}}20]{duan2020curriculum}
\textsc{Duan Y., Zhu H., Wang H., Yi L., Nevatia R., Guibas L.~J.}:
\newblock Curriculum deepsdf.
\newblock In \emph{European Conference on Computer Vision} (2020), Springer,
  pp.~51--67.

\bibitem[DZY{\etalchar{*}}21]{du2021nerflow}
\textsc{Du Y., Zhang Y., Yu H.-X., Tenenbaum J.~B., Wu J.}:
\newblock Neural radiance flow for 4d view synthesis and video processing.
\newblock In \emph{Proceedings of the IEEE/CVF International Conference on
  Computer Vision} (2021).

\bibitem[EGO{\etalchar{*}}20]{erler2020points2surf}
\textsc{Erler P., Guerrero P., Ohrhallinger S., Mitra N.~J., Wimmer M.}:
\newblock Points2surf learning implicit surfaces from point clouds.
\newblock In \emph{Proceedings of the European Conference on Computer Vision}
  (2020), Springer, pp.~108--124.

\bibitem[ERB{\etalchar{*}}18]{eslami2018neural}
\textsc{Eslami S.~A., Rezende D.~J., Besse F., Viola F., Morcos A.~S., Garnelo
  M., Ruderman A., Rusu A.~A., Danihelka I., Gregor K., et~al.}:
\newblock Neural scene representation and rendering.
\newblock \emph{Science 360}, 6394 (2018), 1204--1210.

\bibitem[FBD{\etalchar{*}}19]{flynn2019deepview}
\textsc{Flynn J., Broxton M., Debevec P., DuVall M., Fyffe G., Overbeck R.,
  Snavely N., Tucker R.}:
\newblock Deepview: View synthesis with learned gradient descent.
\newblock In \emph{Proc. Proceedings of the IEEE/CVF Conference on Computer
  Vision and Pattern Recognition} (2019), pp.~2367--2376.

\bibitem[FNPS16]{flynn2016deepstereo}
\textsc{Flynn J., Neulander I., Philbin J., Snavely N.}:
\newblock Deep stereo: Learning to predict new views from the
  world{\textquoteright}s imagery.
\newblock In \emph{Proc. Proceedings of the IEEE/CVF Conference on Computer
  Vision and Pattern Recognition} (2016).

\bibitem[FXW{\etalchar{*}}21]{fang2021neusample}
\textsc{Fang J., Xie L., Wang X., Zhang X., Liu W., Tian Q.}:
\newblock Neusample: Neural sample field for efficient view synthesis, 2021.
\newblock \href {http://arxiv.org/abs/2111.15552} {\path{arXiv:2111.15552}}.

\bibitem[GCL{\etalchar{*}}21]{guo2021adnerf}
\textsc{Guo Y., Chen K., Liang S., Liu Y., Bao H., Zhang J.}:
\newblock Ad-nerf: Audio driven neural radiance fields for talking head
  synthesis.
\newblock In \emph{IEEE/CVF International Conference on Computer Vision (ICCV)}
  (2021).

\bibitem[GCS{\etalchar{*}}20]{genova2020local}
\textsc{Genova K., Cole F., Sud A., Sarna A., Funkhouser T.}:
\newblock Local deep implicit functions for 3d shape.
\newblock In \emph{Proceedings of the IEEE/CVF Conference on Computer Vision
  and Pattern Recognition} (2020), pp.~4857--4866.

\bibitem[GCV{\etalchar{*}}19]{genova2019learning}
\textsc{Genova K., Cole F., Vlasic D., Sarna A., Freeman W.~T., Funkhouser T.}:
\newblock Learning shape templates with structured implicit functions.
\newblock In \emph{Proceedings of the International Conference on Computer
  Vision} (2019), pp.~7154--7164.

\bibitem[GKB{\etalchar{*}}21]{guo2021nerfren}
\textsc{Guo Y.-C., Kang D., Bao L., He Y., Zhang S.-H.}:
\newblock Nerfren: Neural radiance fields with reflections, 2021.
\newblock \href {http://arxiv.org/abs/2111.15234} {\path{arXiv:2111.15234}}.

\bibitem[GKJ{\etalchar{*}}21]{garbin2021fastnerf}
\textsc{Garbin S.~J., Kowalski M., Johnson M., Shotton J., Valentin J.}:
\newblock Fastnerf: High-fidelity neural rendering at 200fps.
\newblock \emph{arXiv preprint arXiv:2103.10380} (2021).

\bibitem[GLWT21a]{anonymous2022stylenerf}
\textsc{Gu J., Liu L., Wang P., Theobalt C.}:
\newblock Stylenerf: A style-based 3d-aware generator for high-resolution image
  synthesis, 2021.
\newblock \href {http://arxiv.org/abs/2110.08985} {\path{arXiv:2110.08985}}.

\bibitem[GLWT21b]{gu2021stylenerf}
\textsc{Gu J., Liu L., Wang P., Theobalt C.}:
\newblock Stylenerf: A style-based 3d-aware generator for high-resolution image
  synthesis, 2021.
\newblock \href {http://arxiv.org/abs/2110.08985} {\path{arXiv:2110.08985}}.

\bibitem[GPAM{\etalchar{*}}14]{Goodfellow_GAN}
\textsc{Goodfellow I., Pouget-Abadie J., Mirza M., Xu B., Warde-Farley D.,
  Ozair S., Courville A., Bengio Y.}:
\newblock Generative adversarial nets.
\newblock In \emph{Advances in Neural Information Processing Systems} (2014),
  Ghahramani Z., Welling M., Cortes C., Lawrence N., Weinberger K.~Q., (Eds.),
  vol.~27, Curran Associates, Inc.
\newblock URL:
  \url{https://proceedings.neurips.cc/paper/2014/file/5ca3e9b122f61f8f06494c97b1afccf3-Paper.pdf}.

\bibitem[GSHG98]{greger1998irradiance}
\textsc{Greger G., Shirley P., Hubbard P.~M., Greenberg D.~P.}:
\newblock The irradiance volume.
\newblock \emph{IEEE Computer Graphics and Applications 18}, 2 (1998), 32--43.

\bibitem[GSKH21]{Gao-freeviewvideo}
\textsc{Gao C., Saraf A., Kopf J., Huang J.-B.}:
\newblock Dynamic view synthesis from dynamic monocular video.
\newblock \emph{Proceedings of the IEEE International Conference on Computer
  Vision} (2021).

\bibitem[GSL{\etalchar{*}}20]{Gao-portraitnerf}
\textsc{Gao C., Shih Y., Lai W.-S., Liang C.-K., Huang J.-B.}:
\newblock Portrait neural radiance fields from a single image.
\newblock \emph{arXiv preprint arXiv:2012.05903} (2020).

\bibitem[GTZN21]{Gafni_2021_CVPR}
\textsc{Gafni G., Thies J., Zollh{\"o}fer M., Nie{\ss}ner M.}:
\newblock Dynamic neural radiance fields for monocular 4d facial avatar
  reconstruction.
\newblock In \emph{Proceedings of the IEEE/CVF Conference on Computer Vision
  and Pattern Recognition (CVPR)} (June 2021), pp.~8649--8658.

\bibitem[GYH{\etalchar{*}}20]{gropp2020implicit}
\textsc{Gropp A., Yariv L., Haim N., Atzmon M., Lipman Y.}:
\newblock Implicit geometric regularization for learning shapes.
\newblock \emph{arXiv preprint arXiv:2002.10099} (2020).

\bibitem[HAL{\etalchar{*}}20]{difftaichi}
\textsc{Hu Y., Anderson L., Li T.-M., Sun Q., Carr N., Ragan-Kelley J., Durand
  F.}:
\newblock Difftaichi: Differentiable programming for physical simulation.
\newblock \emph{ICLR} (2020).

\bibitem[Har96]{hart1996sphere}
\textsc{Hart J.~C.}:
\newblock Sphere tracing: A geometric method for the antialiased ray tracing of
  implicit surfaces.
\newblock \emph{The Visual Computer 12}, 10 (1996), 527--545.

\bibitem[HCC{\etalchar{*}}14]{deepspeech}
\textsc{Hannun A., Case C., Casper J., Catanzaro B., Diamos G., Elsen E.,
  Prenger R., Satheesh S., Sengupta S., Coates A., Y.~Ng A.}:
\newblock {DeepSpeech}: Scaling up end-to-end speech recognition.

\bibitem[HDD{\etalchar{*}}92]{hoppe1992surface}
\textsc{Hoppe H., DeRose T., Duchamp T., McDonald J., Stuetzle W.}:
\newblock Surface reconstruction from unorganized points.
\newblock \emph{SIGGRAPH} (1992).

\bibitem[HLA{\etalchar{*}}19]{hu2019taichi}
\textsc{Hu Y., Li T.-M., Anderson L., Ragan-Kelley J., Durand F.}:
\newblock Taichi: a language for high-performance computation on spatially
  sparse data structures.
\newblock \emph{ACM Transactions on Graphics (TOG) 38}, 6 (2019), 201.

\bibitem[HPX{\etalchar{*}}21]{hong2021headnerf}
\textsc{Hong Y., Peng B., Xiao H., Liu L., Zhang J.}:
\newblock Headnerf: A real-time nerf-based parametric head model, 2021.
\newblock \href {http://arxiv.org/abs/2112.05637} {\path{arXiv:2112.05637}}.

\bibitem[HRRR18]{henzler18}
\textsc{Henzler P., Rasche V., Ropinski T., Ritschel T.}:
\newblock Single-image tomography: 3d volumes from 2d cranial x-rays.
\newblock In \emph{Eurographics} (2018).

\bibitem[HSM{\etalchar{*}}21]{hedman2021snerg}
\textsc{Hedman P., Srinivasan P.~P., Mildenhall B., Barron J.~T., Debevec P.}:
\newblock Baking neural radiance fields for real-time view synthesis.
\newblock \emph{arXiv} (2021).

\bibitem[HSW89]{HORNIK1989359}
\textsc{Hornik K., Stinchcombe M., White H.}:
\newblock Multilayer feedforward networks are universal approximators.
\newblock \emph{Neural Networks 2}, 5 (1989), 359--366.
\newblock URL:
  \url{https://www.sciencedirect.com/science/article/pii/0893608089900208},
  \href {https://doi.org/https://doi.org/10.1016/0893-6080(89)90020-8}
  {\path{doi:https://doi.org/10.1016/0893-6080(89)90020-8}}.

\bibitem[HYZ{\etalchar{*}}21]{hu2021hvtr}
\textsc{Hu T., Yu T., Zheng Z., Zhang H., Liu Y., Zwicker M.}:
\newblock Hvtr: Hybrid volumetric-textural rendering for human avatars.
\newblock \href {http://arxiv.org/abs/2112.10203} {\path{arXiv:2112.10203}}.

\bibitem[HZF{\etalchar{*}}21]{huang2021hdrnerf}
\textsc{Huang X., Zhang Q., Feng Y., Li H., Wang X., Wang Q.}:
\newblock Hdr-nerf: High dynamic range neural radiance fields.
\newblock \emph{arXiv} (December 2021).

\bibitem[ID18]{insafutdinov2018unsupervised}
\textsc{Insafutdinov E., Dosovitskiy A.}:
\newblock Unsupervised learning of shape and pose with differentiable point
  clouds.
\newblock In \emph{Proceedings of the IEEE International Conference on Neural
  Information Processing Systems (NeurIPS)} (2018), pp.~2802--2812.

\bibitem[IKH{\etalchar{*}}11]{izadi2011kinectfusion}
\textsc{Izadi S., Kim D., Hilliges O., Molyneaux D., Newcombe R., Kohli P.,
  Shotton J., Hodges S., Freeman D., Davison A., Fitzgibbon A.}:
\newblock Kinectfusion: Real-time 3d reconstruction and interaction using a
  moving depth camera.
\newblock In \emph{UIST '11 Proceedings of the 24th annual ACM symposium on
  User interface software and technology} (October 2011), ACM, pp.~559--568.

\bibitem[JA21]{Jang_2021_ICCV}
\textsc{Jang W., Agapito L.}:
\newblock Codenerf: Disentangled neural radiance fields for object categories.
\newblock In \emph{Proceedings of the IEEE/CVF International Conference on
  Computer Vision (ICCV)} (October 2021), pp.~12949--12958.

\bibitem[JAC{\etalchar{*}}21]{Jeong_2021_ICCV_Self_calibrating}
\textsc{Jeong Y., Ahn S., Choy C., Anandkumar A., Cho M., Park J.}:
\newblock Self-calibrating neural radiance fields.
\newblock In \emph{Proceedings of the IEEE/CVF International Conference on
  Computer Vision (ICCV)} (October 2021), pp.~5846--5854.

\bibitem[JAFF16]{perceptual_loss}
\textsc{Johnson J., Alahi A., Fei-Fei L.}:
\newblock Perceptual losses for real-time style transfer and super-resolution.
\newblock In \emph{Computer Vision -- ECCV 2016} (Cham, 2016), Leibe B., Matas
  J., Sebe N., Welling M., (Eds.), Springer International Publishing,
  pp.~694--711.

\bibitem[Jak19]{enoki}
\textsc{Jakob W.}:
\newblock Enoki: structured vectorization and differentiation on modern
  processor architectures, 2019.
\newblock https://github.com/mitsuba-renderer/enoki.

\bibitem[Jar08]{jarosz08thesis}
\textsc{Jarosz W.}:
\newblock \emph{Efficient Monte Carlo Methods for Light Transport in Scattering
  Media}.
\newblock PhD thesis, UC San Diego, September 2008.

\bibitem[JDV{\etalchar{*}}14]{Jensen2014}
\textsc{Jensen R., Dahl A., Vogiatzis G., Tola E., Aanæs H.}:
\newblock Large scale multi-view stereopsis evaluation.
\newblock In \emph{Computer Vision and Pattern Recognition (CVPR)} (2014).

\bibitem[JJHZ20]{jiang2020sdfdiff}
\textsc{Jiang Y., Ji D., Han Z., Zwicker M.}:
\newblock Sdfdiff: Differentiable rendering of signed distance fields for 3d
  shape optimization.
\newblock In \emph{Proceedings of the IEEE/CVF Conference on Computer Vision
  and Pattern Recognition} (2020).

\bibitem[JLF21]{johari2021geonerf}
\textsc{Johari M.~M., Lepoittevin Y., Fleuret F.}:
\newblock Geonerf: Generalizing nerf with geometry priors, 2021.
\newblock \href {http://arxiv.org/abs/2111.13539} {\path{arXiv:2111.13539}}.

\bibitem[JMB{\etalchar{*}}21]{jain2021dreamfields}
\textsc{Jain A., Mildenhall B., Barron J.~T., Abbeel P., Poole B.}:
\newblock Zero-shot text-guided object generation with dream fields.
\newblock \emph{arXiv} (December 2021).

\bibitem[JSM{\etalchar{*}}20]{jiang2020local}
\textsc{Jiang C.~M., Sud A., Makadia A., Huang J., Nie{\ss}ner M., Funkhouser
  T.}:
\newblock Local implicit grid representations for 3d scenes.
\newblock In \emph{Proceedings IEEE Conf. on Computer Vision and Pattern
  Recognition (Proceedings of the IEEE/CVF Conference on Computer Vision and
  Pattern Recognition)} (2020).

\bibitem[JXX{\etalchar{*}}21]{zhang2021stnerf}
\textsc{Jiakai Z., Xinhang L., Xinyi Y., Fuqiang Z., Yanshun Z., Minye W.,
  Yingliang Z., Lan X., Jingyi Y.}:
\newblock Editable free-viewpoint video using a layered neural representation.
\newblock In \emph{ACM SIGGRAPH} (2021).

\bibitem[Kaj86]{kajiya1986rendering}
\textsc{Kajiya J.~T.}:
\newblock The rendering equation.
\newblock In \emph{Proceedings of the 13th annual conference on Computer
  graphics and interactive techniques} (1986), pp.~143--150.

\bibitem[KB04]{KOBBELT2004801}
\textsc{Kobbelt L., Botsch M.}:
\newblock A survey of point-based techniques in computer graphics.
\newblock \emph{Computers and Graphics 28}, 6 (2004), 801--814.
\newblock URL:
  \url{https://www.sciencedirect.com/science/article/pii/S0097849304001487},
  \href {https://doi.org/https://doi.org/10.1016/j.cag.2004.08.009}
  {\path{doi:https://doi.org/10.1016/j.cag.2004.08.009}}.

\bibitem[KB14]{adam}
\textsc{Kingma D.~P., Ba J.}:
\newblock Adam: {A} method for stochastic optimization.
\newblock \emph{CoRR abs/1412.6980} (2014).
\newblock URL: \url{http://arxiv.org/abs/1412.6980}, \href
  {http://arxiv.org/abs/1412.6980} {\path{arXiv:1412.6980}}.

\bibitem[KBS15]{LBF}
\textsc{Kalantari N.~K., Bako S., Sen P.}:
\newblock {A Machine Learning Approach for Filtering Monte Carlo Noise}.
\newblock \emph{ACM Transactions on Graphics (TOG) (Proceedings of SIGGRAPH
  2015) 34}, 4 (2015).

\bibitem[KHM17]{kar17}
\textsc{Kar A., H\"ane C., Malik J.}:
\newblock Learning a multi-view stereo machine.
\newblock In \emph{NeurIPS} (2017).

\bibitem[KIT{\etalchar{*}}21]{kondo2021vaxnerf}
\textsc{Kondo N., Ikeda Y., Tagliasacchi A., Matsuo Y., Ochiai Y., Gu S.~S.}:
\newblock Vaxnerf: Revisiting the classic for voxel-accelerated neural radiance
  field, 2021.
\newblock \href {http://arxiv.org/abs/2111.13112} {\path{arXiv:2111.13112}}.

\bibitem[KJJ{\etalchar{*}}21]{kellnhofer2021neural}
\textsc{Kellnhofer P., Jebe L., Jones A., Spicer R., Pulli K., Wetzstein G.}:
\newblock Neural lumigraph rendering.
\newblock In \emph{CVPR} (2021).

\bibitem[KOC{\etalchar{*}}21]{kuang2021neroic}
\textsc{Kuang Z., Olszewski K., Chai M., Huang Z., Achlioptas P., Tulyakov S.}:
\newblock Neroic: Neural object capture and rendering from online image
  collections.
\newblock In \emph{arXiv} (2021).

\bibitem[KPLD21]{KPLD21}
\textsc{Kopanas G., Philip J., Leimkühler T., Drettakis G.}:
\newblock Point-based neural rendering with per-view optimization.
\newblock \emph{Computer Graphics Forum (Proceedings of the Eurographics
  Symposium on Rendering) 40}, 4 (June 2021).
\newblock URL: \url{http://www-sop.inria.fr/reves/Basilic/2021/KPLD21}.

\bibitem[KSW20]{kohli2020inferring}
\textsc{Kohli A., Sitzmann V., Wetzstein G.}:
\newblock {Semantic Implicit Neural Scene Representations with Semi-supervised
  Training}.
\newblock In \emph{International Conference on 3D Vision (3DV)} (2020).

\bibitem[KSZ{\etalchar{*}}21]{Kosiorek2021}
\textsc{Kosiorek A.~R., Strathmann H., Zoran D., Moreno P., Schneider R.,
  Mokr{\'{a}} S., Rezende D.~J.}:
\newblock {NeRF-VAE: A Geometry Aware 3D Scene Generative Model}.
\newblock URL: \url{http://arxiv.org/abs/2104.00587}, \href
  {http://arxiv.org/abs/2104.00587} {\path{arXiv:2104.00587}}.

\bibitem[KTEM18]{kanazawa2018learning}
\textsc{Kanazawa A., Tulsiani S., Efros A.~A., Malik J.}:
\newblock Learning category-specific mesh reconstruction from image
  collections.
\newblock In \emph{Proceedings of the European Conference on Computer Vision}
  (2018), pp.~371--386.

\bibitem[KUH18]{kato2018neural}
\textsc{Kato H., Ushiku Y., Harada T.}:
\newblock Neural 3{D} mesh renderer.
\newblock In \emph{Proceedings of the IEEE/CVF Conference on Computer Vision
  and Pattern Recognition} (2018), pp.~3907--3916.

\bibitem[KYK{\etalchar{*}}21]{kania2021conerf}
\textsc{Kania K., Yi K.~M., Kowalski M., Trzciński T., Tagliasacchi A.}:
\newblock Conerf: Controllable neural radiance fields, 2021.
\newblock \href {http://arxiv.org/abs/2112.01983} {\path{arXiv:2112.01983}}.

\bibitem[LADL18a]{li2018differentiable}
\textsc{Li T.-M., Aittala M., Durand F., Lehtinen J.}:
\newblock Differentiable monte carlo ray tracing through edge sampling.
\newblock In \emph{ACM Transactions on Graphics (proceedings of ACM SIGGRAPH
  ASIA)} (2018), ACM, p.~222.

\bibitem[LADL18b]{redner}
\textsc{Li T.-M., Aittala M., Durand F., Lehtinen J.}:
\newblock Differentiable monte carlo ray tracing through edge sampling.
\newblock \emph{ACM Trans. Graph. (Proc. SIGGRAPH Asia) 37}, 6 (2018),
  222:1--222:11.

\bibitem[LB14]{loper2014opendr}
\textsc{Loper M.~M., Black M.~J.}:
\newblock Opendr: An approximate differentiable renderer.
\newblock In \emph{Proceedings of the European Conference on Computer Vision}
  (2014), Springer, pp.~154--169.

\bibitem[LFS{\etalchar{*}}21]{Li_2021_ICCV}
\textsc{Li J., Feng Z., She Q., Ding H., Wang C., Lee G.~H.}:
\newblock Mine: Towards continuous depth mpi with nerf for novel view
  synthesis.
\newblock In \emph{International Conference on Computer Vision (ICCV)} (2021).

\bibitem[LGA{\etalchar{*}}18]{halide}
\textsc{Li T.-M., Gharbi M., Adams A., Durand F., Ragan-Kelley J.}:
\newblock Differentiable programming for image processing and deep learning in
  {Halide}.
\newblock \emph{ACM Trans. Graph. (Proc. SIGGRAPH) 37}, 4 (2018),
  139:1--139:13.

\bibitem[LGL{\etalchar{*}}20]{liu2020neural}
\textsc{Liu L., Gu J., Lin K.~Z., Chua T.-S., Theobalt C.}:
\newblock Neural sparse voxel fields.
\newblock \emph{Proceedings of the IEEE International Conference on Neural
  Information Processing Systems (NeurIPS)} (2020).

\bibitem[LH96]{levoy_lightfield_rendering}
\textsc{Levoy M., Hanrahan P.}:
\newblock Light field rendering.
\newblock In \emph{Proceedings of the 23rd Annual Conference on Computer
  Graphics and Interactive Techniques} (New York, NY, USA, 1996), SIGGRAPH '96,
  Association for Computing Machinery, p.~31–42.
\newblock URL: \url{https://doi.org/10.1145/237170.237199}, \href
  {https://doi.org/10.1145/237170.237199} {\path{doi:10.1145/237170.237199}}.

\bibitem[LHL{\etalchar{*}}21]{lyu2021efficient}
\textsc{Lyu L., Habermann M., Liu L., Tewari A., Theobalt C., et~al.}:
\newblock Efficient and differentiable shadow computation for inverse problems.
\newblock \emph{arXiv preprint arXiv:2104.00359} (2021).

\bibitem[LHR{\etalchar{*}}21]{liu2021neural}
\textsc{Liu L., Habermann M., Rudnev V., Sarkar K., Gu J., Theobalt C.}:
\newblock Neural actor: Neural free-view synthesis of human actors with pose
  control.
\newblock \emph{ACM Trans. Graph.(ACM SIGGRAPH Asia)} (2021).

\bibitem[LJR{\etalchar{*}}20]{asha}
\textsc{Li L., Jamieson K., Rostamizadeh A., Gonina E., Ben-Tzur J., Hardt M.,
  Recht B., Talwalkar A.}:
\newblock {A SYSTEM FOR MASSIVELY PARALLEL HYPERPARAMETER TUNING}.
\newblock \emph{MLSys 2} (2020).
\newblock \href {http://arxiv.org/abs/1810.05934v5}
  {\path{arXiv:1810.05934v5}}.

\bibitem[LJRT18]{hyperband}
\textsc{Li L., Jamieson K., Rostamizadeh A., Talwalkar A.}:
\newblock {Hyperband: A Novel Bandit-Based Approach to Hyperparameter
  Optimization}.
\newblock \emph{Journal of Machine Learning Research 18} (2018), 1--52.
\newblock URL: \url{http://jmlr.org/papers/v18/16-558.html.}, \href
  {http://arxiv.org/abs/1603.06560v4} {\path{arXiv:1603.06560v4}}.

\bibitem[LK10]{laine2010efficient}
\textsc{Laine S., Karras T.}:
\newblock Efficient sparse voxel octrees--analysis, extensions, and
  implementation.
\newblock \emph{NVIDIA Corporation 2} (2010).

\bibitem[LK21]{LangeKutz2021}
\textsc{{Lange} H., {Kutz} J.~N.}:
\newblock Fc2t2: The fast continuous convolutional taylor transform with
  applications in vision and graphics.
\newblock \emph{arXiv e-prints} (2021).

\bibitem[LKL18]{lin2018learning}
\textsc{Lin C.-H., Kong C., Lucey S.}:
\newblock Learning efficient point cloud generation for dense 3d object
  reconstruction.
\newblock In \emph{AAAI Conference on Artificial Intelligence} (2018).

\bibitem[LLCL19]{liu2019soft}
\textsc{Liu S., Li T., Chen W., Li H.}:
\newblock Soft rasterizer: A differentiable renderer for image-based 3{D}
  reasoning.
\newblock In \emph{Proceedings of the International Conference on Computer
  Vision} (2019), pp.~7708--7717.

\bibitem[LLN{\etalchar{*}}18]{tune}
\textsc{Liaw R., Liang E., Nishihara R., Moritz P., Gonzalez J.~E., Stoica I.}:
\newblock Tune: A research platform for distributed model selection and
  training.
\newblock \emph{arXiv preprint arXiv:1807.05118} (2018).

\bibitem[LLYX21]{liu2021nelf}
\textsc{Liu C., Li Z., Yuan J., Xu Y.}:
\newblock Nelf: Practical novel view synthesis with neural light field.
\newblock \emph{arXiv preprint arXiv:2105.07112} (2021).

\bibitem[LMR{\etalchar{*}}15]{SMPL2015}
\textsc{Loper M., Mahmood N., Romero J., Pons-Moll G., Black M.~J.}:
\newblock {SMPL}: A skinned multi-person linear model.
\newblock \emph{ACM Trans. Graphics (Proc. SIGGRAPH Asia) 34}, 6 (2015),
  248:1--248:16.

\bibitem[LMTL21]{lin2021barf}
\textsc{Lin C.-H., Ma W.-C., Torralba A., Lucey S.}:
\newblock Barf: Bundle-adjusting neural radiance fields.
\newblock In \emph{IEEE International Conference on Computer Vision ({ICCV})}
  (2021).

\bibitem[LMW21]{lindell2020autoint}
\textsc{Lindell D.~B., Martel J.~N., Wetzstein G.}:
\newblock Autoint: Automatic integration for fast neural volume rendering.
\newblock In \emph{Proceedings of the IEEE/CVF Conference on Computer Vision
  and Pattern Recognition} (2021).

\bibitem[LNSW21]{li2021neural}
\textsc{Li Z., Niklaus S., Snavely N., Wang O.}:
\newblock Neural scene flow fields for space-time view synthesis of dynamic
  scenes.
\newblock In \emph{Proceedings of the IEEE/CVF Conference on Computer Vision
  and Pattern Recognition} (2021), pp.~6498--6508.

\bibitem[LPX{\etalchar{*}}21]{lin2021efficient}
\textsc{Lin H., Peng S., Xu Z., Bao H., Zhou X.}:
\newblock Efficient neural radiance fields with learned depth-guided sampling.
\newblock In \emph{arXiv} (2021).

\bibitem[LSCL19]{liu2019learning}
\textsc{Liu S., Saito S., Chen W., Li H.}:
\newblock Learning to infer implicit surfaces without supervision.
\newblock In \emph{Proceedings of the IEEE/CVF Conference on Computer Vision
  and Pattern Recognition} (2019), pp.~8295--8306.

\bibitem[LSS{\etalchar{*}}19]{Lombardi_2019_NeuralVolumes}
\textsc{Lombardi S., Simon T., Saragih J., Schwartz G., Lehrmann A., Sheikh
  Y.}:
\newblock Neural volumes: Learning dynamic renderable volumes from images.
\newblock \emph{ACM Trans. Graph. 38}, 4 (July 2019), 65:1--65:14.

\bibitem[LSS{\etalchar{*}}21]{Lombardi_2021_MVP}
\textsc{Lombardi S., Simon T., Schwartz G., Zollhoefer M., Sheikh Y., Saragih
  J.}:
\newblock Mixture of volumetric primitives for efficient neural rendering.
\newblock \emph{ACM Trans. Graph. 40}, 4 (July 2021).
\newblock URL: \url{https://doi.org/10.1145/3450626.3459863}, \href
  {https://doi.org/10.1145/3450626.3459863}
  {\path{doi:10.1145/3450626.3459863}}.

\bibitem[LSZ{\etalchar{*}}21]{Li2021}
\textsc{Li T., Slavcheva M., Zollhoefer M., Green S., Lassner C., Kim C.,
  Schmidt T., Lovegrove S., Goesele M., Lv Z.}:
\newblock {Neural 3D Video Synthesis}.
\newblock URL: \url{http://arxiv.org/abs/2103.02597}, \href
  {http://arxiv.org/abs/2103.02597} {\path{arXiv:2103.02597}}.

\bibitem[LTJ18]{liu2018paparazzi}
\textsc{Liu H.-T.~D., Tao M., Jacobson A.}:
\newblock Paparazzi: surface editing by way of multi-view image processing.
\newblock \emph{ACM Transactions on Graphics (proceedings of ACM SIGGRAPH ASIA)
  37}, 6 (2018), 221--1.

\bibitem[LVVPW22]{lindell2021bacon}
\textsc{Lindell D.~B., Van~Veen D., Park J.~J., Wetzstein G.}:
\newblock Bacon: Band-limited coordinate networks for multiscale scene
  representation.
\newblock In \emph{Proceedings of the IEEE/CVF Conference on Computer Vision
  and Pattern Recognition} (2022).

\bibitem[LZ21]{Lassner_pulsar}
\textsc{Lassner C., Zollh\"ofer M.}:
\newblock Pulsar: Efficient sphere-based neural rendering.
\newblock In \emph{IEEE/CVF Conference on Computer Vision and Pattern
  Recognition (CVPR)} (June 2021).

\bibitem[LZBD21]{psdrcuda}
\textsc{Luan F., Zhao S., Bala K., Dong Z.}:
\newblock {Unified Shape and SVBRDF Recovery using Differentiable Monte Carlo
  Rendering}.
\newblock \emph{Computer Graphics Forum 40}, 4 (2021), 101--113.
\newblock URL: \url{https://onlinelibrary.wiley.com/doi/abs/10.1111/cgf.14344},
  \href {https://doi.org/https://doi.org/10.1111/cgf.14344}
  {\path{doi:https://doi.org/10.1111/cgf.14344}}.

\bibitem[LZP{\etalchar{*}}20]{liu2020dist}
\textsc{Liu S., Zhang Y., Peng S., Shi B., Pollefeys M., Cui Z.}:
\newblock Dist: Rendering deep implicit signed distance function with
  differentiable sphere tracing.
\newblock In \emph{Proceedings of the IEEE/CVF Conference on Computer Vision
  and Pattern Recognition} (2020).

\bibitem[LZZ{\etalchar{*}}21]{liu2021editing}
\textsc{Liu S., Zhang X., Zhang Z., Zhang R., Zhu J.-Y., Russell B.}:
\newblock Editing conditional radiance fields.
\newblock In \emph{Proceedings of the IEEE/CVF International Conference on
  Computer Vision (ICCV)} (2021).

\bibitem[Max95]{max_direct_vol_rendering}
\textsc{Max N.}:
\newblock Optical models for direct volume rendering.
\newblock \emph{IEEE Transactions on Visualization and Computer Graphics 1}, 2
  (1995), 99--108.
\newblock \href {https://doi.org/10.1109/2945.468400}
  {\path{doi:10.1109/2945.468400}}.

\bibitem[MBRS{\etalchar{*}}21]{martinbrualla2020nerfw}
\textsc{Martin-Brualla R., Radwan N., Sajjadi M. S.~M., Barron J.~T.,
  Dosovitskiy A., Duckworth D.}:
\newblock {NeRF in the Wild: Neural Radiance Fields for Unconstrained Photo
  Collections}.
\newblock In \emph{Proceedings of the IEEE/CVF Conference on Computer Vision
  and Pattern Recognition} (2021).

\bibitem[MC]{enzyme}
\textsc{Moses W.~S., Churavy V.}:
\newblock {Instead of Rewriting Foreign Code for Machine Learning,
  Automatically Synthesize Fast Gradients}.
\newblock URL: \url{https://enzyme.mit.edu.}

\bibitem[MC10]{Max2010LocalAG}
\textsc{Max N.~L., Chen M.~S.}:
\newblock Local and global illumination in the volume rendering integral.
\newblock In \emph{Scientific Visualization: Advanced Concepts} (2010).

\bibitem[MCL{\etalchar{*}}21]{Meng_2021_ICCV_GNeRF}
\textsc{Meng Q., Chen A., Luo H., Wu M., Su H., Xu L., He X., Yu J.}:
\newblock Gnerf: Gan-based neural radiance field without posed camera.
\newblock In \emph{Proceedings of the IEEE/CVF International Conference on
  Computer Vision (ICCV)} (October 2021), pp.~6351--6361.

\bibitem[MCP{\etalchar{*}}21]{enzymegpu}
\textsc{Moses W.~S., Churavy V., Paehler L., H\"{u}ckelheim J., Narayanan S.
  H.~K., Schanen M., Doerfert J.}:
\newblock Reverse-mode automatic differentiation and optimization of gpu
  kernels via enzyme.
\newblock In \emph{Proceedings of the International Conference for High
  Performance Computing, Networking, Storage and Analysis} (New York, NY, USA,
  2021), SC '21, Association for Computing Machinery.
\newblock URL: \url{https://doi.org/10.1145/3458817.3476165}, \href
  {https://doi.org/10.1145/3458817.3476165}
  {\path{doi:10.1145/3458817.3476165}}.

\bibitem[MESK22]{mueller2022ngp}
\textsc{Müller T., Evans A., Schied C., Keller A.}:
\newblock Instant neural graphics primitives with a multiresolution hash
  encoding, 2022.
\newblock URL:
  \url{https://nvlabs.github.io/instant-ngp/assets/mueller2022instant.pdf}.

\bibitem[MGK{\etalchar{*}}19]{meshry2019neural}
\textsc{Meshry M., Goldman D.~B., Khamis S., Hoppe H., Pandey R., Snavely N.,
  Martin-Brualla R.}:
\newblock Neural rerendering in the wild.
\newblock In \emph{Proceedings of the IEEE/CVF Conference on Computer Vision
  and Pattern Recognition} (2019), pp.~6878--6887.

\bibitem[MHMB{\etalchar{*}}21]{mildenhall2021rawnerf}
\textsc{Mildenhall B., Hedman P., Martin-Brualla R., Srinivasan P., Barron
  J.~T.}:
\newblock Nerf in the dark: High dynamic range view synthesis from noisy raw
  images.
\newblock \emph{arXiv} (December 2021).

\bibitem[MLL{\etalchar{*}}21a]{ma2021deblurnerf}
\textsc{Ma L., Li X., Liao J., Zhang Q., Wang X., Wang J., Sander P.~V.}:
\newblock Deblur-nerf: Neural radiance fields from blurry images.
\newblock \emph{arXiv} (December 2021).

\bibitem[MLL{\etalchar{*}}21b]{martel2021acorn}
\textsc{Martel J.~N., Lindell D.~B., Lin C.~Z., Chan E.~R., Monteiro M.,
  Wetzstein G.}:
\newblock Acorn: Adaptive coordinate networks for neural representation.
\newblock \emph{ACM Trans. Graph. (SIGGRAPH)} (2021).

\bibitem[MON{\etalchar{*}}19]{mescheder2019occupancy}
\textsc{Mescheder L., Oechsle M., Niemeyer M., Nowozin S., Geiger A.}:
\newblock Occupancy networks: Learning 3d reconstruction in function space.
\newblock In \emph{CVPR} (2019).

\bibitem[MPJ{\etalchar{*}}19]{michalkiewicz2019implicit}
\textsc{Michalkiewicz M., Pontes J.~K., Jack D., Baktashmotlagh M., Eriksson
  A.}:
\newblock Implicit surface representations as layers in neural networks.
\newblock In \emph{Proceedings of the International Conference on Computer
  Vision} (2019), pp.~4743--4752.

\bibitem[MSOC{\etalchar{*}}19]{Mildenhall:2019}
\textsc{Mildenhall B., Srinivasan P.~P., Ortiz-Cayon R., Kalantari N.~K.,
  Ramamoorthi R., Ng R., Kar A.}:
\newblock Local light field fusion: Practical view synthesis with prescriptive
  sampling guidelines.
\newblock \emph{ACM Trans. Graph. (SIGGRAPH) 38}, 4 (2019).

\bibitem[MST{\etalchar{*}}20]{Mildenhall_2020_NeRF}
\textsc{Mildenhall B., Srinivasan P.~P., Tancik M., Barron J.~T., Ramamoorthi
  R., Ng R.}:
\newblock Nerf: Representing scenes as neural radiance fields for view
  synthesis.
\newblock In \emph{ECCV} (2020).

\bibitem[NDVZJ19]{mitsuba2}
\textsc{Nimier-David M., Vicini D., Zeltner T., Jakob W.}:
\newblock Mitsuba 2: A retargetable forward and inverse renderer.
\newblock \emph{Transactions on Graphics (Proceedings of SIGGRAPH Asia) 38}, 6
  (Dec. 2019).
\newblock \href {https://doi.org/10.1145/3355089.3356498}
  {\path{doi:10.1145/3355089.3356498}}.

\bibitem[NFS15]{newcombe2015dynamicfusion}
\textsc{Newcombe R.~A., Fox D., Seitz S.~M.}:
\newblock Dynamicfusion: Reconstruction and tracking of non-rigid scenes in
  real-time.
\newblock In \emph{Proceedings of the IEEE Conference on Computer Vision and
  Pattern Recognition (CVPR)} (2015), pp.~343--352.

\bibitem[NG20]{Niemeyer2020dvr}
\textsc{Niemeyer M., Geiger A.}:
\newblock {GIRAFFE: Representing Scenes as Compositional Generative Neural
  Feature Fields}.
\newblock URL: \url{http://arxiv.org/abs/2011.12100}, \href
  {http://arxiv.org/abs/2011.12100} {\path{arXiv:2011.12100}}.

\bibitem[NG21a]{Niemeyer2021campari}
\textsc{Niemeyer M., Geiger A.}:
\newblock {CAMPARI: Camera-Aware Decomposed Generative Neural Radiance Fields}.
\newblock 46--48.
\newblock URL: \url{http://arxiv.org/abs/2103.17269}, \href
  {http://arxiv.org/abs/2103.17269} {\path{arXiv:2103.17269}}.

\bibitem[NG21b]{niemeyer2020giraffe}
\textsc{Niemeyer M., Geiger A.}:
\newblock Giraffe: Representing scenes as compositional generative neural
  feature fields.
\newblock In \emph{Computer Vision and Pattern Recognition (CVPR)} (2021).

\bibitem[NMOG20]{Niemeyer2020CVPR}
\textsc{Niemeyer M., Mescheder L., Oechsle M., Geiger A.}:
\newblock Differentiable volumetric rendering: Learning implicit 3d
  representations without 3d supervision.
\newblock In \emph{CVPR} (2020).

\bibitem[NPLT{\etalchar{*}}19]{nguyen2019hologan}
\textsc{Nguyen-Phuoc T., Li C., Theis L., Richardt C., Yang Y.-L.}:
\newblock Hologan: Unsupervised learning of 3d representations from natural
  images.
\newblock In \emph{Proceedings of the IEEE/CVF International Conference on
  Computer Vision} (2019), pp.~7588--7597.

\bibitem[NSLH21]{Noguchi2021}
\textsc{Noguchi A., Sun X., Lin S., Harada T.}:
\newblock Neural articulated radiance field.
\newblock In \emph{International Conference on Computer Vision (ICCV)} (2021).

\bibitem[NSP{\etalchar{*}}21]{Neff2021}
\textsc{Neff T., Stadlbauer P., Parger M., Kurz A., Mueller J.~H., Chaitanya
  C.~R., Kaplanyan A., Steinberger M.}:
\newblock {DONeRF: Towards Real-Time Rendering of Compact Neural Radiance
  Fields using Depth Oracle Networks}.
\newblock \emph{Computer Graphics Forum 40}, 4 (2021), 45--59.
\newblock \href {http://arxiv.org/abs/2103.03231} {\path{arXiv:2103.03231}},
  \href {https://doi.org/10.1111/cgf.14340} {\path{doi:10.1111/cgf.14340}}.

\bibitem[NZIS13]{niessner2013hashing}
\textsc{Nie{\ss}ner M., Zollh\"ofer M., Izadi S., Stamminger M.}:
\newblock Real-time 3d reconstruction at scale using voxel hashing.
\newblock \emph{ACM Transactions on Graphics (TOG)} (2013).

\bibitem[OELS{\etalchar{*}}21]{orel2021styleSDF}
\textsc{Or-El R., Luo X., Shan M., Shechtman E., Park J.~J.,
  Kemelmacher-Shlizerman I.}:
\newblock Stylesdf: High-resolution 3d-consistent image and geometry
  generation.
\newblock \emph{arXiv preprint arXiv:2112.11427} (2021).

\bibitem[OLN{\etalchar{*}}21]{ost2021neural}
\textsc{Ost J., Laradji I., Newell A., Bahat Y., Heide F.}:
\newblock Neural point light fields.
\newblock \emph{arXiv preprint arXiv:2112.01473} (2021).

\bibitem[OMN{\etalchar{*}}19]{Oechsle2019ICCV}
\textsc{Oechsle M., Mescheder L., Niemeyer M., Strauss T., Geiger A.}:
\newblock Texture fields: Learning texture representations in function space.
\newblock In \emph{ICCV} (2019).

\bibitem[OMT{\etalchar{*}}21]{Ost2021}
\textsc{Ost J., Mannan F., Thuerey N., Knodt J., Heide F.}:
\newblock {Neural Scene Graphs for Dynamic Scenes}.
\newblock In \emph{Conference on Computer Vision and Pattern Recognition
  (CVPR)} (2021).

\bibitem[OPG21]{oechsle2021unisurf}
\textsc{Oechsle M., Peng S., Geiger A.}:
\newblock Unisurf: Unifying neural implicit surfaces and radiance fields for
  multi-view reconstruction.
\newblock \emph{arXiv preprint arXiv:2104.10078} (2021).

\bibitem[PBDCO19]{petersen2019pix2vex}
\textsc{Petersen F., Bermano A.~H., Deussen O., Cohen-Or D.}:
\newblock Pix2vex: Image-to-geometry reconstruction using a smooth
  differentiable renderer.
\newblock \emph{arXiv preprint arXiv:1903.11149} (2019).

\bibitem[PC21]{piala2021terminerf}
\textsc{Piala M., Clark R.}:
\newblock Terminerf: Ray termination prediction for efficient neural rendering,
  2021.
\newblock \href {http://arxiv.org/abs/2111.03643} {\path{arXiv:2111.03643}}.

\bibitem[PCPMMN21]{pumarola2020d}
\textsc{Pumarola A., Corona E., Pons-Moll G., Moreno-Noguer F.}:
\newblock {D-NeRF: Neural Radiance Fields for Dynamic Scenes}.
\newblock In \emph{Proceedings of the IEEE/CVF Conference on Computer Vision
  and Pattern Recognition} (2021).

\bibitem[PD84]{porterduff}
\textsc{Porter T., Duff T.}:
\newblock Compositing digital images.
\newblock \emph{SIGGRAPH Comput. Graph. 18}, 3 (Jan. 1984), 253–259.
\newblock URL: \url{https://doi.org/10.1145/964965.808606}, \href
  {https://doi.org/10.1145/964965.808606} {\path{doi:10.1145/964965.808606}}.

\bibitem[PDW{\etalchar{*}}21]{peng2021animatable}
\textsc{Peng S., Dong J., Wang Q., Zhang S., Shuai Q., Bao H., Zhou X.}:
\newblock Animatable neural radiance fields for human body modeling.
\newblock \emph{arXiv preprint arXiv:2105.02872} (2021).

\bibitem[PFAK20]{poursaeed2020coupling}
\textsc{Poursaeed O., Fisher M., Aigerman N., Kim V.~G.}:
\newblock Coupling explicit and implicit surface representations for generative
  3d modeling.
\newblock In \emph{European Conference on Computer Vision} (2020), Springer,
  pp.~667--683.

\bibitem[PFS{\etalchar{*}}19]{park2019deepsdf}
\textsc{Park J.~J., Florence P., Straub J., Newcombe R., Lovegrove S.}:
\newblock Deepsdf: Learning continuous signed distance functions for shape
  representation.
\newblock \emph{CVPR} (2019).

\bibitem[PGM{\etalchar{*}}19]{pytorch}
\textsc{Paszke A., Gross S., Massa F., Lerer A., Bradbury J., Chanan G.,
  Killeen T., Lin Z., Gimelshein N., Antiga L., Desmaison A., Kopf A., Yang E.,
  DeVito Z., Raison M., Tejani A., Chilamkurthy S., Steiner B., Fang L., Bai
  J., Chintala S.}:
\newblock Pytorch: An imperative style, high-performance deep learning library.
\newblock In \emph{Advances in Neural Information Processing Systems} (2019),
  Wallach H., Larochelle H., Beygelzimer A., d\textquotesingle Alch\'{e}-Buc
  F., Fox E., Garnett R., (Eds.), vol.~32, Curran Associates, Inc.
\newblock URL:
  \url{https://proceedings.neurips.cc/paper/2019/file/bdbca288fee7f92f2bfa9f7012727740-Paper.pdf}.

\bibitem[PNM{\etalchar{*}}20]{peng2020convolutional}
\textsc{Peng S., Niemeyer M., Mescheder L., Pollefeys M., Geiger A.}:
\newblock Convolutional occupancy networks.
\newblock In \emph{European Conference on Computer Vision (Proceedings of the
  European Conference on Computer Vision)} (2020).

\bibitem[PSB{\etalchar{*}}21]{park2021nerfies}
\textsc{Park K., Sinha U., Barron J.~T., Bouaziz S., Goldman D.~B., Seitz
  S.~M., Martin-Brualla R.}:
\newblock Nerfies: Deformable neural radiance fields.
\newblock \emph{ICCV} (2021).

\bibitem[PSDV{\etalchar{*}}18]{perez2018film}
\textsc{Perez E., Strub F., De~Vries H., Dumoulin V., Courville A.}:
\newblock Film: Visual reasoning with a general conditioning layer.
\newblock In \emph{Proceedings of the AAAI Conference on Artificial
  Intelligence} (2018), vol.~32.

\bibitem[PSH{\etalchar{*}}21]{park2021hypernerf}
\textsc{Park K., Sinha U., Hedman P., Barron J.~T., Bouaziz S., Goldman D.~B.,
  Martin-Brualla R., Seitz S.~M.}:
\newblock Hypernerf: A higher-dimensional representation for topologically
  varying neural radiance fields.
\newblock \emph{arXiv preprint arXiv:2106.13228} (2021).

\bibitem[PXL{\etalchar{*}}21]{pan2021shadegan}
\textsc{Pan X., Xu X., Loy C.~C., Theobalt C., Dai B.}:
\newblock A shading-guided generative implicit model for shape-accurate
  3d-aware image synthesis.
\newblock In \emph{Advances in Neural Information Processing Systems (NeurIPS)}
  (2021).

\bibitem[PZvBG00]{pfister2000surfelssurface}
\textsc{Pfister H., Zwicker M., van Baar J., Gross M.}:
\newblock Surfels-surface elements as rendering primitives.
\newblock In \emph{ACM Transactions on Graphics (Proc. ACM SIGGRAPH)} (7/2000
  2000), pp.~335--342.

\bibitem[PZX{\etalchar{*}}21]{Peng_2021_CVPR}
\textsc{Peng S., Zhang Y., Xu Y., Wang Q., Shuai Q., Bao H., Zhou X.}:
\newblock Neural body: Implicit neural representations with structured latent
  codes for novel view synthesis of dynamic humans.
\newblock In \emph{Proceedings of the IEEE/CVF Conference on Computer Vision
  and Pattern Recognition (CVPR)} (June 2021), pp.~9054--9063.

\bibitem[RBM{\etalchar{*}}21]{roessle2021dense}
\textsc{Roessle B., Barron J.~T., Mildenhall B., Srinivasan P.~P., Nießner
  M.}:
\newblock Dense depth priors for neural radiance fields from sparse input
  views.
\newblock \emph{arXiv} (December 2021).

\bibitem[RCV{\etalchar{*}}19]{roessler2019faceforensics++}
\textsc{R\"ossler A., Cozzolino D., Verdoliva L., Riess C., Thies J.,
  Nie{\ss}ner M.}:
\newblock Faceforensics++: Learning to detect manipulated facial images.
\newblock In \emph{ICCV 2019} (2019).

\bibitem[RES{\etalchar{*}}21]{rudnev2021nerfosr}
\textsc{Rudnev V., Elgharib M., Smith W., Liu L., Golyanik V., Theobalt C.}:
\newblock Neural radiance fields for outdoor scene relighting, 2021.
\newblock \href {http://arxiv.org/abs/2112.05140} {\path{arXiv:2112.05140}}.

\bibitem[RFS21a]{rueckert2021adop}
\textsc{R{\"u}ckert D., Franke L., Stamminger M.}:
\newblock Adop: Approximate differentiable one-pixel point rendering.
\newblock \href {http://arxiv.org/abs/2110.06635} {\path{arXiv:2110.06635}}.

\bibitem[RFS21b]{ruckert2021adop}
\textsc{R{\"u}ckert D., Franke L., Stamminger M.}:
\newblock Adop: Approximate differentiable one-pixel point rendering.
\newblock \emph{arXiv preprint arXiv:2110.06635} (2021).

\bibitem[RKH{\etalchar{*}}21]{radford2021learning}
\textsc{Radford A., Kim J.~W., Hallacy C., Ramesh A., Goh G., Agarwal S.,
  Sastry G., Askell A., Mishkin P., Clark J., Krueger G., Sutskever I.}:
\newblock Learning transferable visual models from natural language
  supervision, 2021.
\newblock \href {http://arxiv.org/abs/2103.00020} {\path{arXiv:2103.00020}}.

\bibitem[RL21]{ramasinghe2021unify}
\textsc{Ramasinghe S., Lucey S.}:
\newblock Beyond periodicity: Towards a unifying framework for activations in
  coordinate-mlps.
\newblock \emph{CoRR abs/2111.15135} (2021).
\newblock URL: \url{https://arxiv.org/abs/2111.15135}, \href
  {http://arxiv.org/abs/2111.15135} {\path{arXiv:2111.15135}}.

\bibitem[RLS{\etalchar{*}}21]{rematas2021urban}
\textsc{Rematas K., Liu A., Srinivasan P.~P., Barron J.~T., Tagliasacchi A.,
  Funkhouser T., Ferrari V.}:
\newblock Urban radiance fields, 2021.
\newblock \href {http://arxiv.org/abs/2111.14643} {\path{arXiv:2111.14643}}.

\bibitem[RMBF21]{Rematas2021}
\textsc{Rematas K., Martin-Brualla R., Ferrari V.}:
\newblock {ShaRF: Shape-conditioned Radiance Fields from a Single View}.
\newblock URL: \url{http://arxiv.org/abs/2102.08860}, \href
  {http://arxiv.org/abs/2102.08860} {\path{arXiv:2102.08860}}.

\bibitem[RMG{\etalchar{*}}21]{Richard2021binaural}
\textsc{Richard A., Markovic D., Gebru I.~D., Krenn S., Butler G., de~la Torre
  F., Sheikh Y.}:
\newblock Neural synthesis of binaural speech from mono audio.
\newblock In \emph{International Conference on Learning Representations (ICLR)}
  (2021).

\bibitem[RMY{\etalchar{*}}21]{rebain2021lolnerf}
\textsc{Rebain D., Matthews M., Yi K.~M., Lagun D., Tagliasacchi A.}:
\newblock Lolnerf: Learn from one look.
\newblock \emph{arXiv preprint arXiv:2111.09996} (2021).

\bibitem[ROUG17]{riegler2017octnet}
\textsc{Riegler G., Osman~Ulusoy A., Geiger A.}:
\newblock Octnet: Learning deep 3d representations at high resolutions.
\newblock In \emph{Proceedings of the IEEE conference on computer vision and
  pattern recognition} (2017), pp.~3577--3586.

\bibitem[RPLG21]{Reiser2021}
\textsc{Reiser C., Peng S., Liao Y., Geiger A.}:
\newblock {KiloNeRF: Speeding up Neural Radiance Fields with Thousands of Tiny
  MLPs}.
\newblock URL: \url{http://arxiv.org/abs/2103.13744}, \href
  {http://arxiv.org/abs/2103.13744} {\path{arXiv:2103.13744}}.

\bibitem[RRN{\etalchar{*}}20]{pytorch3d}
\textsc{Ravi N., Reizenstein J., Novotny D., Gordon T., Lo W.-Y., Johnson J.,
  Gkioxari G.}:
\newblock Accelerating 3d deep learning with pytorch3d.
\newblock \emph{arXiv:2007.08501} (2020).

\bibitem[RROG18]{roveri2018network}
\textsc{Roveri R., Rahmann L., Oztireli C., Gross M.}:
\newblock A network architecture for point cloud classification via automatic
  depth images generation.
\newblock In \emph{Proceedings of the IEEE/CVF Conference on Computer Vision
  and Pattern Recognition} (2018), pp.~4176--4184.

\bibitem[RSH{\etalchar{*}}21]{reizenstein21co3d}
\textsc{Reizenstein J., Shapovalov R., Henzler P., Sbordone L., Labatut P.,
  Novotny D.}:
\newblock Common objects in 3d: Large-scale learning and evaluation of
  real-life 3d category reconstruction.
\newblock In \emph{International Conference on Computer Vision} (2021).

\bibitem[RZS{\etalchar{*}}20]{raj2020pva}
\textsc{Raj A., Zollhoefer M., Simon T., Saragih J., Saito S., Hays J.,
  Lombardi S.}:
\newblock Pva: Pixel-aligned volumetric avatars.
\newblock In \emph{arXiv:2101.02697} (2020).

\bibitem[SCT{\etalchar{*}}20]{sitzmann2019metasdf}
\textsc{Sitzmann V., Chan E.~R., Tucker R., Snavely N., Wetzstein G.}:
\newblock Metasdf: Meta-learning signed distance functions.
\newblock In \emph{NeurIPS} (2020).

\bibitem[SDZ{\etalchar{*}}21]{srinivasan2021nerv}
\textsc{Srinivasan P.~P., Deng B., Zhang X., Tancik M., Mildenhall B., Barron
  J.~T.}:
\newblock {NeRV}: Neural reflectance and visibility fields for relighting and
  view synthesis.
\newblock \emph{CVPR} (2021).

\bibitem[SESM21]{suhail2021light}
\textsc{Suhail M., Esteves C., Sigal L., Makadia A.}:
\newblock Light field neural rendering.
\newblock \emph{arXiv preprint arXiv:2112.09687} (2021).

\bibitem[SHN{\etalchar{*}}19]{saito2019pifu}
\textsc{Saito S., Huang Z., Natsume R., Morishima S., Kanazawa A., Li H.}:
\newblock Pifu: Pixel-aligned implicit function for high-resolution clothed
  human digitization.
\newblock In \emph{Proceedings of the International Conference on Computer
  Vision} (2019), pp.~2304--2314.

\bibitem[SK00]{shum2000review}
\textsc{Shum H., Kang S.~B.}:
\newblock Review of image-based rendering techniques.
\newblock In \emph{Visual Communications and Image Processing 2000} (2000),
  vol.~4067, International Society for Optics and Photonics, pp.~2--13.

\bibitem[SLNG20]{Schwarz2020GRAF}
\textsc{Schwarz K., Liao Y., Niemeyer M., Geiger A.}:
\newblock {GRAF: Generative radiance fields for 3D-aware image synthesis}.
\newblock \emph{Advances in Neural Information Processing Systems
  2020-December}, NeurIPS (2020), 1--13.
\newblock \href {http://arxiv.org/abs/2007.02442} {\path{arXiv:2007.02442}}.

\bibitem[SLOD21]{Sucar2021}
\textsc{Sucar E., Liu S., Ortiz J., Davison A.~J.}:
\newblock {iMAP: Implicit Mapping and Positioning in Real-Time}.
\newblock URL: \url{http://arxiv.org/abs/2103.12352}, \href
  {http://arxiv.org/abs/2103.12352} {\path{arXiv:2103.12352}}.

\bibitem[SLPS20]{Schops_2020_CVPR}
\textsc{Schops T., Larsson V., Pollefeys M., Sattler T.}:
\newblock Why having 10,000 parameters in your camera model is better than
  twelve.
\newblock In \emph{Proceedings of the IEEE/CVF Conference on Computer Vision
  and Pattern Recognition (CVPR)} (June 2020).

\bibitem[SMB{\etalchar{*}}20]{sitzmann2020siren}
\textsc{Sitzmann V., Martel J.~N., Bergman A.~W., Lindell D.~B., Wetzstein G.}:
\newblock Implicit neural representations with periodic activation functions.
\newblock In \emph{Conference on Neural Information Processing Systems
  (NeurIPS)} (2020).

\bibitem[SMP{\etalchar{*}}21]{sajjadi2021scene}
\textsc{Sajjadi M.~S., Meyer H., Pot E., Bergmann U., Greff K., Radwan N., Vora
  S., Lucic M., Duckworth D., Dosovitskiy A., et~al.}:
\newblock Scene representation transformer: Geometry-free novel view synthesis
  through set-latent scene representations.
\newblock \emph{arXiv preprint arXiv:2111.13152} (2021).

\bibitem[SP04]{SAINZ2004869}
\textsc{Sainz M., Pajarola R.}:
\newblock Point-based rendering techniques.
\newblock \emph{Computers and Graphics 28}, 6 (2004), 869--879.
\newblock URL:
  \url{https://www.sciencedirect.com/science/article/pii/S0097849304001530},
  \href {https://doi.org/https://doi.org/10.1016/j.cag.2004.08.014}
  {\path{doi:https://doi.org/10.1016/j.cag.2004.08.014}}.

\bibitem[SRF{\etalchar{*}}21]{sitzmann2021lfns}
\textsc{Sitzmann V., Rezchikov S., Freeman W.~T., Tenenbaum J.~B., Durand F.}:
\newblock Light field networks: Neural scene representations with
  single-evaluation rendering.
\newblock In \emph{arXiv} (2021).

\bibitem[SS10]{voxelization}
\textsc{Schwarz M., Seidel H.-P.}:
\newblock Fast parallel surface and solid voxelization on gpus.
\newblock \emph{ACM Trans. Graph. 29}, 6 (Dec. 2010).
\newblock URL: \url{https://doi.org/10.1145/1882261.1866201}, \href
  {https://doi.org/10.1145/1882261.1866201}
  {\path{doi:10.1145/1882261.1866201}}.

\bibitem[SSC21]{sun2021direct}
\textsc{Sun C., Sun M., Chen H.-T.}:
\newblock Direct voxel grid optimization: Super-fast convergence for radiance
  fields reconstruction, 2021.
\newblock \href {http://arxiv.org/abs/2111.11215} {\path{arXiv:2111.11215}}.

\bibitem[SSS06]{phototourism}
\textsc{Snavely N., Seitz S.~M., Szeliski R.}:
\newblock Photo tourism: Exploring photo collections in 3d.
\newblock In \emph{SIGGRAPH Conference Proceedings} (New York, NY, USA, 2006),
  ACM Press, pp.~835--846.

\bibitem[SSSJ20]{Saito2020}
\textsc{Saito S., Simon T., Saragih J., Joo H.}:
\newblock Pifuhd: Multi-level pixel-aligned implicit function for
  high-resolution 3d human digitization.
\newblock In \emph{Computer Vision and Pattern Recognition (CVPR)} (2020).

\bibitem[STB{\etalchar{*}}19]{srinivasan19}
\textsc{Srinivasan P.~P., Tucker R., Barron J.~T., Ramamoorthi R., Ng R.,
  Snavely N.}:
\newblock Pushing the boundaries of view extrapolation with multiplane images.
\newblock In \emph{CVPR} (2019).

\bibitem[STH{\etalchar{*}}19]{sitzmann2019deepvoxels}
\textsc{Sitzmann V., Thies J., Heide F., Nie{\ss}ner M., Wetzstein G.,
  Zollh{\"o}fer M.}:
\newblock Deepvoxels: Learning persistent 3d feature embeddings.
\newblock In \emph{CVPR} (2019).

\bibitem[SWZ{\etalchar{*}}21]{sun2021fenerf}
\textsc{Sun J., Wang X., Zhang Y., Li X., Zhang Q., Liu Y., Wang J.}:
\newblock Fenerf: Face editing in neural radiance fields, 2021.
\newblock \href {http://arxiv.org/abs/2111.15490} {\path{arXiv:2111.15490}}.

\bibitem[SYZR21]{su2021anerf}
\textsc{Su S.-Y., Yu F., Zollhoefer M., Rhodin H.}:
\newblock A-nerf: Surface-free human 3d pose refinement via neural rendering.
\newblock In \emph{Conference on Neural Information Processing Systems
  (NeurIPS)} (2021).

\bibitem[SZW19]{sitzmann2019srns}
\textsc{Sitzmann V., Zollh{\"o}fer M., Wetzstein G.}:
\newblock Scene representation networks: Continuous 3d-structure-aware neural
  scene representations.
\newblock In \emph{NeurIPS} (2019).

\bibitem[TET{\etalchar{*}}20]{thies2020nvp}
\textsc{Thies J., Elgharib M., Tewari A., Theobalt C., Nie{\ss}ner M.}:
\newblock Neural voice puppetry: Audio-driven facial reenactment.
\newblock \emph{ECCV 2020} (2020).

\bibitem[TFT{\etalchar{*}}20]{tewari2020neuralrendering}
\textsc{Tewari A., Fried O., Thies J., Sitzmann V., Lombardi S., Sunkavalli K.,
  Martin-Brualla R., Simon T., Saragih J., Nie{\ss}ner M., Pandey R., Fanello
  S., Wetzstein G., Zhu J.-Y., Theobalt C., Agrawala M., Shechtman E., Goldman
  D.~B., Zollh{\"o}fer M.}:
\newblock State of the art on neural rendering.
\newblock \emph{EG} (2020).

\bibitem[TFT{\etalchar{*}}21]{siggraph_course_21}
\textsc{Tewari A., Fried O., Thies J., Sitzmann V., Lombardi S., Xu Z., Simon
  T., Nie\ss{}ner M., Tretschk E., Liu L., Mildenhall B., Srinivasan P., Pandey
  R., Orts-Escolano S., Fanello S., Guo M., Wetzstein G., Zhu J.-Y., Theobalt
  C., Agrawala M., Goldman D.~B., Zollh\"{o}fer M.}:
\newblock Advances in neural rendering.
\newblock In \emph{ACM SIGGRAPH 2021 Courses} (New York, NY, USA, 2021),
  SIGGRAPH '21, Association for Computing Machinery.
\newblock URL: \url{https://doi.org/10.1145/3450508.3464573}, \href
  {https://doi.org/10.1145/3450508.3464573}
  {\path{doi:10.1145/3450508.3464573}}.

\bibitem[TLY{\etalchar{*}}21]{takikawa2021nglod}
\textsc{Takikawa T., Litalien J., Yin K., Kreis K., Loop C., Nowrouzezahrai D.,
  Jacobson A., McGuire M., Fidler S.}:
\newblock Neural geometric level of detail: Real-time rendering with implicit
  {3D} shapes.
\newblock In \emph{Proceedings of the IEEE/CVF Conference on Computer Vision
  and Pattern Recognition} (2021).

\bibitem[TMW{\etalchar{*}}21]{tancik2020learned}
\textsc{Tancik M., Mildenhall B., Wang T., Schmidt D., Srinivasan P.~P., Barron
  J.~T., Ng R.}:
\newblock Learned initializations for optimizing coordinate-based neural
  representations.
\newblock In \emph{CVPR} (2021).

\bibitem[TRS21]{turki2021meganerf}
\textsc{Turki H., Ramanan D., Satyanarayanan M.}:
\newblock Mega-nerf: Scalable construction of large-scale nerfs for virtual
  fly-throughs, 2021.
\newblock \href {http://arxiv.org/abs/2112.10703} {\path{arXiv:2112.10703}}.

\bibitem[TS20]{tucker2020single}
\textsc{Tucker R., Snavely N.}:
\newblock Single-view view synthesis with multiplane images.
\newblock In \emph{Proceedings of the IEEE/CVF Conference on Computer Vision
  and Pattern Recognition} (2020), pp.~551--560.

\bibitem[TSM{\etalchar{*}}20]{tancik2020fourfeat}
\textsc{Tancik M., Srinivasan P.~P., Mildenhall B., Fridovich-Keil S., Raghavan
  N., Singhal U., Ramamoorthi R., Barron J.~T., Ng R.}:
\newblock Fourier features let networks learn high frequency functions in low
  dimensional domains.
\newblock \emph{NeurIPS} (2020).

\bibitem[TTG{\etalchar{*}}20]{tretschk2020patchnets}
\textsc{Tretschk E., Tewari A., Golyanik V., Zollh{\"o}fer M., Stoll C.,
  Theobalt C.}:
\newblock Patchnets: Patch-based generalizable deep implicit 3d shape
  representations.
\newblock In \emph{European Conference on Computer Vision} (2020), Springer,
  Springer International Publishing, pp.~293--309.

\bibitem[TTG{\etalchar{*}}21]{tretschk2021nonrigid}
\textsc{Tretschk E., Tewari A., Golyanik V., Zollh\"{o}fer M., Lassner C.,
  Theobalt C.}:
\newblock Non-rigid neural radiance fields: Reconstruction and novel view
  synthesis of a dynamic scene from monocular video.
\newblock In \emph{{IEEE} International Conference on Computer Vision ({ICCV})}
  (2021), {IEEE}.

\bibitem[TY20]{Trevithick2020}
\textsc{Trevithick A., Yang B.}:
\newblock {GRF: Learning a General Radiance Field for 3D Representation and
  Rendering}.
\newblock URL: \url{http://arxiv.org/abs/2010.04595}, \href
  {http://arxiv.org/abs/2010.04595} {\path{arXiv:2010.04595}}.

\bibitem[TZEM17]{tulsiani17}
\textsc{Tulsiani S., Zhou T., Efros A.~A., Malik J.}:
\newblock Multi-view supervision for single-view reconstruction via
  differentiable ray consistency.
\newblock In \emph{CVPR} (2017).

\bibitem[TZN19]{thies2019deferred}
\textsc{Thies J., Zollh{\"o}fer M., Nie{\ss}ner M.}:
\newblock Deferred neural rendering: Image synthesis using neural textures.
\newblock \emph{ACM Trans. Graph. 38}, 4 (2019), 1--12.

\bibitem[TZS{\etalchar{*}}16]{thies2016face}
\textsc{Thies J., Zollh{\"o}fer M., Stamminger M., Theobalt C., Nie{\ss}ner
  M.}:
\newblock Face2face: Real-time face capture and reenactment of rgb videos.
\newblock In \emph{Proc. Computer Vision and Pattern Recognition (CVPR), IEEE}
  (2016).

\bibitem[VHM{\etalchar{*}}21]{verbin2021refnerf}
\textsc{Verbin D., Hedman P., Mildenhall B., Zickler T., Barron J.~T.,
  Srinivasan P.~P.}:
\newblock Ref-nerf: Structured view-dependent appearance for neural radiance
  fields.
\newblock In \emph{arXiv} (2021).

\bibitem[VKP{\etalchar{*}}19]{tensorflowgraphics}
\textsc{Valentin J., Keskin C., Pidlypenskyi P., Makadia A., Sud A., Bouaziz
  S.}:
\newblock Tensorflow graphics: Computer graphics meets deep learning.

\bibitem[Vla09]{Vladsinger2009}
\textsc{Vladsinger}:
\newblock Surface control point diagram used in freeform modeling, 2009.
\newblock URL:
  \url{https://en.wikipedia.org/wiki/B-spline##/media/File:Surface_modelling.svg}.

\bibitem[VSP{\etalchar{*}}17]{transformers}
\textsc{Vaswani A., Shazeer N., Parmar N., Uszkoreit J., Jones L., Gomez A.~N.,
  Kaiser L.~u., Polosukhin I.}:
\newblock Attention is all you need.
\newblock In \emph{Advances in Neural Information Processing Systems} (2017),
  Guyon I., Luxburg U.~V., Bengio S., Wallach H., Fergus R., Vishwanathan S.,
  Garnett R., (Eds.), vol.~30, Curran Associates, Inc.
\newblock URL:
  \url{https://proceedings.neurips.cc/paper/2017/file/3f5ee243547dee91fbd053c1c4a845aa-Paper.pdf}.

\bibitem[WBL{\etalchar{*}}21]{wang2020learning}
\textsc{Wang Z., Bagautdinov T., Lombardi S., Simon T., Saragih J., Hodgins J.,
  Zollh{\"{o}}fer M.}:
\newblock {Learning Compositional Radiance Fields of Dynamic Human Heads}.
\newblock \emph{Proceedings of the IEEE Conference on Computer Vision and
  Pattern Recognition} (2021).

\bibitem[WCH{\etalchar{*}}21]{wang2021clipnerf}
\textsc{Wang C., Chai M., He M., Chen D., Liao J.}:
\newblock Clip-nerf: Text-and-image driven manipulation of neural radiance
  fields, 2021.
\newblock \href {http://arxiv.org/abs/2112.05139} {\path{arXiv:2112.05139}}.

\bibitem[WCS{\etalchar{*}}22]{weng2022humannerf}
\textsc{Weng C.-Y., Curless B., Srinivasan P.~P., Barron J.~T.,
  Kemelmacher-Shlizerman I.}:
\newblock Humannerf: Free-viewpoint rendering of moving people from monocular
  video, 2022.
\newblock \href {http://arxiv.org/abs/2201.04127} {\path{arXiv:2201.04127}}.

\bibitem[WGSJ20]{Wiles_2020_CVPR}
\textsc{Wiles O., Gkioxari G., Szeliski R., Johnson J.}:
\newblock Synsin: End-to-end view synthesis from a single image.
\newblock In \emph{Proceedings of the IEEE/CVF Conference on Computer Vision
  and Pattern Recognition} (6 2020).

\bibitem[WLB{\etalchar{*}}21]{wu2021diver}
\textsc{Wu L., Lee J.~Y., Bhattad A., Wang Y., Forsyth D.}:
\newblock Diver: Real-time and accurate neural radiance fields with
  deterministic integration for volume rendering, 2021.
\newblock \href {http://arxiv.org/abs/2111.10427} {\path{arXiv:2111.10427}}.

\bibitem[WLG{\etalchar{*}}17]{wang2017cnn}
\textsc{Wang P.-S., Liu Y., Guo Y.-X., Sun C.-Y., Tong X.}:
\newblock O-cnn: Octree-based convolutional neural networks for 3d shape
  analysis.
\newblock \emph{ACM Transactions On Graphics (TOG) 36}, 4 (2017), 1--11.

\bibitem[WLL{\etalchar{*}}21]{wang2021neus}
\textsc{Wang P., Liu L., Liu Y., Theobalt C., Komura T., Wang W.}:
\newblock Neus: Learning neural implicit surfaces by volume rendering for
  multi-view reconstruction.
\newblock \emph{NeurIPS} (2021).

\bibitem[WLR{\etalchar{*}}21]{wei2021nerfingmvs}
\textsc{Wei Y., Liu S., Rao Y., Zhao W., Lu J., Zhou J.}:
\newblock Nerfingmvs: Guided optimization of neural radiance fields for indoor
  multi-view stereo.
\newblock In \emph{ICCV} (2021).

\bibitem[WPYS21]{Wizadwongsa2021NeX}
\textsc{Wizadwongsa S., Phongthawee P., Yenphraphai J., Suwajanakorn S.}:
\newblock Nex: Real-time view synthesis with neural basis expansion.
\newblock In \emph{Proceedings of the IEEE/CVF Conference on Computer Vision
  and Pattern Recognition} (2021).

\bibitem[WWG{\etalchar{*}}21a]{wang2021nerfsr}
\textsc{Wang C., Wu X., Guo Y.-C., Zhang S.-H., Tai Y.-W., Hu S.-M.}:
\newblock Nerf-sr: High-quality neural radiance fields using super-sampling.
\newblock \emph{arXiv} (December 2021).

\bibitem[WWG{\etalchar{*}}21b]{Wang2021ibrnet}
\textsc{Wang Q., Wang Z., Genova K., Srinivasan P., Zhou H., Barron J.~T., Noah
  R. M.-b., Funkhouser T., Tech C.}:
\newblock {IBRNet : Learning Multi-View Image-Based Rendering}.
\newblock \emph{Proceedings of the IEEE/CVF Conference on Computer Vision and
  Pattern Recognition} (2021), 4690----4699.

\bibitem[WWX{\etalchar{*}}21]{wang2021nerfmm}
\textsc{Wang Z., Wu S., Xie W., Chen M., Prisacariu V.~A.}:
\newblock Ne{RF}$--$: Neural radiance fields without known camera parameters.
\newblock \emph{arXiv preprint arXiv:2102.07064} (2021).

\bibitem[XAS21]{Xu2021}
\textsc{Xu H., Alldieck T., Sminchisescu C.}:
\newblock H-nerf: Neural radiance fields for rendering and temporal
  reconstruction of humans in motion.
\newblock In \emph{Advances in Neural Information Processing Systems (NeurIPS)}
  (2021).

\bibitem[XFYS20]{xu2020ladybird}
\textsc{Xu Y., Fan T., Yuan Y., Singh G.}:
\newblock Ladybird: Quasi-{M}onte {C}arlo sampling for deep implicit field
  based 3{D} reconstruction with symmetry.
\newblock arXiv preprint arXiv:2007.13393, 2020.

\bibitem[XHKK21]{xian2021space}
\textsc{Xian W., Huang J.-B., Kopf J., Kim C.}:
\newblock Space-time neural irradiance fields for free-viewpoint video.
\newblock In \emph{Proceedings of the IEEE/CVF Conference on Computer Vision
  and Pattern Recognition (CVPR)} (2021), pp.~9421--9431.

\bibitem[XPLD21]{xu2021generative}
\textsc{Xu X., Pan X., Lin D., Dai B.}:
\newblock Generative occupancy fields for 3d surface-aware image synthesis.
\newblock In \emph{Advances in Neural Information Processing Systems(NeurIPS)}
  (2021).

\bibitem[XPMBB21]{xie2021fignerf}
\textsc{Xie C., Park K., Martin-Brualla R., Brown M.}:
\newblock Fig-nerf: Figure-ground neural radiance fields for 3d object category
  modelling.
\newblock \emph{arXiv preprint arXiv:2104.08418} (2021).

\bibitem[XPY{\etalchar{*}}21]{xu2021volumegan}
\textsc{Xu Y., Peng S., Yang C., Shen Y., Zhou B.}:
\newblock 3d-aware image synthesis via learning structural and textural
  representations.

\bibitem[XWC{\etalchar{*}}19]{neurips2019_39059724}
\textsc{Xu Q., Wang W., Ceylan D., Mech R., Neumann U.}:
\newblock Disn: Deep implicit surface network for high-quality single-view 3d
  reconstruction.
\newblock In \emph{Proceedings of the IEEE International Conference on Neural
  Information Processing Systems (NeurIPS)} (2019), vol.~32, Curran Associates,
  Inc.

\bibitem[XXH{\etalchar{*}}21]{xiang2021neutex}
\textsc{Xiang F., Xu Z., Hašan M., Hold-Geoffroy Y., Sunkavalli K., Su H.}:
\newblock {NeuTex}: Neural texture mapping for volumetric neural rendering.
\newblock \emph{CVPR} (2021).

\bibitem[XXP{\etalchar{*}}21]{xiangli2021citynerf}
\textsc{Xiangli Y., Xu L., Pan X., Zhao N., Rao A., Theobalt C., Dai B., Lin
  D.}:
\newblock Citynerf: Building nerf at city scale, 2021.
\newblock \href {http://arxiv.org/abs/2112.05504} {\path{arXiv:2112.05504}}.

\bibitem[Yad19]{hydra}
\textsc{Yadan O.}:
\newblock Hydra - a framework for elegantly configuring complex applications.
\newblock Github, 2019.
\newblock URL: \url{https://github.com/facebookresearch/hydra}.

\bibitem[YAK{\etalchar{*}}20]{yifan2020iso}
\textsc{Yifan W., Aigerman N., Kim V.~G., Chaudhuri S., Sorkine-Hornung O.}:
\newblock Neural cages for detail-preserving 3d deformations.
\newblock In \emph{Proceedings of the IEEE/CVF Conference on Computer Vision
  and Pattern Recognition (Proceedings of the IEEE/CVF Conference on Computer
  Vision and Pattern Recognition)} (6 2020).

\bibitem[YFKT{\etalchar{*}}21]{yu2021plenoxels}
\textsc{Yu A., Fridovich-Keil S., Tancik M., Chen Q., Recht B., Kanazawa A.}:
\newblock Plenoxels: Radiance fields without neural networks, 2021.
\newblock \href {http://arxiv.org/abs/2112.05131} {\path{arXiv:2112.05131}}.

\bibitem[YGKL21a]{yariv2021volume_sdf_figure}
\textsc{Yariv L., Gu J., Kasten Y., Lipman Y.}:
\newblock Volume rendering of neural implicit surfaces, 2021.
\newblock \href {http://arxiv.org/abs/2106.12052} {\path{arXiv:2106.12052}}.

\bibitem[YGKL21b]{yariv2021volume}
\textsc{Yariv L., Gu J., Kasten Y., Lipman Y.}:
\newblock Volume rendering of neural implicit surfaces.
\newblock \emph{arXiv preprint arXiv:2106.12052} (2021).

\bibitem[YKG{\etalchar{*}}20]{yoon2020dynamic}
\textsc{Yoon J.~S., Kim K., Gallo O., Park H.~S., Kautz J.}:
\newblock Novel view synthesis of dynamic scenes with globally coherent depths
  from a monocular camera.
\newblock In \emph{Computer Vision and Pattern Recognition (CVPR)} (2020).

\bibitem[YKM{\etalchar{*}}20]{yariv2020multiview}
\textsc{Yariv L., Kasten Y., Moran D., Galun M., Atzmon M., Basri R., Lipman
  Y.}:
\newblock Multiview neural surface reconstruction by disentangling geometry and
  appearance.
\newblock In \emph{NeurIPS} (2020).

\bibitem[YLT{\etalchar{*}}21]{yu2021plenoctrees}
\textsc{Yu A., Li R., Tancik M., Li H., Ng R., Kanazawa A.}:
\newblock {PlenOctrees} for real-time rendering of neural radiance fields.
\newblock In \emph{arXiv} (2021).

\bibitem[YRSH21]{yifan2021geometryconsistent}
\textsc{Yifan W., Rahmann L., Sorkine-Hornung O.}:
\newblock Geometry-consistent neural shape representation with implicit
  displacement fields, 2021.
\newblock \href {http://arxiv.org/abs/2106.05187} {\path{arXiv:2106.05187}}.

\bibitem[YSW{\etalchar{*}}19a]{Yifan2019DSS}
\textsc{Yifan W., Serena F., Wu S., {\"{O}}ztireli C., Sorkine{-}Hornung O.}:
\newblock Differentiable surface splatting for point-based geometry processing.
\newblock \emph{ACM Transactions on Graphics (proceedings of ACM SIGGRAPH ASIA)
  38}, 6 (2019).

\bibitem[YSW{\etalchar{*}}19b]{Yifan:DSS:2019}
\textsc{Yifan W., Serena F., Wu S., {\"{O}}ztireli C., Sorkine{-}Hornung O.}:
\newblock Differentiable surface splatting for point-based geometry processing.
\newblock \emph{ACM Transactions on Graphics (proceedings of ACM SIGGRAPH ASIA)
  38}, 6 (2019).

\bibitem[YTB{\etalchar{*}}21]{yenamandra2021i3dmm}
\textsc{Yenamandra T., Tewari A., Bernard F., Seidel H.-P., Elgharib M.,
  Cremers D., Theobalt C.}:
\newblock i3dmm: Deep implicit 3d morphable model of human heads.
\newblock In \emph{Proceedings of the IEEE/CVF Conference on Computer Vision
  and Pattern Recognition} (2021), pp.~12803--12813.

\bibitem[YWYZ21]{yao2021ddnerf}
\textsc{Yao G., Wu H., Yuan Y., Zhou K.}:
\newblock Dd-nerf: Double-diffusion neural radiance field as a generalizable
  implicit body representation, 2021.
\newblock \href {http://arxiv.org/abs/2112.12390} {\path{arXiv:2112.12390}}.

\bibitem[YYTK21]{yu2020pixelnerf}
\textsc{Yu A., Ye V., Tancik M., Kanazawa A.}:
\newblock pixelnerf: Neural radiance fields from one or few images.
\newblock In \emph{Proceedings of the IEEE/CVF Conference on Computer Vision
  and Pattern Recognition} (2021).

\bibitem[ZAC{\etalchar{*}}21]{zheng2021i}
\textsc{Zheng Y., Abrevaya V.~F., Chen X., Bühler M.~C., Black M.~J., Hilliges
  O.}:
\newblock I m avatar: Implicit morphable head avatars from videos, 2021.
\newblock \href {http://arxiv.org/abs/2112.07471} {\path{arXiv:2112.07471}}.

\bibitem[ZLLD21]{zhi2021place}
\textsc{Zhi S., Laidlow T., Leutenegger S., Davison A.~J.}:
\newblock In-place scene labelling and understanding with implicit scene
  representation.
\newblock \emph{Proc. ICCV} (2021).

\bibitem[ZLW{\etalchar{*}}21]{zhang2021physg}
\textsc{Zhang K., Luan F., Wang Q., Bala K., Snavely N.}:
\newblock {PhySG}: Inverse rendering with spherical gaussians for physics-based
  material editing and relighting.
\newblock \emph{CVPR} (2021).

\bibitem[ZPVBG01]{zwicker2001surface}
\textsc{Zwicker M., Pfister H., Van~Baar J., Gross M.}:
\newblock Surface splatting.
\newblock In \emph{Proc. Conf. on Computer Graphics and Interactive techniques}
  (2001), ACM, pp.~371--378.

\bibitem[ZPVBG02]{zwicker2002ewa}
\textsc{Zwicker M., Pfister H., Van~Baar J., Gross M.}:
\newblock Ewa splatting.
\newblock \emph{IEEE Transactions on Visualization and Computer Graphics 8}, 3
  (2002), 223--238.

\bibitem[ZRSK20]{zhang2020nerf}
\textsc{Zhang K., Riegler G., Snavely N., Koltun V.}:
\newblock Nerf++: Analyzing and improving neural radiance fields.
\newblock \emph{arXiv preprint arXiv:2010.07492} (2020).

\bibitem[ZSD{\etalchar{*}}21]{zhang2021nerfactor}
\textsc{Zhang X., Srinivasan P.~P., Deng B., Debevec P., Freeman W.~T., Barron
  J.~T.}:
\newblock {NeRFactor}: Neural factorization of shape and reflectance under an
  unknown illumination.
\newblock \emph{SIGGRAPH Asia} (2021).

\bibitem[ZTF{\etalchar{*}}18]{zhou2018}
\textsc{Zhou T., Tucker R., Flynn J., Fyffe G., Snavely N.}:
\newblock Stereo magnification: Learning view synthesis using multiplane
  images.
\newblock \emph{ACM Trans. Graph. (SIGGRAPH)} (2018).

\bibitem[ZXNT21]{zhou2021cips3d}
\textsc{Zhou P., Xie L., Ni B., Tian Q.}:
\newblock Cips-3d: A 3d-aware generator of gans based on
  conditionally-independent pixel synthesis, 2021.
\newblock \href {http://arxiv.org/abs/2110.09788} {\path{arXiv:2110.09788}}.

\bibitem[ZYQ21]{zhang2021learning}
\textsc{Zhang J., Yao Y., Quan L.}:
\newblock Learning signed distance field for multi-view surface reconstruction.
\newblock \emph{arXiv preprint arXiv:2108.09964} (2021).

\bibitem[ZYZ{\etalchar{*}}21]{zhao2021humannerf}
\textsc{Zhao F., Yang W., Zhang J., Lin P., Zhang Y., Yu J., Xu L.}:
\newblock Humannerf: Generalizable neural human radiance field from sparse
  inputs, 2021.
\newblock \href {http://arxiv.org/abs/2112.02789} {\path{arXiv:2112.02789}}.

\bibitem[ZZSC21]{zhuang2021mofanerf}
\textsc{Zhuang Y., Zhu H., Sun X., Cao X.}:
\newblock Mofanerf: Morphable facial neural radiance field, 2021.
\newblock \href {http://arxiv.org/abs/2112.02308} {\path{arXiv:2112.02308}}.

\end{thebibliography}

\end{document}